\documentclass[preprint,12pt,authoryear]{elsarticle}
\usepackage[a4paper,outer=1.4cm,inner=1.4cm,top=1.25cm,bottom=1.35cm,includeheadfoot]{geometry}

\usepackage{amsthm}
\usepackage{amsfonts}
\usepackage{amsmath}
\usepackage{t1enc}
\usepackage{textcomp}
\usepackage{amssymb}
\usepackage[figuresright]{rotating}
\usepackage{subfig}
\usepackage{float}
\usepackage{graphicx,afterpage,rotating}

\def\vec#1{\ensuremath{\mathchoice
                     {\mbox{\boldmath$\displaystyle\mathbf{#1}$}}
                     {\mbox{\boldmath$\textstyle\mathbf{#1}$}}
                     {\mbox{\boldmath$\scriptstyle\mathbf{#1}$}}
                     {\mbox{\boldmath$\scriptscriptstyle\mathbf{#1}$}}}}

\def\tens#1{\relax\ifmmode\mathsf{#1}\else\textsf{#1}\fi}

\newcommand\sym{\operatorname*{sym}}
\newcommand\sgn{\operatorname*{sgn}}
\newcommand\conv{\operatorname*{conv}}

\renewcommand\div{\operatorname*{div}}
\newcommand\Span{\operatorname*{span}}

\newcommand{\R}{\mathbb{R}}
\newcommand{\N}{\mathbb{N}}
\newcommand{\alphaII}{\ensuremath{{\vec \alpha}^\text{I\!I}}}

\usepackage{color}
\usepackage[normalem]{ulem} 

\newcommand{\Replace}[2]{#2}

\newcommand{\Delete}[1]{}

\newcommand{\Insert} [1]{#1}

\newcommand{\Figref}[1]{Fig.~\ref{#1}}
\newcommand{\secref}[1]{Sect.~\ref{#1}}

\journal{International Journal of Plasticity}

\begin{document}

\begin{frontmatter}



\title{A link between microstructure evolution and macroscopic response in
  elasto-plasticity: formulation and numerical approximation 
  of the higher-dimensional Continuum Dislocation Dynamics theory}

\author[FAU]{Stefan Sandfeld\corref{cor1}}
\cortext[cor1]{Corresponding author.}
\ead{stefan.sandfeld@fau.de}
\ead[url]{http://www.matsim.techfak.uni-erlangen.de}
\address[FAU]{Institute for Materials Simulation, 
              Department of Materials Science,
              Friedrich-Alexander University Erlangen-N\"urnberg (FAU), 
              Dr.-Mack-Str. 77, 90762 F\"urth, Germany}

\author[KIT]{Ekkachai Thawinan}
\author[KIT]{Christian Wieners}
\address[KIT]{Institute for Applied and Numerical Mathematics, 
              Karlsruhe Institute of Technology (KIT), 
              Kaiserstr. 12, 
              76128 Karlsruhe, Germany}

  \begin{abstract}\footnotesize
    Micro-plasticity theories and models are suitable to explain and predict
    mechanical response of devices on length scales where the influence of the
    carrier of plastic deformation -- the dislocations -- cannot be neglected or
    completely averaged out. To consider these effects without 
    resolving each single dislocation a large variety of continuum descriptions
    has been developed, amongst which the higher-dimensional continuum
    dislocation dynamics (hdCDD) theory by Hochrainer et al. (Phil.~Mag.~87,
    pp. 1261-1282) takes a different, statistical approach and contains
    information that are usually only contained in 
    discrete dislocation models.
    We present a concise formulation of hdCDD in a general single-crystal
    plasticity context together with a discontinuous Galerkin scheme for the
    numerical implementation which we evaluate by numerical examples: a
    thin film under tensile and shear loads. We study the influence of different
    realistic boundary conditions and demonstrate that dislocation fluxes and
    their lines' curvature are key features in small-scale plasticity.
  \end{abstract}
\end{frontmatter}


\section{Introduction}

Plastic deformation of metals has been utilized by man since the copper age
and the knowledge of how to process metals (by e.g.\ forging) has constantly
grown. However, the physical mechanisms underlying the empirical procedures
could not be understood until the crystalline structure of metals was
investigated by Rutherford at the beginning of the 20th century. Subsequent
attempts to explain the discrepancy between the theoretically predicted shear
strength of a metal and the experimentally observed yield stresses lead to the
concept of the {\lq}dislocation{\rq} -- a linear crystal defect -- which was
proposed in the 1930s independently by \citet{Orowan1934_ZPhys_p605},
\citet{Polanyi1934_ZPhys_p660} and \citet{Taylor1934_ProcRoySocA_p362}.  Two
decades later \citet{Kondo1952_Proc2JapanNatCongressofApplMech_p41},
\citet{Nye1953_ActaMetall}, \citet{Bilby1955_ProcRoySocLondonSerA_p263} and
\citet{Kroner1958} independently introduced equivalent measures for the
average plastic deformation state of a crystal in the form of a second-rank
dislocation density tensor. This 'Kr\"oner-Nye tensor' is introduced to link
the microscopically discontinuous to a macroscopically continuous deformation
state and is the fundamental quantity in Kr\"oner's continuum theory of
dislocations. It has been used widely until today (see e.g. the related works
in \citet{Acharya2012,Taupin2013370,Le2014164}).  This tensor, however, only
captures inhomogeneous plastic deformation states associated with so-called
geometrically necessary dislocations (GNDs) and does not account for the
accumulation of so-called statistically stored dislocations (SSDs) in
homogeneous plasticity. This limits the applicability of the classical
dislocation density measure within continuum theories of plasticity.

Phenomenological continuum models for plasticity which are not based on
dislocation mechanics have been successful in a wide range of engineering
applications. They operate on length scales where the properties of materials
and systems are scale invariant. The scale-invariance, however, breaks down at
dimensions below a few micro-meters, which is also a scale of growing
technological interest. These microstructural effects become more and more
pronounced in small systems and lead to so-called 'size effects'
\citep[e.g.][]{Ashby1970_PhilMag_p399,
  Arzt1998_ActaMater_p5611,Stolken1998_ActaMater_p5109, Greer:2011bb}.
Phenomenological continuum theories incorporate internal length scales by
introducing strain gradient terms -- sometimes based on the consideration of
GND densities -- into their constitutive equations
\citep[e.g.][]{Fleck1994_ActaMetallMater_p475,
  Nix1998_JMechandPhysSolids_p411, Gurtin2002_JMPS50_p5,
  Gao2003_ScrMater_p113, Zhang201438} but are not able to consider fluxes of
dislocations or the conversion of SSDs into GNDs and vice versa. Refined
continuum formulations (with or without strain gradients) include additional
information in a mechanism-based approach \citep{Engels2012159,Li20143} or
take multi-scale approaches by directly including information from lower scale
models \citep{Wallin20083167,Xiong2012899}.

Discrete dislocation dynamics (DDD) models \citep[e.g.][]{Kubin1992,
  Devincre1997,fivel1997,Ghoniem2000,weygand2002,bulatov2002,
  arsenlis2007,Zhou2010,Po:2014tc} contain very detailed information
about the dislocation microstructure and the interaction and evolution of
dislocations and have been successful over the last two decades in
predicting plasticity at the micro-meter scale. DDD simulations allow to
investigate complex plastic deformation mechanisms but are, however, due
to their high computational cost limited to small system
sizes/small densities. 

A different approach which is closely related to DDD and which
generalizes the classical continuum theory of dislocations was undertaken by
Groma et al.\
\citep{Groma1997_PhysRevB_p5807,Groma2003_ActaMater}. They used methods from
statistical physics to describe systems of positive and negative straight edge
dislocations in analogy to densities of charged point particles. The resulting
evolution equations are able to faithfully describe fluxes of
signed edge dislocations \Insert{and the conversion of SSDs into GNDs (and vice versa)}. 
This approach has been successfully used by a number of groups  \citep[e.g.][]{Yefimov2004_JMechandPhysSolids52,Kratochvil2007_PhysRevB,Hirschberger2011_MSMSE19,Scardia2014_JMPS70}.
A generalization to systems of curved dislocation loops, however, is not straightforward. Pioneering steps into that direction have been undertaken by \citet{Kosevich79, El-Azab2000_PhysRevB,Sedlacek2003_PhilMag}. Furthermore, 'screw-edge' representations have been introduced as an approximation by  \citet{Arsenlis2004_JMPS52, Zaiser2006_ScriptaMater_p717, Reuber2014333} and \citet{Leung20151}; 
\Insert{Xiang and co-workers developed a model for the evolution of curved systems of geometrically necessary dislocations \citep{Xiang2009728}. Their model also includes line tension effects and was e.g. applied to model Frank-Read sources \citep{Zhu201419}}.
A new approach based on statistical averages of differential geometrical formulations of dislocation
lines has been done by Hochrainer 
\citep{Hochrainer2006_Dissertation,Hochrainer2007_PhilMag,sandfeld_etal10}
who generalized the statistical approach of Groma towards systems of
dislocations with arbitrary line orientation and line curvature introducing
the higher-dimensional \emph{Continuum Dislocation Dynamics} (hdCDD) theory.
The key idea of hdCDD is based on mapping spatial, parameterized dislocation
lines into a higher-dimensional configuration space, which contains the local
line orientation as additional information. This is particularly useful in situations with complex microstructure \citep{sandfeld_etal10, Sandfeld2015_MRS}. 
In order to avoid the high
computational cost of the higher-dimensional configuration space, 'integrated'
variants of hdCDD -- denoted by CDD -- have also been developed recently
 and their simplifying assumptions already have been benchmarked for a number of
situation \citep{Hochrainer2009_ICNAAM,Sandfeld_JMR,Hochrainer2013_JMPS,Zaiser2014_ModSimMat,Monavari2014}. Furthermore, very recently one CDD variant has been coupled to a strain gradient plasticity model \citep{Wulfinghoff2015} and was used to study size effects of a composite material. Until now hdCDD nonetheless serves as reference method for all CDD formulations, since it can be considered as an almost exact continuum
representation of ensembles of curved dislocations and allows to access many relevant microstructural information.\\

In this article, we derive and formulate the governing equations for hdCDD in
a general crystal plasticity context. We start by introducing the formulation
for single-crystal plasticity and the elasto-plastic boundary value problem
and connect this to the dislocation system in a staggered scheme.  The
evolution of the dislocation density and the dislocation curvature density is
computed in representative slip planes depending on the dislocation
velocity. This is approximated by a discontinuous Galerkin \Insert{(DG)}
scheme introduced in Sect.~\ref{sec:dG} with a local Fourier ansatz suitable
for the higher dimensional configuration space.  Numerical examples in
Sect.~\ref{sec:numerics} -- tension and shearing of a thin film -- demonstrate
the influence of passivated and non-passivated surfaces on the microstructure
evolution. Furthermore we study the influence of the lines' curvature on the
plastic deformation behavior and also on the stress-strain response of the
specimen and compare with the formulation of Groma.

\section{Crystal elasto-plasticity based on the hdCDD theory}
\label{hdCDD}

In this section we introduce the elastic eigenstrain problem and its
variational formulation (\secref{sec:elastic} and \secref{sec:variational})
and two different models for representing slip planes
(\secref{sec:intro_phys_SP_model}) in a continuum framework. Then we outline
the hdCDD theory extending the classical quantities from Kr\"oner's continuum
theory: the governing equations for the evolution of dislocation density and
curvature density as defined by the higher dimensional continuum dislocation
dynamics theory are introduced in Sect.~\ref{sec:Hochrainer_theory}; our model
for the dislocation velocity governing the dislocation dynamics is determined
by the constitutive setting explained in Sect.~\ref{sec:velocity}. 

\subsection{A continuum model for single-crystal plasticity}
\label{sec:elastic}

Let the reference configuration $\mathcal B$ be a bounded Lipschitz
domain in $\mathbb R^3$ and let $\partial_\text{D} \mathcal
B\cup \partial_\text{N} \mathcal B = \partial \mathcal B$ be
a non-overlapping decomposition into Dirichlet boundary
$\partial_\text{D} \mathcal B$ and Neumann boundary $\partial_\text{N}
\mathcal B$. The position of a material point is denoted by $\vec x$
and the displacement of the body from its reference configuration by
$\vec u(\vec x, t)$.
The \Insert{distortion} tensor 
\begin{eqnarray}
  \mathrm D\vec u &=& \vec \beta^\text{el} + \vec \beta^\text{pl}
\end{eqnarray}
is decomposed additively
into elastic and plastic parts $\vec \beta^\text{el}$ and $\vec
\beta^\text{pl}$, respectively.
Small deformations are assumed so that the infinitesimal strain is given by
\begin{eqnarray}
  \vec \varepsilon = \vec \varepsilon(\vec u) = \sym (\mathrm D\vec u)
  \,.
\end{eqnarray}

Plastic slip is assumed to take place on $N$ slip systems determined
by a unit normal $\vec m_s$ and slip direction $\vec d_s= \frac1{b_s}
\vec b_s$ on the $s$-th system, where $\vec b_s$ is the Burgers vector
of length $b_s = |\vec b_s|$.
As an example, the face-centered cubic (FCC) crystal has $N=12$ slip
systems. In special situations symmetry can be exploited and the case
$N=1$ or $N=2$ can be considered over the crystal of thin film via
shearing and bending situations.

The plastic shear strain in the slip system $s$ is denoted by
$\gamma_s$.  In the single crystal, we assume that the plastic part of
the displacement gradient is given by the sum over the contributions
from all active slip systems
\begin{eqnarray}
  \vec \beta^\text{pl} &=& \sum\nolimits_s \gamma_s \vec M_s
\end{eqnarray}
where $\vec M_s = \frac1{b_s} \vec b_s \otimes \vec m_s = \vec d_s
\otimes \vec m_s $ is the projection tensor accounting for the
orientation of the slip system~$s$.
Depending on the vector of plastic shear strains $\vec \gamma =
(\gamma_1,\dots,\gamma_N)^\top$, the plastic strain is given by
\begin{eqnarray}
  \vec \varepsilon^\text{pl} = \vec \varepsilon^\text{pl}(\vec \gamma) 
  = \text{sym}(\vec \beta^\text{pl})
  =  \sum\nolimits_s \gamma_s \vec M_s^\text{sym} 
  \,,
  \qquad
  \vec M_s^\text{sym} = \sym (\vec M_s)
  \,.
\end{eqnarray}
This defines the elastic strain
\begin{eqnarray}
  \vec \varepsilon^\text{el} 
  &=& \vec \varepsilon^\text{el}(\vec u, \vec \gamma)
  \quad = \quad \vec \varepsilon(\vec u) 
  - \vec \varepsilon^\text{pl}(\vec \gamma)
  \,.
\end{eqnarray}

\subsection{Variational balance equations}
\label{sec:variational}

The macroscopic equilibrium equation is given by
\begin{eqnarray}
  -\div \vec \sigma &=& \vec f_{\mathcal B} 
  \quad \text{ in } \, \mathcal B\,,
\end{eqnarray}
with \Insert{the body force $\vec f_{\mathcal B}$ and} the constitutive relation for the Cauchy stress tensor 
\begin{eqnarray}
  \vec \sigma = \mathbb C : \vec \varepsilon^\text{el} 
  = \mathbb C :(\vec\varepsilon - \vec\varepsilon^\text{pl})
\end{eqnarray}
depending on the elasticity tensor $\mathbb C$.  The macroscopic boundary
conditions are
\begin{eqnarray}
  \vec u = \vec u_\text{D} 
  \ \text{ on } \ \partial_\text{D} \mathcal B, 
  &\qquad& 
  \vec \sigma \vec n = \vec g_\text{N} \ \text{ on } \ 
  \partial_\text{N} \mathcal B\,,
\end{eqnarray}
\Insert{where $\vec u_\text{D}$ is a prescribed displacement for Dirichlet boundary and $\vec g_\text{N}$ is an applied traction for Neumann boundary.}
Let $U = \big\{ \vec u \in \mathrm H^1(\mathcal B,\mathbb R^3)\colon
\vec u = \vec 0$ on $\partial_\text{D} \mathcal B\big\}$, and assume
that $\vec u_\text{D} $ extends to $\mathcal B$. Then, 
for given $\vec\varepsilon^\text{pl}=\vec\varepsilon^\text{pl}(\vec \gamma)$ , 
we have in weak form: find $\vec u\in \vec u_\text{D} + U$ such that
\begin{eqnarray}
  \label{eq:linear}
  \int_{\mathcal B}
  \big(\vec\varepsilon (\vec u) - \vec\varepsilon^\text{pl}\big)
  : \mathbb C : \vec\varepsilon (\delta \vec u)\, \mathrm d \vec x
  = 
  \int_{\mathcal B} \vec f_{\mathcal B}  \cdot \delta \vec u \, \mathrm d \vec x
  + \int_{\partial_\text{N}\mathcal B}  
  \vec g_\text{N} \cdot \delta \vec u \, \mathrm d \vec a
  \,,
  \qquad
  \delta \vec u \in U
  \,.
\end{eqnarray}
This is complemented by an evolution equation for the plastic shear strain
(the Orowan relation \citep{Orowan1940_ProcPhysSoc52}) depending on the 
dislocation density and the dislocation velocity.

\subsection{Dislocation densities, plastic shear strain, and the representation of physical slip planes}
\label{sec:intro_phys_SP_model}

Since continuum dislocation theories operate with averaged measures one has
to consider slip planes in a consistent sense (i.e., consistent with the 
averaging). In the slip system $s$, we use a discrete set of
'crystallographic' slip planes (SP) of distance ${\vartriangle} s>0$
\begin{eqnarray}
  \Gamma_{s,g} = \big\{ \vec z_{s,g} + \eta
  \vec d_s + \xi \vec d_s \times \vec m_s\colon (\eta,\xi)\in\mathbb
  R^2\big\}
  \,,
\end{eqnarray} 
where $\vec z_{s,g}\in \Gamma_{s,g }$ denotes the origin of the local
$(\eta,\xi)$ coordinate system which is aligned such that the Burgers
vector $\vec b_s$ points into positive $\eta$ direction and $\xi$
points into the line direction of a positive edge dislocation; a position
in the slip plane is denoted by $\vec r\in \Gamma_{s,g }$.  Each SP
is expanded to a thin layer of width $h\leq{\vartriangle} s$
(collecting a small number of physical slip planes)
\begin{eqnarray}
  \mathcal{B}_{s,g} =
  \Bigl\{ \vec z \in \mathcal{B}\colon 
  \vec z = \vec r + \zeta\vec m_s \text{ with }
  \vec r \in \Gamma_{s,g} \text{ and }|\zeta|\leq h/2 \Big\}      
  \,.
\end{eqnarray}
In our model, the dislocation density in the slip system $s$ is represented 
by the average $\rho_{s,g}$ in the layer $\mathcal{B}_{s,g}$ such that the sum of all $\rho_{s,g}$ integrated over each layer equals the total dislocation line length in $\mathcal{B}$.
\begin{figure}
	\centering
   \footnotesize
	\begin{minipage}{0.35\textwidth}
    {\bf Experimental observation}
    of surface slip traces. This illustrative example shows   
    a plastically deformed (cadmium) single crystal 
    (height $\approx 800$\textmu m). \\
    \phantom{.}\hfill {\scriptsize Permission granted by DoITPoMS}
	\end{minipage}
	\hfill
	\begin{minipage}{0.64\textwidth}
	\centering\qquad
	\includegraphics[viewport=0 50 900 660, clip, height= 0.24\textwidth,width=0.42\textwidth]{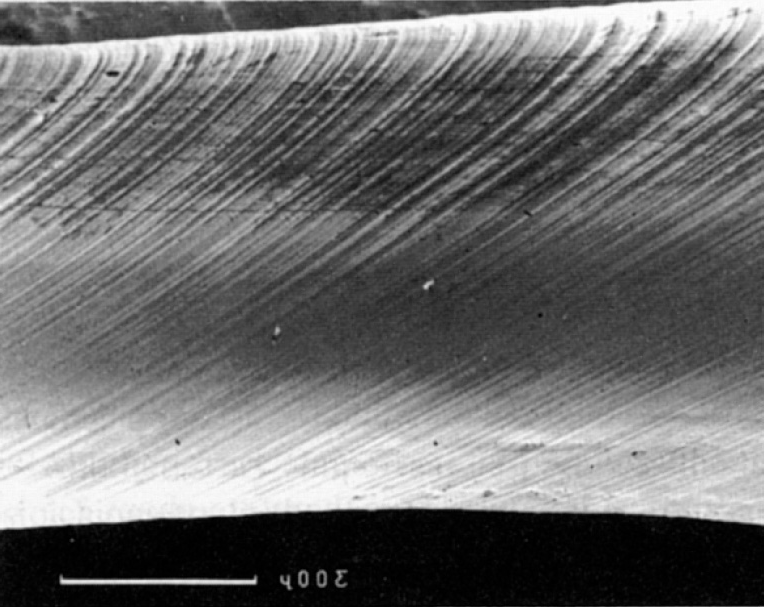}
	\end{minipage}
	\vspace{5mm}

	\begin{minipage}{0.35\textwidth}
    The \textbf{physical model}
    idealizes the real specimen:  The 'crystallographic' 
    slip planes  are evenly spaced  with distance ${\vartriangle} s$
    which is typically much smaller than the simulated distance in 
    the numerical model below.
	\end{minipage}
	\hfill
	\begin{minipage}{0.64\textwidth}
	\centering
	\includegraphics[width=0.4\textwidth]{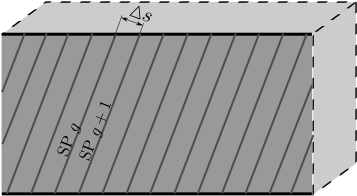}
	\end{minipage}
	\vspace{4mm}
	
	\begin{minipage}{0.35\textwidth}
    \textbf{Numerical model}\\
    The 'numerical slip planes'  represent a  number of
    physical slip planes either in a quasi-discrete setting (Case~1, left)
    or in a fully continuous setting (Case~2, right).
	\end{minipage}
	\hfill
	\begin{minipage}{0.64\textwidth}
	\centering
	\includegraphics[width=0.4\textwidth]{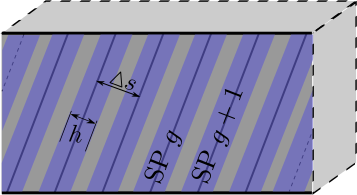}
	\includegraphics[width=0.4\textwidth]{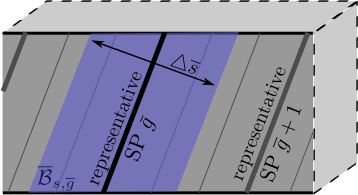}\\
	\hspace{5mm}  Case~1: direct SP representation \hspace{2mm}
                 Case~2: averaged SP representation \hspace{1cm}\hbox{}
	\end{minipage}
  \caption{\label{fig:SPs}From the real system to the numerical model:
  representation of crystallographic slip planes for one slip system. }
\end{figure}
Since the evolution of the dislocation density $\rho_{s,g}$ and
the Orowan relation of the plastic shear strain are evaluated only in
the crystallographic slip planes $\Gamma_{s,g}$, the continuum
approach requires to extend the values to the body $\mathcal B$. For
this purpose, we introduce the orthogonal projection $P_{s,g}\colon
\mathcal{B}\longrightarrow \Gamma_{s,g}$, and for $\vec r \in
\Gamma_{s,g}\setminus\mathcal{B}$ the plastic shear
strain $\gamma_{s,g}$ is extended by constant continuation.  We consider two cases \Insert{(compare Fig. \ref{fig:SPs})}:

\begin{description}
\item[Case 1: Direct representation of crystallographic
  SPs.]${}$\newline We set
  \begin{eqnarray}
    \gamma_s(\vec r) 
    &=&
    \begin{cases}
     \frac{{\vartriangle}s}{h} \gamma_{s,g}({\vec r}) & \vec r \in \mathcal{B}_{s,g}
      \text{ for some } g\,,\\
      0 & \text{else} \,
    \end{cases}
  \end{eqnarray}	 
  where the factor of ${{\vartriangle}s}/{h}$ accounts for the fact that the
  plastic strain will be defined  for the elastic BVP only in the blue regions
  in Case~1. The objective of Case~1 is to analyze how to choose
  ${\vartriangle} s$ and $h$ for our density-based micro-structure
  representation. \Insert{This case also serves as a reference for Case~2 and was set up following the idea of discrete dislocation dynamics  simulations: for the limit $h\rightarrow 0$ we can approach a discrete SP representation; this provides benchmark data for Case~2.}

\item[Case 2: Averaged representation of crystallographic
  SPs.]${}$\newline Alternatively, we average over multiple
  crystallographical SPs in order to arrive at a representation of,
  e.g., dislocation density or plastic strain in which they are
  distributed field quantities -- not only within the SP but also in
  direction of the slip system normal.  Therefore, we first collapse a
  number of crystallographic SPs that are contained within a region of
  width ${\vartriangle}\overline{s}\gg h$ into one
  \emph{representative} SP by summing up the respective dislocation
  field variables over ${\vartriangle}\overline{s}$.  The
  representative SPs are numbered by~$\overline g$. All points $\vec
  r$ of this layer belong to the domain
  \begin{eqnarray}
    \overline{\mathcal{B}}_{s,\overline g}  = 
    \Bigl\{ \vec z \in \mathcal{B}\colon 
    \vec z = \vec r_{\overline g} + \zeta\vec m_s \text{ with }
    \vec r_{\overline g} \in \Gamma_{s,\overline g} 
    \text{ and }|\zeta|\leq {\vartriangle}\overline{s}/2 \Big\}.  
  \end{eqnarray}
  \Delete{averaging over representative slip planes.}
  The domains
  $\overline{\mathcal{B}}_{s,\overline g}$ are non-overlapping with
  $\bigcup\limits_{\overline g} \overline{\mathcal{B}}_{s,\overline
    g}=\mathcal{B}$.  For the plastic shear strain in
  $\overline{\mathcal{B}}_{s,\overline g}$ we define the average  \begin{eqnarray}
    \overline{\gamma}_{s,\overline g}
    = 
    \frac{{\vartriangle} s}{{\vartriangle}\overline{s}}
    \sum\limits_{\Gamma_{s,g}\cap \mathcal{B}\subset \overline{\mathcal{B}}_{s,\overline g} } 
    \gamma_{s,g}
    \,.
  \end{eqnarray}
  At points between the representative slip planes
  \begin{eqnarray*}
    \vec r = 
    \frac{{\vartriangle}\overline{s} - \zeta}{{\vartriangle}\overline{s}} P_{s,\overline g}\vec r +
    \frac{\zeta}{{\vartriangle}\overline{s}}P_{s,\overline g+1}\vec r\,,
    \qquad 
    \zeta \in [0,{\vartriangle}\overline{•}{s}]
  \end{eqnarray*}
  we define the plastic shear strain in the body $\mathcal B$ by
  linear interpolation
  \begin{eqnarray}
    \label{eq:Case2}
    \gamma_s(\vec r) = 
    \frac{{\vartriangle}\overline{s} - \zeta}{{\vartriangle}\overline{s}}
    \overline\gamma_{s,\overline g}(P_{s,\overline g}\vec r) +
    \frac{\zeta}{{\vartriangle}\overline{s}}
    \overline\gamma_{s,\overline g + 1}(P_{s,\overline g+1}\vec r) \,.
  \end{eqnarray}
\end{description}


\subsection{Kr\"oner's continuum dislocation dynamics theory}
\label{sec:Kroner_theory}

The hdCDD theory extends Kr\"oner's continuum dislocation model
\citep{Kroner1958}, which is based on the dislocation density tensor
(the so-called Kr\"oner-Nye tensor)
\begin{eqnarray} 
  \vec \alpha &=& \nabla \times \vec
  \beta^\text{pl}
\end{eqnarray}
relating the plastic distortion in an averaging volume to a dislocation density%
\footnote{We note, that this density measure is dependent on the
  spatial resolution with which the plastic distortion is defined
  \citep{sandfeld_etal10}.}. 
Thus, we have
\begin{eqnarray}
  \nabla \cdot \vec \alpha &=& 0
\end{eqnarray}
which implies that dislocation lines cannot end or start within the
crystal. 

In the special case that all dislocations in an averaging volume form smooth
bundles of non-intersecting lines which all have the same line orientation,
they are geometrically necessary; the respective dislocation density is
referred to as the GND density $\rho^\text{G}$. Based on this volume density,
the dislocation density tensor depends on the average line orientation 
$\vec l$ and the Burgers vector $\vec b$ and is given by
\begin{align}
  \label{eq:alpha} 
  \vec\alpha=\rho^\text{G}\vec l\otimes\vec b
  \,.
\end{align}
Now we consider a specific slip system $s$. If all GNDs 
move with the same velocity $\vec v_s=v_s \vec l_s\times \vec d_s$, one
obtains the evolution equation
\begin{eqnarray} \label{eq:dalpha}
  \partial_t\vec \alpha_s &=& -\nabla\times(\vec v_s \times\vec\alpha_s)
  \,.
\end{eqnarray}
The scalar velocity $v_s = |\vec v_s|$ is a
constitutive function depending on, e.g., stresses of the solution of
the elastic boundary value problem \eqref{eq:linear}.

\subsection{A higher-dimensional model for continuum dislocation
  dynamics of curved dislocations}
\label{sec:Hochrainer_theory}

In order to represent systems of curved dislocations and their evolution in a
representative slip plane $\Gamma_{s,g}$, Hochrainer generalized Kr\"oner's theory \citep{Kroner1958} and the statistical approach
of Groma \citep{Groma1997_PhysRevB_p5807,Groma2003_ActaMater} towards systems
of dislocations with arbitrary line orientation and line curvature introducing
the higher-dimensional \emph{Continuum Dislocation Dynamics} (hdCDD) theory
\citep{Hochrainer2006_Dissertation,
  Hochrainer2007_PhilMag,sandfeld_etal10}. hdCDD is based on
mapping spatial, parameterized dislocation lines into a higher-dimensional
configuration space $\Gamma_{s,g} \times S^1$, where $S^1 = \mathbb
R/2\pi\mathbb Z \equiv [0,2\pi)$ is the orientation space containing the local
line orientation as additional information. The continuum representation of
lines in this configuration space requires the notion of a so-called
generalized line direction $\vec L_{s,g}$ and generalized velocity $\vec
V_{s,g}$ together with the dislocation density tensor of second order
$\alphaII_{s,g}$, which is also defined in the configuration space.  This
density tensor contains the Kr\"oner-Nye tensor as a special case but is
furthermore able to describe the evolution of very general systems of curved
dislocations with arbitrary orientation; in particular the common
differentiation between GND and SSD density becomes dispensable. Similar to
Kr\"oner's or Groma's frameworks, this continuum theory again also describes
only the kinematics, i.e., the evolution of dislocation density in a
\emph{given} velocity field. The additionally available information of hdCDD,
e.g.\ line orientation and curvature, however, is crucial for determining
dislocation interaction stresses and modeling physically-based boundary
conditions in a realistic manner.

In the following, $\vec r = (\eta,\xi)$ is a point in the slip plane
  and $(\vec r, \varphi)$ denotes a point in $\Gamma_{s,g} \times S^1$.
The dislocation density on $\Gamma_{s,g}$ must be understood as a volume
density, and thus to obtain the total line length in the SP we have to
integrate $\rho_{s,g}^\mathcal{B}$ over $\mathcal{B}_{s,g}$. Let $\vec
l_s(\varphi) = \cos\varphi \,\vec d_s + \sin\varphi\, \vec d_s\times \vec m_s$
be the canonical spatial line direction, and $\vec L_{s,g}(\vec r, \varphi)
= (\vec l_s, k_{s,g})^\top$ defines the generalized line direction in the
higher-order configuration space, where $k_{s,g}$ is the average line curvature.
The dislocation density tensor of second order is given by 
\begin{eqnarray} \label{eq:alphaII} \alphaII_{s,g}(\vec r,\varphi)=
  \rho_{s,g}(\vec r, \varphi) \vec L_{s,g}(\vec r, \varphi) \otimes
  \vec b_s
  \,,
\end{eqnarray}
and the evolution equation for this tensor has the form
\begin{eqnarray} \label{eq:dalphaII}
  \partial_t \alphaII_{s,g} (\vec r, \varphi)
  = -\hat{\nabla} \times \left(\vec V_{s,g}(\vec r, \varphi) \times
  \alphaII_{s,g}(\vec r, \varphi)\right)
  \,,
\end{eqnarray}
with $\hat \nabla = (\partial_\eta, \partial_\xi, \partial_\varphi)$,
where the vector $\vec V_{s,g} = ( -v_{s,g} \partial_\varphi\vec l_s,
-\vec L_{s,g} \cdot \hat{\nabla} v_{s,g})$ denotes the generalized
velocity in configuration space. For detailed
information on derivations and implications of these equations refer
to~\citep{Hochrainer2007_PhilMag,sandfeld_etal10}. 

Since the dislocation density tensor of second order is determined by the
dislocation density and the average line curvature, inserting
\eqref{eq:alphaII} in the evolution equation \eqref{eq:dalphaII} results into
the equivalent system for $\rho_{s,g}$ and $k_{s,g}$.  Introducing the
curvature density $q_{s,g}= \rho_{s,g}k_{s,g}$, this takes the form
\begin{subequations}
  \label{eq:cdd}
  \begin{eqnarray}
    \partial_t \rho_{s,g}  
    &=& 
    - \hat{\nabla} \cdot \left(\rho_{s,g} \vec V_{s,g} \right) + q_{s,g} v_{s,g}
    \,,
    \\
    \partial_tq_{s,g} 
    &=& 
    - \hat \nabla \cdot \left( q_{s,g} \vec V_{s,g} \right) 
    -\rho_{s,g} \Big(\vec L_{s,g}\cdot \hat\nabla 
    (\vec L_{s,g} \cdot \hat{\nabla} v_{s,g})\Big)
  \end{eqnarray}
\end{subequations}
(see \citep{Ebrahimi2014} for an interpretation of these two equations as
integrated total line length in $\mathcal{B}_{s,g}$ and the number of closed
dislocation loops, respectively).  The system \eqref{eq:cdd} is complemented
by boundary conditions which define, e.g., whether a dislocation can leave the
crystal (free surface) or not (impenetrable surface). In the latter case, the
density flux through the boundary is set to zero, e.g.\ $\rho_{s,g}\vec
v_{s,g}\cdot \vec n=0$, where $\vec n$ is the outward unit normal at the
boundary $\partial \mathcal B$. 

\subsection{Averaged quantities in the slip planes}

From a formal point of view the Kr\"oner-Nye framework with equations
\eqref{eq:alpha} and \eqref{eq:dalpha} corresponds to the 
higher-dimensional Hochrainer framework 
\eqref{eq:alphaII} and \eqref{eq:dalphaII} where $\vec l_s$ and $\vec
v_s$ are replaced by their higher-dimensional counterparts.
One can retrieve the Kr\"oner-Nye tensor for the slip plane $s$ from
the density function by
\begin{eqnarray} 
  \vec \alpha_{s,g}(\vec r)
  = \int\nolimits_0^{2\pi} \rho_{s,g}(\vec r,\varphi) \vec
  l_s(\varphi)\otimes \vec b_s \,\mathrm d \varphi.
\end{eqnarray}
Other classical measures can be derived as well, e.g.\ the total scalar density
is obtained by integrating $\rho_{s,g}(\vec r,\varphi)$ over all orientations
and the GND density is obtained as the norm of the GND vector $\vec
\kappa_{s,g}=[(\vec\alpha_{s,g})_{11}, (\vec\alpha_{s,g})_{12}]$.
%
The GND vector and the GND vector rotated by $\pi /2$ derive as
\begin{eqnarray}
  \vec \kappa_{s,g}(\vec r) = 
  \int\nolimits_0^{2\pi}\rho_{s,g}(\vec r,\varphi)
  \vec l_s(\varphi) 
  \,\text{d}\varphi
  \,,
  \qquad
  \vec \kappa_{s,g}^\perp(\vec r) = 
  \int\nolimits_0^{2\pi}\rho_{s,g}(\vec r,\varphi)
  \partial_\varphi \vec l_s(\varphi) 
  \,\text{d}\varphi
  \,,
\end{eqnarray}
where $\partial_\varphi\vec l_s(\varphi) = -\sin\varphi \,\vec d_s +
\cos\varphi\, \vec d_s\times \vec m_s$ is the orthogonal line direction, and
the total scalar densities are given by
    \begin{eqnarray} \label{eq:rhotot} 
      \rho^\text{tot}_{s,g}(\vec r) =
      \int\nolimits_0^{2\pi} \rho_{s,g}(\vec r,\varphi) \,\mathrm d
      \varphi\,,
      \qquad
      q^\text{tot}_{s,g}(\vec r) =
      \int\nolimits_0^{2\pi} q_{s,g}(\vec r,\varphi) \,\mathrm d
      \varphi\,.
    \end{eqnarray}
The plastic shear strain 
is determined integrating the Orowan relation over the orientation space
\begin{eqnarray}
  \label{eqn:Orowan2}
  \partial_t \gamma_{s,g}(\vec r) 
  &=& b_{s}\int\nolimits_{0}^{2\pi}\rho_{s,g}(\vec r,\varphi) 
  v_{s,g}(\vec r,\varphi) \, {\rm d}\varphi 
  \,.
\end{eqnarray}
Thus, for the case that the dislocation velocity is independent of the
orientation this simplifies to the classical plastic strain rate equation
$\partial_t \gamma_{s,g}= b_{s}\rho_{s,g}^\text{tot} v_{s,g}$
  \citep{Orowan1940_ProcPhysSoc52}\footnote{\Insert{Note that it is the total density which governs the plastic strain rate and not the GND density; the latter would not be able to account for plastic strain due to expansion of a statistically homogeneous distribution of dislocation loops.}}.

\subsection{The dislocation velocity}
\label{sec:velocity}

Continuum dislocation models are 'kinematic' theories predicting the flux of
density depending on the dislocation velocity $v_{s,g}$ as a constitutive
ingredient.  Hence, in Kr\"oner's, Groma's and Hochrainer's theories alike, stresses
from dislocation interactions are \emph{not} a priori included in these
theories and need to be determined separately.  How to derive physically
meaningful dislocation interaction stress components is a topic of ongoing
research; details about rigorous analysis of some dislocation systems in a
continuum framework can be found in \citep[e.g.][]{Zaiser2001_PhysRevB,
  Groma2003_ActaMater,El-Azab_PhilMag, Sandfeld2013, Schulz2013, Scardia2014_JMPS70}. Here, we base the
dynamics of our dislocation systems on the following assumptions for the
velocity function:
\begin{enumerate}
\item The scalar velocity $v_{s,g}$ in $\Gamma_{s,g}$ is assumed to
  depend linearly on the stresses acting on dislocations.
\item The velocity function is decomposed into different 
  stress contributions of two separate 
  classes: the
  projection of the resolved stress $\tau_{s,g}= \vec M_s:\vec
  \sigma$ computed from the solution of the elastic
  BVP \eqref{eq:linear},
  and stresses governing short-range elastic dislocation
  interactions (the back stress $\tau_{s,g}^{\text{b}}$, 
  line tension $\tau_{s,g}^{\text{lt}}$, and the 
  yield stress $\tau_{s,g}^{\text{y}}$).
\item We assume that dislocations move only if the yield stress 
  is exceeded by the other stress contributions 
  $\tau^{\text{0}}_{s,g} = \tau_{s,g} - \tau_{s,g}^{\text{b}} -
  \tau_{s,g}^{\text{lt}}$, and the flow direction is determined by the sign of 
  $\tau^{\text{0}}_{s,g}$.
\end{enumerate}
With these assumptions the velocity function takes the form
\begin{eqnarray}
  v_{s,g} = \left\{ 
    \begin{array}{ll} 
      \frac{b_s}{B} \sgn(\tau_{s,g}^0)(|\tau_{s,g}^0| - \tau_{s,g}^{\text{y}})& \;\;
      \text{if}\;\;  |\tau_{s,g}^0| >\tau_{s,g}^{\text{y}},\\
      \vspace{-1mm}\\
      0                                 & \;\;\text{otherwise},
    \end{array}\right.
  \label{eq:bending:eqofmotion}
\end{eqnarray} 
\Insert{which assumes an overdamped dislocation motion neglecting inertial effects. $B>0$ is the respective drag coefficient.} 

The resolved stress $\tau_{s,g}$ depends on the global boundary value problem
and includes the contribution from the eigenstrain, which governs the
long-range interaction between dislocations \citep[see e.g.][]{Kroener1955_ZPhys,El-Azab_PhilMag,Sandfeld2013}.

The back stress is an approximation for the repelling forces between parallel
dislocations with the same line direction \citep[see e.g.][]{Groma2003_ActaMater,Hirschberger2011_MSMSE19,Schulz2013}.  For a system of
curved dislocations we adapt a formulation which was suggested and implemented
in the related works \citep{Hochrainer2006_Dissertation, Sandfeld_diss}, and we
define
\begin{align}
  \tau_{s,g}^\text{b} &= \frac{D\mu b_s}{\rho_{s,g}^\text{tot}} \nabla
  \cdot \vec\kappa_{s,g}^\perp\,.
\end{align}
This implies that the back stress at $\vec r$ in direction of $\vec
l_s(\varphi)$ is proportional to the gradient of GNDs perpendicular to the
line direction. Here, \Insert{$\mu$ is the shear modulus}, $D\in [0.4,1]$ is a constant and \Insert{was obtained in \cite{Groma2003_ActaMater} by statistical analysis of discrete dislocation systems. 
	}

The line tension $\tau_{s,g}^{\text{lt}}$ describes the self-interaction
of a dislocation loop (a loop subjected to no other stress would contract
due to the line tension). In general this depends on the local character of the line
\Insert{(e.g., whether it is a screw or edge dislocation, cf.\ \cite{Foreman1967_PhilMag15_p1011})}. For simplicity we use an
approximation by a constant line tension which is independent of the line
orientation
    \begin{align}
      \tau^{\text{lt}}_{s,g} 
      = \frac{T_s}{b_s} \frac{q^\text{tot}_{s,g}}{\rho^\text{tot}_{s,g}}
    \end{align}
where the constant $T_s\in [0.5\mu b_s^2,\, 1.0\mu b_s^2]$ describes the
strength of the interaction and $q_{s,g}^\text{tot}$ is the average curvature
density.
Finally, the yield stress is governed by a Taylor-type term
 in the case of only one active slip system
    \begin{equation}
      \tau^{\text{y}}_{s,g}=  a \mu b_s \sqrt{\rho^\text{tot}_{s,g}}
    \end{equation}
with a constant $a \in [0.2,0.4]$ ,
\Insert{see e.g.\ \cite{Taylor1934_ProcRoySocA145} or the overview in \cite{Basinski79}. In general, the yield stress can depend on the
dislocation density of all slip systems \citep{Kubin2008_ActaMater56}}; this will be considered as well in
Study~3 below.

\section{A numerical scheme for the reduced 2D system}
\label{sec:dG}

For the numerical evaluation of the physical behavior of our model we
consider a 2D reduction of the fully coupled system, where 
we assume to have a homogeneous distribution over the 
\Replace{$y$}{$z$} 
direction and slip plane normals $\vec m_s$ in the $x-y$ plane. This leads to
$\partial_\eta \rho_{s,g} \equiv \partial_\eta q_{s,g} \equiv \partial_\eta
v_{s,g} \equiv 0$ everywhere in the system. In this section we describe the
discretization for a single slip plane $\Gamma = \Gamma_{s,g}$; for simplicity
we skip the indices $s$~and~$g$.

\paragraph{Reduced hdCDD over 1D slip plane}

Rewriting \eqref{eq:cdd} using $\partial_\eta\rho\equiv \partial_\eta q \equiv
\partial_\eta v \equiv 0$ yields
\begin{subequations}
  \label{eqn:model_rho_rhok_1D}
  \begin{eqnarray}
    \label{eqn:model_rho_rhok_1D_1}
    \partial_t \rho 
    &=& 
    -\widetilde \nabla \cdot (\rho 
    \vec {\widetilde V}) + qv\\
    \label{eqn:model_rho_rhok_1D_2}
    \partial_t q 
    &=& 
    -\widetilde \nabla \cdot (q 
    \vec {\widetilde V}) 
    - \rho\cos\varphi \partial_\xi \Big( \cos\varphi\partial_\xi v + 
    k \partial_\varphi v \Big)
    - q \partial_\varphi \Big( \cos\varphi\partial_\xi v 
    + k \partial_\varphi v \Big)    
  \end{eqnarray}
\end{subequations}
with $ \widetilde \nabla = (\partial_\xi, \partial_\varphi)^\top$ and $ \vec
{\widetilde V} = (v\sin\varphi, 
\Insert{-}\cos\varphi \partial_\xi v \Insert{-} 
k\partial_\varphi v)^\top$. Assuming that the velocity does not exhibit any
angular anisotropy the system \eqref{eqn:model_rho_rhok_1D} reduces further to
\begin{subequations}
  \label{eqn:Model_rho_rhok_1D}
  \begin{eqnarray}
    \label{eqn:Model_rho_rhok_1D_1}
    \partial_t \rho 
    &=& 
    -\widetilde \nabla \cdot (\rho 
    \vec {\widetilde V}) + qv\\
    \label{eqn:Model_rho_rhok_1D_2}
    \partial_t q 
    &=& 
    -\widetilde \nabla \cdot (q 
    \vec {\widetilde V}) 
    - \rho (\cos\varphi)^2\partial_\xi^2 v 
    + q \sin\varphi\partial_\xi v     
    \,.
  \end{eqnarray}
\end{subequations}
We rewrite \eqref{eqn:Model_rho_rhok_1D} in the equivalent form of a linear
conservation law for $\vec w = (\rho, q)^\top$, i.e.,
\begin{eqnarray}
  \label{eqn:conserv_rho_rhok}
  \partial_t \vec  w + 
  \widetilde \nabla
  \cdot \vec  F(\vec  w)  + \vec  B \vec  w &=& \vec  0 
  \quad \, \text{ in }\, \Gamma \times S^1,
\end{eqnarray}
where $ \widetilde \nabla \cdot \vec F(\vec w)=
\partial_\xi(\vec F_1 \vec w) +
\partial_\varphi(\vec F_2 \vec w)$ with
\begin{eqnarray*}
  &&\vec  F_1 = 
  \begin{pmatrix}
    v \sin \varphi & 0 \\
    0 & v \sin \varphi
  \end{pmatrix}
  \,,
  \quad
  \vec  F_2  =  
  \begin{pmatrix}
    -\cos\varphi \partial_\xi v  & 0 \\
    0 & -\cos\varphi \partial_\xi v
  \end{pmatrix}
  \,,
  \quad
  \vec  B 
   =  
  \begin{pmatrix}
    0 & - v \\
    (\cos\varphi)^2 \partial_\xi^2 v & -\sin\varphi \partial_\xi v
  \end{pmatrix}
\end{eqnarray*}
depending on the dislocation velocity $v$.

\Insert{
For this first-order system we now construct a discontinuous Galerkin 
discretization in space, since this discretization is well adapted to
conservation laws, leads to a better approximation for discontinuous weak solutions
and is significantly less diffusive than standard continuous Lagrange elements
\citep{HesW08}.}

\paragraph{The \Insert{Runge--Kutta discontinuous Galerkin} (RKDG) method}

The system \eqref{eqn:conserv_rho_rhok} reads as follows:
find $\vec w : \Gamma\times S^1 \times [0,T] \to \R^2$ solving
\begin{eqnarray*}
  \label{eqn:model_eq}
    \partial_t \vec  w + \vec  A \vec  w &=& \vec  0 \quad \text{ in }  \, 
    \Gamma\times S^1
\end{eqnarray*}
subject to the initial condition
$\vec  w(\vec{r}, \varphi, 0) = \vec  w_0(\vec{r}, \varphi) $,
periodic boundary conditions $
  \vec  w(\xi,\eta,0,t) = \vec  w(\xi,\eta,2\pi,t) $
in $\varphi$, and where
$\vec A \vec w =\widetilde \nabla \cdot \vec F(\vec w) +\vec B \vec w =
\vec F_1 \partial_\xi\vec w + \vec F_2 \partial_\varphi \vec w +
(\partial_\xi\vec F_1 + \partial_\varphi \vec F_2+\vec B )\vec w $.

In the first step we derive a semi-discrete discontinuous Galerkin scheme.
Let $\mathcal{T}_h= \big\{K\} $ be a triangulation of $\mathcal B$,
and assume that this triangulation is aligned with the slip plane
$\Gamma$, i.e., $\Gamma = \bigcup_{f\in \mathcal F_\Gamma} f $ with
faces $f\in \mathcal F_\Gamma =
\big\{\partial K\cap\Gamma\colon K\in \mathcal T_h\big\}$. 
Let $\mathcal Z_h$ be the set of all vertices of the triangulation.
For a fixed face $\bar f = \conv\{\vec z_{j-1},\vec z_j\}$ with $\vec
z_{j-1},\vec z_j\in \Gamma$ we observe
\begin{eqnarray*}
  \vec  A(\vec  w,\vec  \psi)_{f\times S^1} 
  &=& 
  \int_{f\times S^1} 
  \big(\widetilde \nabla \cdot \vec F(\vec  w) 
  +\vec  B  \vec  w\big)\cdot \vec \psi \,\mathrm d\xi
  \mathrm d\varphi
  \\
  &=&
  \int_{f\times S^1} 
  \big(-\vec F(\vec  w) \cdot \widetilde \nabla \vec \psi 
  +\vec  B  \vec  w \cdot \vec \psi \big)
  \,\mathrm d\xi \mathrm d\varphi
  +
  \int_{S^1} 
  \vec {\widetilde n} \cdot \vec F(\vec  w) 
  \cdot \vec \psi \Big|_{\vec z_{j-1}}^{\vec z_j} \mathrm d\varphi
\end{eqnarray*}
for all smooth functions $\vec \psi\colon f \times S^1\longrightarrow
\mathbb R^2$.
For the discretization we choose an ansatz space $X_f$ on every face
defining a discontinuous ansatz space $X_h = \big\{\vec \psi_h \in \mathrm
L_2(\Gamma,\mathbb R^2)\colon \vec \psi_h|_f \in X_f\big \}$ and the
numerical flux
\begin{eqnarray*}
  \vec{F}_{f,e}^*(\vec \psi_h) 
  &=& \{\!\{\vec{F}_f \vec \psi_h \}\!\}_{f,e} + 
  \frac{C_f}{2}[[\vec\psi_h]]_{f,e} \vec {\widetilde n}
\end{eqnarray*}
on the face intersections $e = \partial f\cap \partial f_e$ (which are single
points in a 1D slip plane). Here, we choose the stabilization constant $C_f =
|\vec {\widetilde V}|$, and the average and the jump along a normal $ \vec
{\widetilde n}$ oriented from $f$ to $f_e$ are given by
\begin{eqnarray*}
  \{\!\{\vec  F \vec{w}_h\}\!\}_{f,e} := 
  \frac12 \Big(\vec  F \vec{w}_h|_{f} + \vec  F \vec{w}_h|_{f_e}\Big)\,, 
  &\quad& 
  [[\vec  w_h]]_{f,e} := \vec  w_h|_{f} - \vec  w_h|_{f_e} 
  \,,
\end{eqnarray*}
respectively. 
%
{On the open boundary, we set $\vec{w}_h|_{f_e} = \vec{w}_h|_{f}$ and $\vec{w}_h|_{f_e} = -\vec{w}_h|_{f}$ for the impenetrable boundary.}
%
Now, defining locally 
\begin{eqnarray*}
  (\vec  A_{f,h}\vec  w_h,\vec \psi_h)_{f\times S^1} 
  &=&
  \int_{f\times S^1} 
  \big(-\vec F(\vec  w_h) \cdot \widetilde \nabla \vec \psi_h 
  +\vec  B  \vec  w_h \cdot \vec \psi_h \big)
  \,\mathrm d\xi \mathrm d\varphi
  + \int_{S^1} 
  \vec {\widetilde n} \cdot \vec F^*_{f,e}(\vec  w_h)
  \cdot \vec \psi_h \Big|_{\vec z_{j-1}}^{\vec z_j} \mathrm d\varphi
\end{eqnarray*}
yields the discrete operator by
$
  (\vec  A_h\vec  w_h,\vec \psi_h)_{\Gamma\times S^1} 
  = 
  \sum_f (\vec  A_{f,h}\vec  w_h,\vec \psi_h)_{f\times S^1} $.
We choose a DG ansatz space with Fourier basis functions $X_f = \mathbb P_k
\otimes \mathbb F_n$, where $\mathbb P_k = \Span\{1,\xi,\dots,\xi^k\}$ are
polynomials, and the truncated Fourier space is given by $\mathbb F_n =
\Span\{1,\cos(\varphi),\sin(\varphi),\dots,\cos(n\varphi),\sin(n\varphi)\}$ . 
Thus, the components of $\vec w_h = (\rho_h,k_h)$ have the form
\begin{eqnarray*}
  \rho_h(\xi,\varphi) = \sum_{l=0}^k 
  \xi^l\Big(a_{l0} +
  \sum_{m=1}^n\big(a_{lm}\cos(m\varphi)+b_{lm}\sin(m\varphi)
  \big)\Big)
\end{eqnarray*}
and similar for $q_h$.
The Runge--Kutta time discretization is now obtained by the method of lines.
Therefore, choosing a basis $\vec \psi^1,\dots,\vec\psi^{N_h}$ of $X_h$ 
yields the matrix formulation
\begin{eqnarray}
  \label{eq:RK}
  \underbar M \partial_t \underbar w(t) + \underbar A 
  \,\underbar w(t) &=& 0
\end{eqnarray}
with $ \underbar M \, = \, \left((\vec  \psi^k, \vec  \psi^m)\right)_{m,k}$
and $\underbar A \, = \, \left(\vec  A_h(\vec  \psi^k,\vec\psi^m)\right)_{m,k}$.
This yields for the time step from $t_n$ to $t_{n+1}$
\begin{eqnarray*}
  \underbar w^{n+1} = 
  \underbar w^{n} - {\vartriangle} t \underbar M^{-1}\underbar A
  \big(\underbar w^{n} - \frac12 
  {\vartriangle} t \underbar M^{-1}\underbar A\big(\underbar w^{n} - \frac13
  {\vartriangle} t \underbar M^{-1}\underbar A\big(\underbar w^{n} - \frac14
  {\vartriangle} t \underbar M^{-1}\underbar A \, \underbar w^{n}\big)\big)\big)
\end{eqnarray*}
for the classical explicit Runge--Kutta schemes of order 4 
(see \cite{Hochbruck13} for alternative time integration methods in
combination with DG schemes).

\paragraph{Finite element discretization of the solid}

Let $V_h = \big\{ \vec v \in \mathrm H^1(\mathcal B)^2\colon \vec v|_K \in
\mathbb P_1(K)$ for $ K\in \mathcal T_h\big\}$ be a standard finite element
space for the displacements and set $V_h(\vec u_\text{D}) = \big\{ \vec v \in
V_h\colon \vec v(\vec z) = \vec u_\text{D}(\vec z)$ for all nodal points $
\vec z\in \mathcal Z_h\cap \partial_\text{D} \mathcal B \big\}$.  For $\vec
u\in V_h$ the strain $\vec\varepsilon (\vec u)$ and the stress $\vec \sigma =
\mathbb C:\vec\varepsilon$ is piecewise constant in $K$.
%
%
Now, the coupled algorithm is defined as follows:

\hbox{}
\fbox{\begin{minipage}{0.95\textwidth}
\begin{itemize}
\item[S0)]
Select ${\vartriangle} t>0$, $t_\text{max}>0$,  
set $n = 0$,\, set initial values for $\rho_{s,g}^0,q_{s,g}^0,\gamma_{s,g}^0$
on $\Gamma_{s,g}$.

\item[S1)] Set $t_n = n{\vartriangle} t$, $\vec u_\text{D}^n=\vec
u_\text{D}(t_n)$, $\vec f_{\mathcal B}^n= \vec f_{\mathcal B}(t_n)$
and $\vec g_\text{N}^n = \vec g_\text{N}(t_n)$.  

Compute $ \gamma_s^n$ in $\mathcal B$ from 
$\gamma_{s,g}^n$ in $\Gamma_{s,g}$ (depending on Case~1 or 2).
 
\item[S2)] 
Evaluate the plastic strain $\vec\varepsilon^{\text{pl},n}|_K = \sum_s
\gamma_s^n|_K \vec M_s^\text{sym}$ and compute $\vec u^n\in V_h(\vec
u_\text{D}^n)$ with
\begin{eqnarray*}
  \hspace*{-5mm}
  \int_{\mathcal B}
  \vec\varepsilon (\vec u^n) 
  : \mathbb C : \vec\varepsilon (\delta \vec u)\, \mathrm d \vec x
  = 
  \int_{\mathcal B}
  \vec\varepsilon^{\text{pl},n}
  : \mathbb C : \vec\varepsilon (\delta \vec u)\, \mathrm d \vec x
  +
  \int_{\mathcal B} \vec f_{\mathcal B}  \cdot \delta \vec u \, \mathrm d \vec x
  + \int_{\partial_\text{N}\mathcal B}  
  \vec g_\text{N} \cdot \delta \vec u \, \mathrm d \vec a\,,
  \ \delta \vec u \in V_h(0).
\end{eqnarray*}
\item[S3)]
On $f = K\cap K'\subset \Gamma_{s,g}$ set 
$  \tau_{s,g}^n |_f  = \frac12  \Big(\vec \sigma^n|_K + \vec
  \sigma^n|_{K'}\Big):  \vec M_s$
and compute   velocities $v_{s,g}^n$ (eqn.~\eqref{eq:bending:eqofmotion}).

\item[S4)]
Compute $(\rho_{s,g}^{n+1},q_{s,g}^{n+1})$ independently on every
$\Gamma_{s,g}$ by $M$ explicit Runge--Kutta steps for
\eqref{eq:RK} with step size ${\vartriangle} t/M$ and fixed velocity
$v_{s,g}^n$.

\item[S5)]
If \Insert{$t_n<t_\text{max}$}, set $n= n+1$ and go to S1).
\end{itemize}
\end{minipage}}\hfill\hbox{}

\bigskip

\noindent
Since this scheme is in step S4) fully explicit, the 
\Insert{Courant-Friedrich-Levy}
condition requires sufficiently small time steps. Also the coupling with the
boundary value problem S2) is explicit. In our numerical tests we choose the
global time step ${\vartriangle} t$ and the local time step ${\vartriangle}
t/M$ small enough to observe convergence by comparing the results with
different mesh resolution $h$ \Insert{and time steps ${\vartriangle} t$}.

\section{Numerical experiments}
\label{sec:numerics}

We evaluate our model for a single-crystal thin film with idealized passivated
and non-passivated surfaces in tensile and shear test settings. These are well
established model tests for crystal plasticity models \citep[see e.g.][]{Liu2011201,Schwarz2005_MSEA,Zaiser2007_PhilMag_p1283,Deshpande2005,Fredriksson20051834,Fertig2009874}. The
novelty of hdCDD, however, is that we can directly link the dislocation
microstructure in almost DDD-like details to the macroscopic response. In the
following, we study in particular the influence of the line curvature and two
different physical boundary conditions in single- and multislip
configurations. Additionally, we numerically evaluate  effects due to
averaging for Case 1 and 2. We note, that especially the line curvature is an important
physical quantity \Insert{that, however, only can be represented by few continuum models \citep[e.g][]{sedlacek_kw03,Xiang2009728} - a fact that makes detailed comparisons with other approaches difficult}.

\begin{figure}[htb]
\centering
\subfloat[Tensile test: Displacements are prescribed at left and right boundaries, top and bottom boundaries are free surfaces.]{\includegraphics[height=0.16\textwidth]{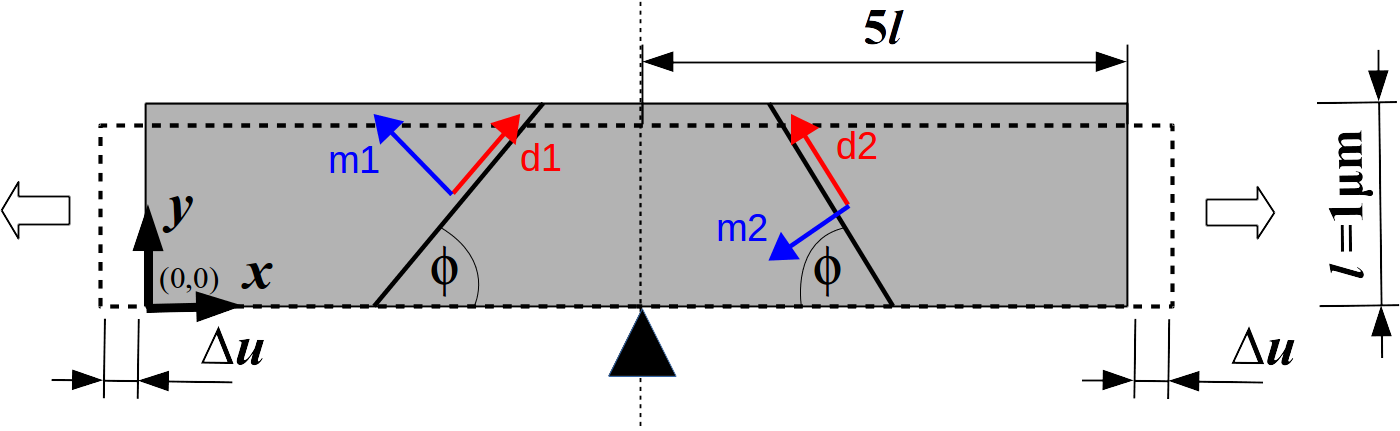}}
\hfill
\subfloat[Shear test: Displacements are prescribed at bottom and top
boundaries, left and right boundaries are free surfaces.]{\includegraphics[height=0.16\textwidth]{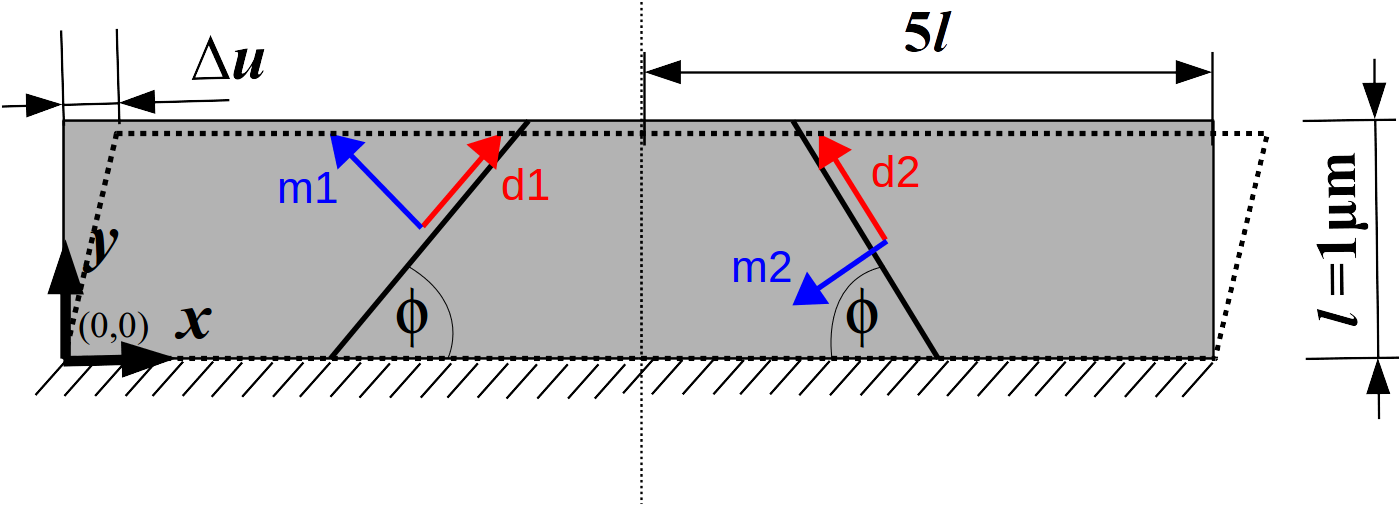}}\\ 
\caption{\label{fig:geometry}Geometry and boundary conditions 
of the investigated model systems for the tensile test (Study~1 and Study~2)
and the shear test (Study~3).}
\end{figure}
\paragraph*{Geometry and slip system}
We consider the configurations \Insert{as shown in Fig.~\ref{fig:geometry}} for the
investigation of the deformation behavior of a thin, single-crystalline Al
film 
assuming plane strain. The film is represented by
a 2-dimensional body $\mathcal B = (0,10 l) \times (0,l)$ with $l= 1$ 
\textmu m. 
We consider one or two active slip systems ($N=1$ or 2) with 1-dimensional
(crystallographic) slip planes determined by
\begin{subequations}
\label{eq:slips}
\begin{align}
  &\vec d^1 = \cos\phi~\vec e_1 + \sin\phi~\vec e_2\,, 
  &&\vec m^1  =  -\sin\phi~\vec e_1 + \cos\phi~\vec e_2\,,\\
  &\vec d^2 = -\cos\phi~\vec e_1 + \sin\phi~\vec e_2\,, 
  &&\vec m^2  =  -\sin\phi~\vec e_1 - \cos\phi~\vec e_2 
\end{align}
\end{subequations}
with the angle $\phi= \pi/3$ between slip planes and the film surfaces.  The
distance between crystallographic slip planes is set to ${\vartriangle} s =
0.05,0.1,0.2$ \textmu m, respectively.  For the thickness of the
crystallographic layer $\mathcal{B}_{s,g}$ we choose $h = 0.5{\vartriangle}
s$. In Case~2, we set ${\vartriangle} \overline s = 0.2,0.4$ \textmu m.
As material we use aluminum with  a Young's modulus of $E=7\cdot10^{10}$\,Pa,
Poisson ratio $\nu=0.3$,  Burgers vector size $b=2.56\cdot10^{-10}$\,m and
drag coefficient $B=2.0\cdot10^{-4}$\,Pa$\cdot$s.
\paragraph{Numerical aspects}
We use two different finite element meshes: the mesh for the elastic problem
consists of triangular linear finite elements with altogether $\approx 223\,000$ degrees of
freedom (dofs), the hdCDD mesh consists of linear Fourier elements with, e.g.
$\approx 262\,000$ dofs for Case~1 with ${\vartriangle} s=100$\,nm and
$h=50$\,nm.  For the time integration we used a step size of ${\vartriangle}
t=10^{-3}$\textmu s with $M=10$ 'micro-time steps' per macroscopic displacement increment, resulting in $6\cdot 10^{3}$ micro-time steps for reaching the total strain of
$\varepsilon^\text{tot}=1.2\%$. For the shear test the same step size is
used with a total number of $5\cdot 10^3$ micro-time steps for obtaining the total strain of
$\varepsilon^\text{tot} = 2.5\%$.
\Insert{During each of the micro-time steps the stresses from the elastic BVP are kept constant and only the dislocation microstructure with the respective short-range interaction stresses evolves.}
%

\paragraph*{Boundary conditions} 

\begin{figure}[H]
\centering
\includegraphics[viewport=80 180 750 420, clip, width=0.95\textwidth]{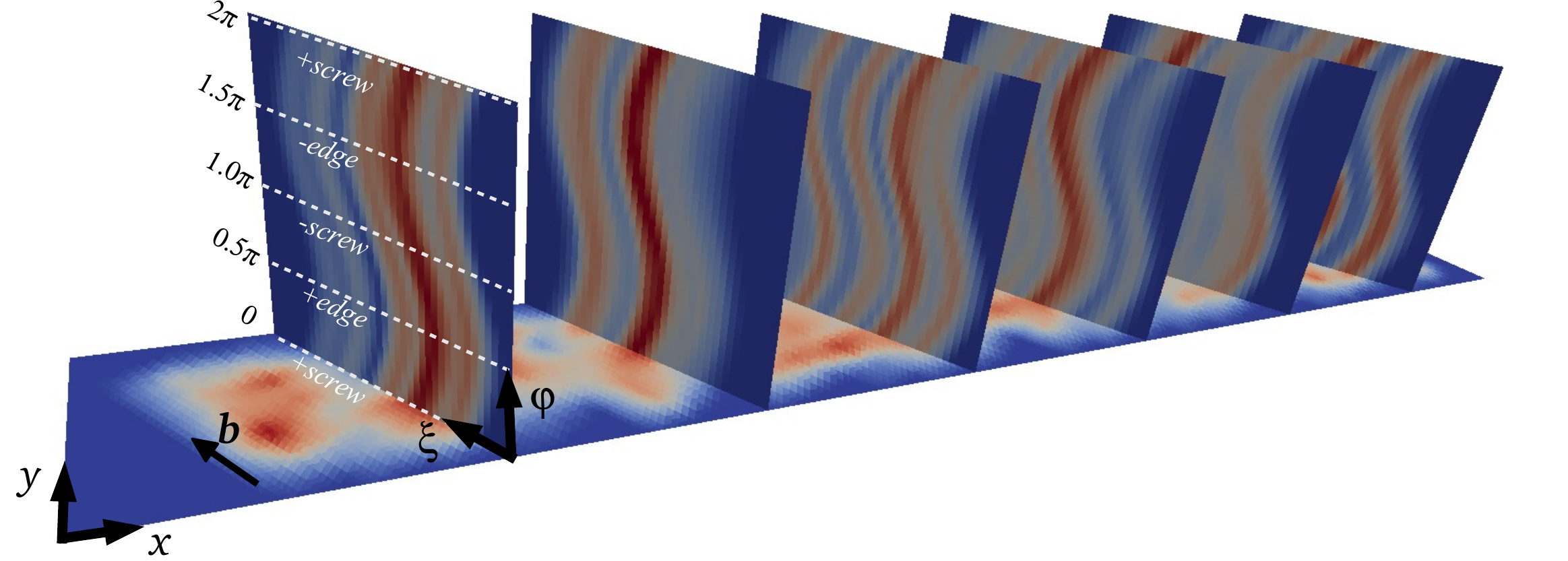}
\caption{\label{fig:IVs}Visualization of the initial dislocation density in the
  higher-dimensional configuration space for a system with 25~representative
  slip planes, each of them containing 28~'smeared-out' randomly positioned
  loops.  In vertical direction the variable $\varphi \in [0,2\pi]$ is
  displayed for some of the 25 slip planes; the dashed white lines indicate
  the line orientation for screw and edge segments (compare Burgers vector
  direction in the spatial plane). Each of the wavy distributions corresponds
  to (a) dislocation loop(s). On the bottom plane the spatial projection of
  the total density $\rho^\text{tot}$ is shown as spatial average as seen in
  'Case 2'. }
\end{figure}

For the two different systems we introduce different boundary
conditions for the elasticity problem:
\begin{itemize}
\item[(1)]
For the symmetric tensile test we consider prescribed boundary displacements
along the left and right boundary face (at $x_1=0$ and $x_1=10l$,
respectively). We increase the displacements with a constant rate $ \dot
u_{1,{\rm right}} = -\dot u_{1,{\rm left}} = 1.0$ m/s for $t\in [0,0.06]$
\textmu s. In order to avoid vertical translations we additionally fix the
displacements at the point $(5l,0)^\top$.
\item[(2)] 
For the shear test we consider prescribed boundary displacements along the
upper surface at $x_2 = l$ and fixed displacements at the lower surface at
$x_2 = 0$. The upper prescribed displacements are increased with a constant
rate $\dot u_{1,\rm{up}} = 1.0$ m/s for $t \in [0,0.05]$ \textmu s.
\end{itemize}
For both systems, also the boundary conditions for the dislocation problem
w.r.t.\ dislocation fluxes have to be considered. In physical terms surfaces
can either be open (dislocations can leave the film) or impenetrable
(dislocations can not leave the film). Open boundaries can simply be modeled
by extrapolating the hdCDD field values. For impenetrable surfaces we require
(i) that the flux of dislocation density normal to the surface vanishes and
that (ii) dislocations directly at the surface must be straight and thus must
have zero curvature. Numerically, we model the impenetrable flux boundary
condition by 
{introducing a numerical inflow defined as the negative outflow of density and curvature densities on the considered boundary}. 
%
\paragraph*{Initial values}
We construct consistent initial values which guarantee that, e.g., the
solenoidality of $\alphaII$ (i.e.\ $\div\alphaII=0$) is not violated and that
the GND density vector comes out as a gradient of the plastic slip. This is done by
superposition of $N_{\rm d}$ randomly distributed discrete dislocation loops in a 2D
slip plane followed by an appropriate 'smearing-out' procedure as described in
detail in the appendix. One-dimensional slip plane data are then obtained by
integrating the CDD field values over the second, homogeneous direction (see
\Figref{fig:IVs}). Depending on the used averaging, i.e.\ Case 1 or 2, we
finally have to consider the distance of the representative slip planes.

\subsection{Study 1: The influence of boundary conditions (Case 1)}
\label{sec:results1}
For this investigation we use 80 representative slip planes and the shear
strain extension of Case 1, starting in each SP with 5 dislocation loops of
radius $r$ between 100\,nm and 200\,nm at random positions. The height of the
quasi-discrete numerical SPs has a relatively small value of $h=50$\,nm below
which no appreciable difference in the system response could be observed (also
see Study 2). For averaging purposes we assume an out-of-plane length
of $L_z=l/\sin(\phi)=1.15$\,\textmu m resulting in an average
dislocation density $\langle \rho\rangle=3.1\times 10^{13}/{\rm m}^2$. For the
analysis of the influence of boundary conditions we study two configurations:
in the first configuration we choose open boundaries (abbreviated as 'open
BCs'), i.e.\ dislocations can leave the volume, and the second imposes
impenetrable boundaries (abbreviated as 'imp.\ BCs'), i.e.\ dislocations can not
leave the volume through the surface $\partial B$. The simulation is driven by
a prescribed constant strain rate $\dot\varepsilon=0.2$ \textmu s$^{-1}$ until the
maximum total strain $\varepsilon^{\rm tot}=1.2\%$ is reached.
\begin{figure}[ht]
\centering
\footnotesize
\hbox{}\hspace{1.5cm}\makebox(15,16)[tl]{\textbf{open boundary condition\;}} \hfill %
\includegraphics[width=0.20\textwidth]{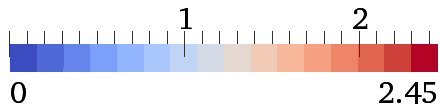}\makebox(15,16)[tl]{$\times 10^{14}/{\rm m}^2$}\hfill%
\makebox(90,16)[tr]{\hspace{0.5cm}\textbf{\quad impenetrable boundary condition}} \hspace{0.5cm}\hbox{}\vspace{-18mm}\\
\begin{sideways}\parbox{35mm}{$\quad t=0\mu $s\\${}\;\varepsilon^{\rm tot}=0.0\%$}\end{sideways}$\;$%
\includegraphics[viewport=0 0 625 270, clip, width=0.46\textwidth]{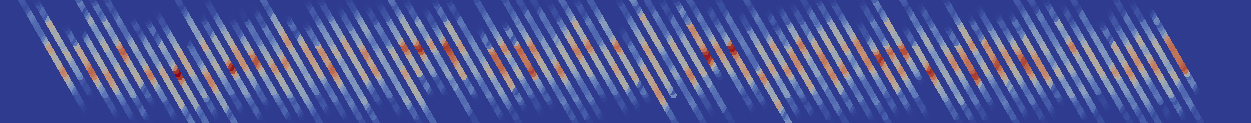}\quad
\includegraphics[viewport=625 0 1250 270, clip, width=0.46\textwidth]{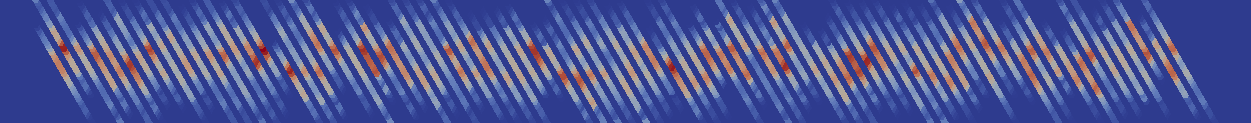}\vspace{-15mm}\\
\begin{sideways}\parbox{35mm}{$\;t=0.03\mu$ s\\${}\varepsilon^{\rm tot}=0.6\%$}\end{sideways}$\;$%
\includegraphics[viewport=0 0 625 270, clip, width=0.46\textwidth]{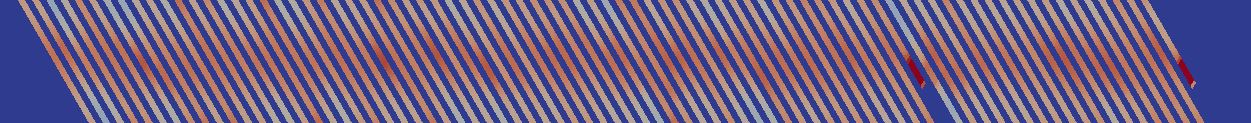}\quad
\includegraphics[viewport=625 0 1250 270, clip, width=0.46\textwidth]{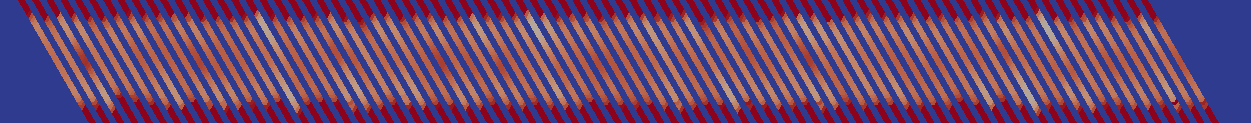}\vspace{-15mm}\\
\begin{sideways}\parbox{35mm}{$t=0.055\mu s$\\${}\varepsilon^{\rm tot}=1.1\%$}\end{sideways}$\;$%
\includegraphics[viewport=0 0 625 270, clip, width=0.46\textwidth]{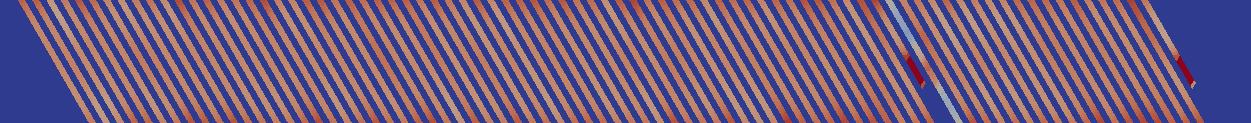}\quad
\includegraphics[viewport=625 0 1250 270, clip, width=0.46\textwidth]{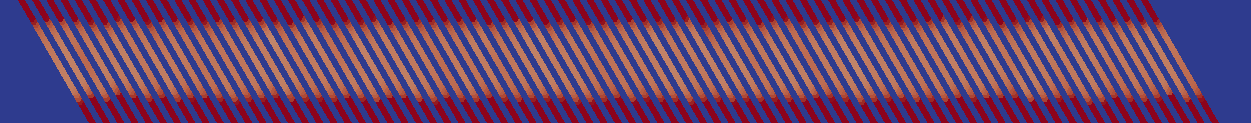}\\
\caption{\label{fig:rhot_open_closed_case1}Initial distribution and time
  evolution of total density $\rho^{\text{tot}}$, Case 1,  with open and impenetrable 
boundary conditions for the dislocation density.}
\end{figure}
In \Figref{fig:rhot_open_closed_case1} the evolution of the total density
$\rho^{\rm tot}$  at three distinct time steps in the $x-y$-plane is illustrated. 

\begin{figure}[ht]
\centering
\footnotesize
\hbox{}\hspace{3cm}\makebox(11,16)[tl]{\textbf{open boundary condition}} \hfill %
\makebox(20,16)[tr]{\textbf{impenetrable boundary condition}} \hspace{1.2cm}\hbox{}\vspace{-6mm}\\
\hbox{}\hspace{1.cm}{$\rho(\xi,\varphi)\;[1/{\rm nm}^2]$ \hspace{0.6cm}$q(\xi,\varphi)\;[1/{\rm nm}^3]$ \hspace{.7cm}$k(\xi,\varphi)\;[1/{\rm nm}]$}  \hfill %
{$\rho(\xi,\varphi)\;[1/{\rm nm}^2]$ \hspace{.6cm}$q(\xi,\varphi)\;[1/{\rm nm}^3]$ \hspace{.8cm}$k(\xi,\varphi)\;[1/{\rm nm}]$} \hspace{0.1cm}\hbox{}\\\vspace{-4mm}
%
\begin{sideways}\parbox{85mm}{%
${}\quad\; t=0.055\,$\textmu$s$ \hspace{1.cm} $t=0.015\,$\textmu$s$ \hspace{1.1cm} $t=0\,$\textmu$s$ \\%
${}\quad \varepsilon^{\rm tot}=1.1\% \hspace{1.2cm} \varepsilon^{\rm tot}=0.3\% \hspace{1.1cm} \varepsilon^{\rm tot}=0.0\%$}\end{sideways}$\;$%
\includegraphics[width=.46\textwidth]{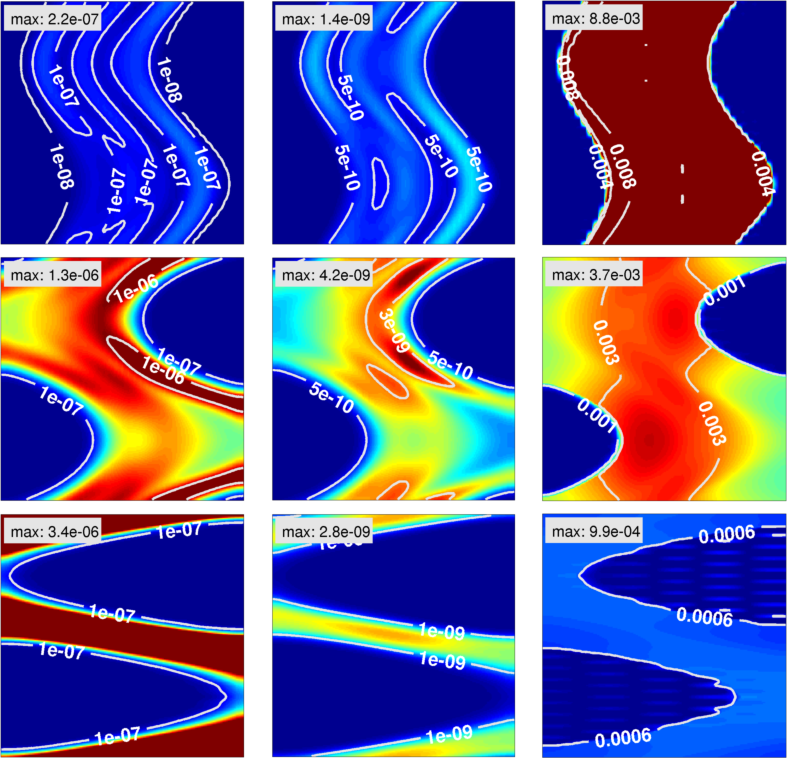}%
\hfill
\includegraphics[width=.46\textwidth]{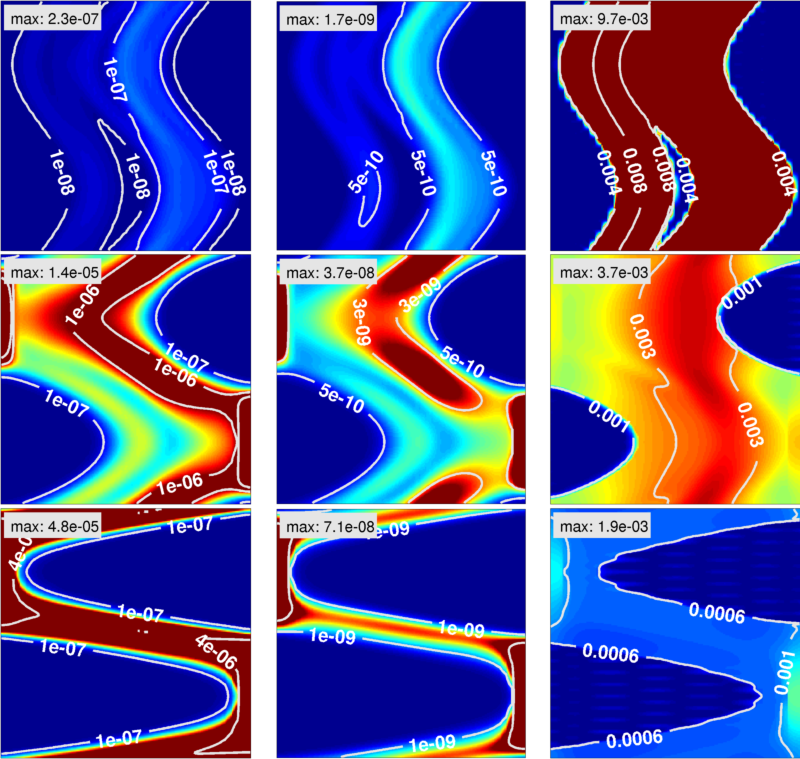}%
\hbox{}\\
\vspace{-1mm}
%
\hbox{}\hspace{7.5mm}
\includegraphics[viewport=0 0 600 60, clip, width=0.45\textwidth]{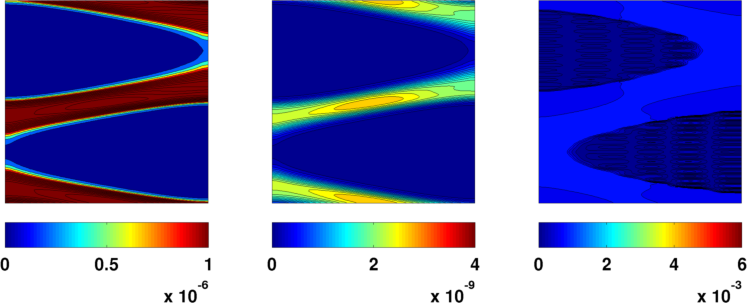}
\hfill%
\includegraphics[viewport=0 0 600 60, clip, width=0.45\textwidth]{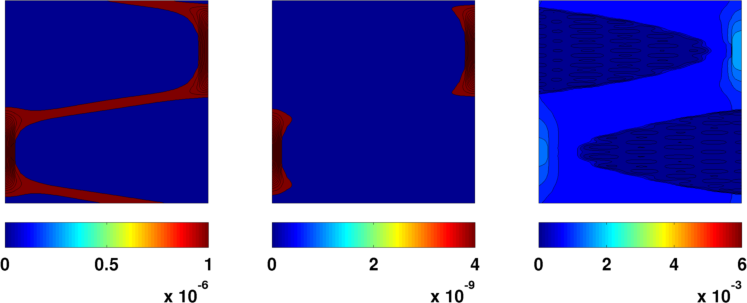}%
\hspace{0.5mm}\hbox{}\\\vspace{-3mm}
\caption{\label{fig:hdrho_open_closed_case1}Time evolution of hdCDD density
  $\rho(\xi,\varphi)$, curvature density $q(\xi,\varphi)$ and curvature
  $k(\xi,\varphi)=q(\xi,\varphi)/\rho(\xi,\varphi)$ for open (left block) and
  impenetrable boundaries (right block) for a slip plane with random initial
  values taken from the center region of
  \Figref{fig:rhot_open_closed_case1}. On the vertical axis is the line
  orientation $\varphi\in [0,2\pi]$, on the horizontal axis is the local $\xi$ coordinate, the small text label indicates the maximum field value.}
\end{figure}

We observe that the configuration with open boundaries approaches a
constant total density distribution along the slip planes while the system with impenetrable
boundaries forms pile-ups of dislocations at the boundary with constant
density values in between. To analyze this behavior we investigate the
dislocation microstructure in the higher-dimensional configuration space for a slip plane in the center of the film in more detail
(\Figref{fig:hdrho_open_closed_case1}).
The higher-dimensional fields show that for both dislocation boundary
conditions after an initial 'incubation' time the center region of the slip
plane approaches a state which is characterized by only 
\Replace{edge}{screw} 
dislocations of
positive and negative orientation: $\rho(x,\varphi)$ is approximately non-zero
only for the orientations $\varphi=\pi/2$ and $\varphi=3\pi/2$. The reason for
this is that dislocation loops expanded and segments with edge orientation
either left the film through the surface or pile up against the surface. In
any case, only screw segments are left behind which in this 2D model thread
the film into the out-of-plane direction. Investigating the curvature
$k=\rho/q$ we also see that the screw segments are nearly straight
(i.e.\ $k\approx 0$); only for the impenetrable boundaries we find a non-zero
curvature shortly before the surface: here, dislocations need to bend strongly
in order to adjust from the threading screw dislocation orientation to the
geometry of the films' surface.
\begin{figure}[htb]
\centering
\subfloat[average stress vs. total strain]{\includegraphics[width=.49\textwidth]{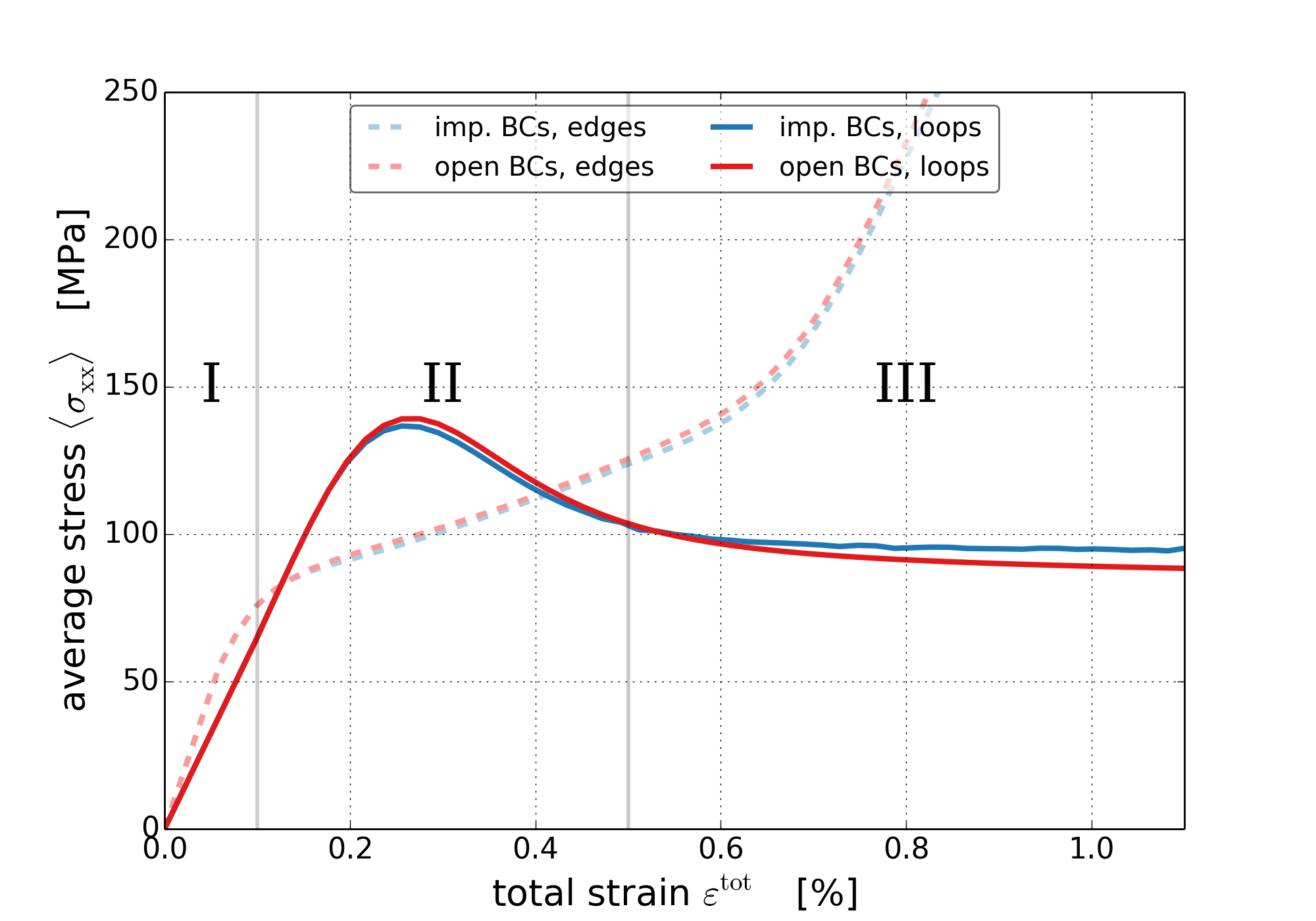}}\hfill
\subfloat[average total density vs. total strain]{\includegraphics[width=.49\textwidth]{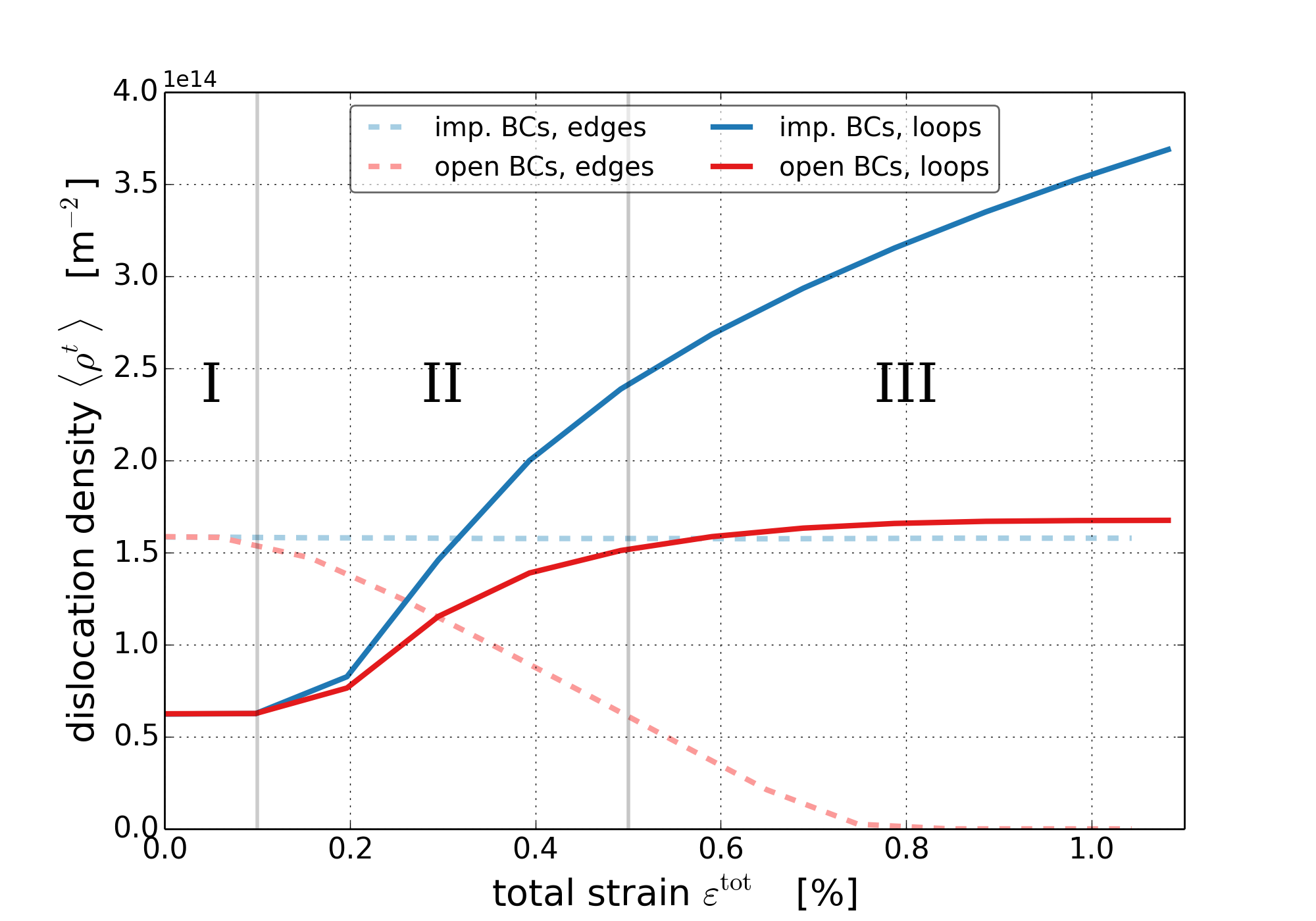}}
\caption{\label{fig:global_case1}Macroscopic quantities (Case 1) for two
  different boundary conditions (open and impenetrable) and two different
  initial configurations (consisting of either only loops or only straight
  edge dislocations). The three different time snapshots in
  \Figref{fig:hdrho_open_closed_case1} correspond to the different
  regimes I, II, and III, respectively.}
\end{figure}
This also suggests that the amount of dislocations inside the film for the
impenetrable system will be significantly higher: to begin with, dislocations
are not 'lost' by out-flux through the surfaces, and an additionally increased
line length production will take place due to the high dislocation curvature
near the surfaces. Plotting the average density evolution in
\Figref{fig:global_case1}~(b) shows that in the elastic regime~I the
dislocation density is constant, i.e.\ the resolved shear stress is not large
enough to overcome the yield stress. This is followed in regime~II by a
transition of 'free loop expansion' which results in a high dislocation
multiplication rate. Towards regime~III, edge components are then lost through
open surfaces, while for impenetrable BCs edge dislocations are deposited at
the surfaces. The open system contains at final strain a density which is
smaller roughly by a factor of 3, and additionally the average density even
reaches a stationary state (threading screw segments are straight and thus
only translate). For the system with impenetrable surfaces the average density
increases approximately linearly (caused by the constant line length increase
of deposited edges). 

What are the consequences for the macroscopic stress-strain response? A higher
dislocation density, on the one hand, obviously comes with a stronger
influence of the Taylor equation for the yield stress. On the other hand, a
higher plastic activity, where the density comes in through the Orowan
relation, causes plastic softening through the solution of the elastic
eigenstrain BVP. The competition of these two effects can be observed in the
macroscopic stress-strain curve in \Figref{fig:global_case1}~(a) where at
larger strains the obtained stress level for the system with open boundaries
is only slightly lower. Interesting to note is also the initial 'hump' right
after the elastic regime~I when softening sets in. This is caused by the fact
that at an early stage dislocation loops are still comparatively small and
thus contribute with a high line tension effect which effectively reduces the
resolved shear stress. Once loops have expanded, this contribution is reduced
and the softening behavior is sustained. A very similar behavior is also
observed in DDD simulations \citep{weygand2005}.

Finally, we compare the evolution of the initial dislocation loop distribution
to a distribution of positive and negative edge dislocations (dashed lines in
\Figref{fig:global_case1}), which would be the equivalent of a 'Groma-type model'. Therefore, we choose the number of edge
dislocations such that the initial density is similar to the saturation
density of loops with open BCs. We observe that after the onset of plastic
yield edge dislocations flow towards the surfaces (regime~II). They either get
lost through the surfaces (open BCs) or pile up against the surface (imp.\ BCs)
in which case the density simply stays constant because the number of edge
dislocations in preserved. This microstructure results in a dramatically
different stress-strain response as compared to the system with curved
dislocations: because we have neither an increase in density nor a line
tension one can only observe a linear hardening (regime~II) which -- regardless
the boundary condition -- is followed by a nearly elastic regime~III where
the stresses are considerably different from those reached for the system with
dislocation loops. The loss of dislocations through surfaces (and ultimately the 'dislocation starvation') has been experimentally observed in nano/micro pillar compression tests \citep{Shan2008,Jerusalem201293} as well. This starvation effect also gave rise to a second elastic regime as observed in our edge dislocation system. In a 3D geometry we also would find this for a distribution of dislocation loops and open BCs.


\subsection{Study 2: Spatial coarsening from Case 1 to Case 2}
\label{sec:results2}

In Case 1 we considered slip planes as quasi-discrete
objects mimicking the situation in DDD simulations. This not only requires a
very high spatial resolution for the finite element scheme, but it is also
somewhat unsatisfying from a conceptual point of view to have two different
resolutions (within the slip plane and perpendicular to it). If this can be
avoided by use of Case~2 and whether it is admissible will be studied
subsequently: for a given initial distribution of dislocations loops we
compare the asymptotic system response for $h\rightarrow 0$ in Case~1 and then
compare with results obtained for different values of ${\vartriangle} \overline s$ in Case~2.

\begin{figure}[htb]
\centering
\hbox{}\quad
\subfloat[Distribution of the stress component $\sigma_{xx}$ 
on the left half part $(0,5l)\times (0,l)$.]{\includegraphics[width=0.38\textwidth]{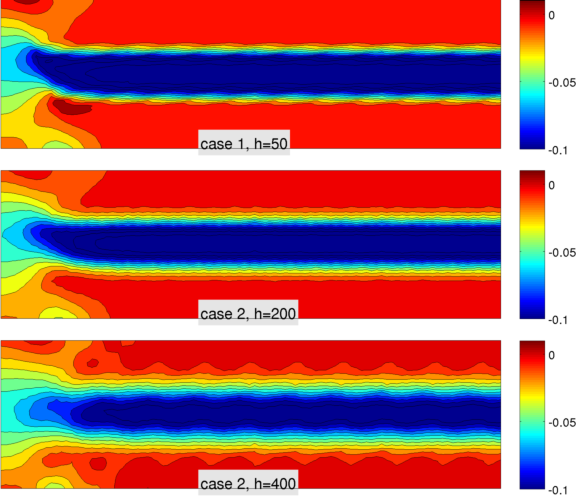}}
\hfill
\subfloat[Averaged stress $\langle \sigma_{xx}\rangle=\frac{1}{L_x}\int_{0}^{L_y}\sigma_{xx}\, {\rm d}x$
at $\varepsilon^{\rm tot}=0$.]{%
\includegraphics[width=0.5\textwidth]{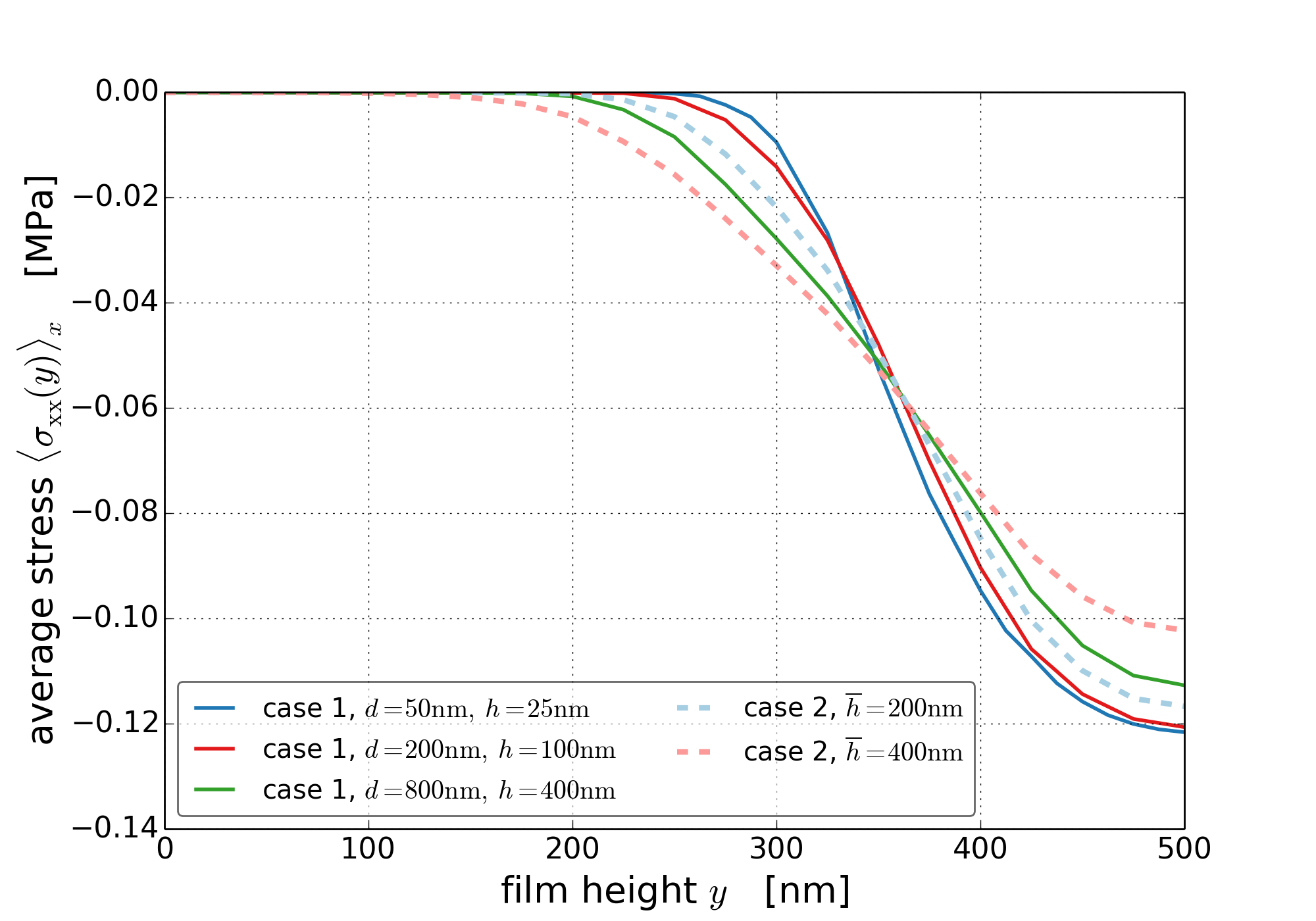}}
\quad\hbox{}
\caption{\label{fig:case1to2} Comparison of Case~1 and Case~2 
for different numerical SP heights at $t=0$ for a homogeneous distribution of
loops of the same radius. (a) shows the spatial stress distribution, (b) shows
averaged stress profiles across the height. Since the configuration is 
symmetric, only half of the profile is shown.}
\end{figure}

All system and geometry properties are the same as in the previous Study~1,
but this time we only compute stresses for a given dislocation
configuration. To make the configurations easier to compare we simply choose a
homogeneous distribution of loops.  \Figref{fig:case1to2} (a) shows the
resulting stresses for the quasi-discrete Case~1 for a SP height of $h=50$\,nm
and two different discretizations for Case~2.
If we average these stress distributions we obtain stress profiles as shown in
\Figref{fig:case1to2} (b). It shows that for values of $h=100$\,nm and below
the stresses do not change appreciably anymore. Running the same simulations
for Case 2 where the plastic strain is coarse grained in between the numerical
SPs and comparing with the results for Case 1 allows for the conclusion that
for this system a ${\vartriangle} \overline s=200$\,nm is sufficient; also the
stress-strain behavior (not shown) does not show any significant
difference. The considered situation of a homogeneous loop distribution is of
course artificial. In fact, differences during time evolution in particular
between the very coarse Case 2 and the fine Case 1 become larger if we start
with the random initial values and as the plastic slip becomes more
heterogeneous (early in regime~II).  Nonetheless, even there, our chosen
approximation of Case 2 with ${\vartriangle} \overline s=200$\,nm is
sufficient.

The advantage of the interpolation approach used in Case 2 becomes obvious
when we take a look at the degrees of freedoms and the computational time used
for the simulations from Study 1 shown in Tab.~\ref{table1}: using Case 2 with ${\vartriangle} \overline s=200$\,nm instead of Case 1 even with  ${\vartriangle} s=100$ nm, $h= 50$\,nm gives already a speedup factor of nearly 2, while at the same time no appreciable differences in the results - also for the time dependent simulations - could be observed.
\renewcommand{\arraystretch}{1.2}
\def\N{\phantom{0}}
\begin{table}[htb]
\begin{center}
\begin{tabular}{|l||c|c|}
\hline            \bf configuration   & \bf dofs (FEM + DG)    & \bf comp.\ time
in h on 8 procs.\\ \hline 
\hline Case 1, ${\vartriangle} s=\N 50$\,nm,  $h=\N 25$\,nm  &  214\,962 + 1\,049\,600 &  7:39:33 \\ 
\hline Case 1, ${\vartriangle} s=100$\,nm, $h=\N 50$\,nm  &   54\,682 + \N 262\,400 &   1:49:01 \\ 
\hline Case 1, ${\vartriangle} s=200$\,nm, $h=100$\,nm &   54\,682 + \N 131\,200 &  1:00:35 \\ 
\hline Case 2, ${\vartriangle} \overline s=200$\,nm      &   54\,682 + \N 131\,200 &  1:00:44 \\ 
\hline Case 2, ${\vartriangle} \overline s=400$\,nm      &   54\,682 + \N\N 65\,600 &  0:46:04 \\ 
\hline 
\end{tabular}
\end{center}
\caption{\label{table1}
Comparison of computational time and degrees of freedom for Case 1 and Case 2.}
\end{table}

\subsection{Study 3: A double slip configuration with Case 2}
\label{sec:study3}

\begin{figure}[htb]
\centering
\textbf{\footnotesize profiles for the single slip configuration}\\\vspace{-4mm}
\hbox{}\hfill
\subfloat[total density] {\includegraphics[width=0.3\textwidth]{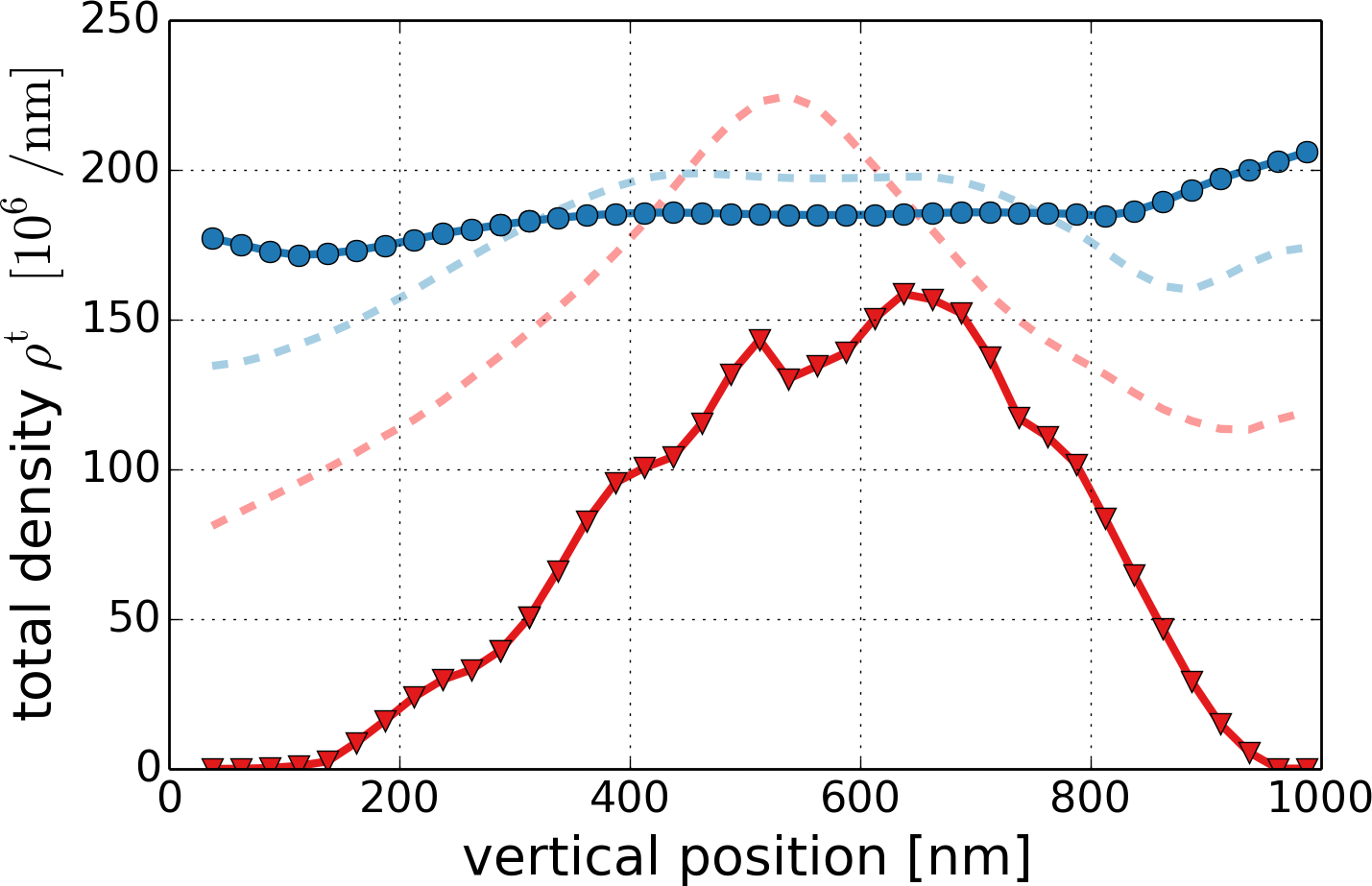}}\hfill
\subfloat[plastic slip] {\includegraphics[width=0.3\textwidth]{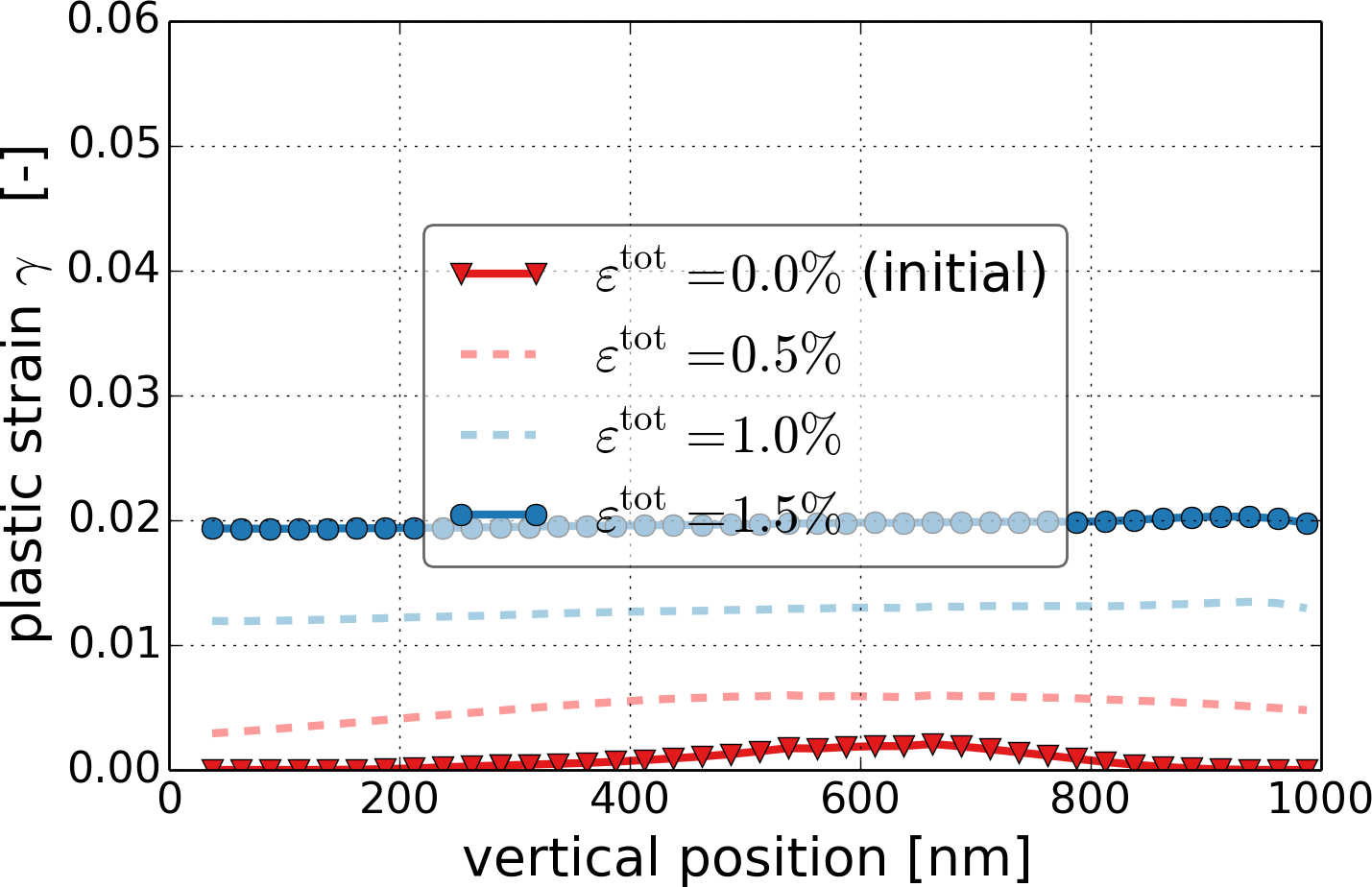}}\hfill
\subfloat[velocity]{\includegraphics[width=0.3\textwidth]{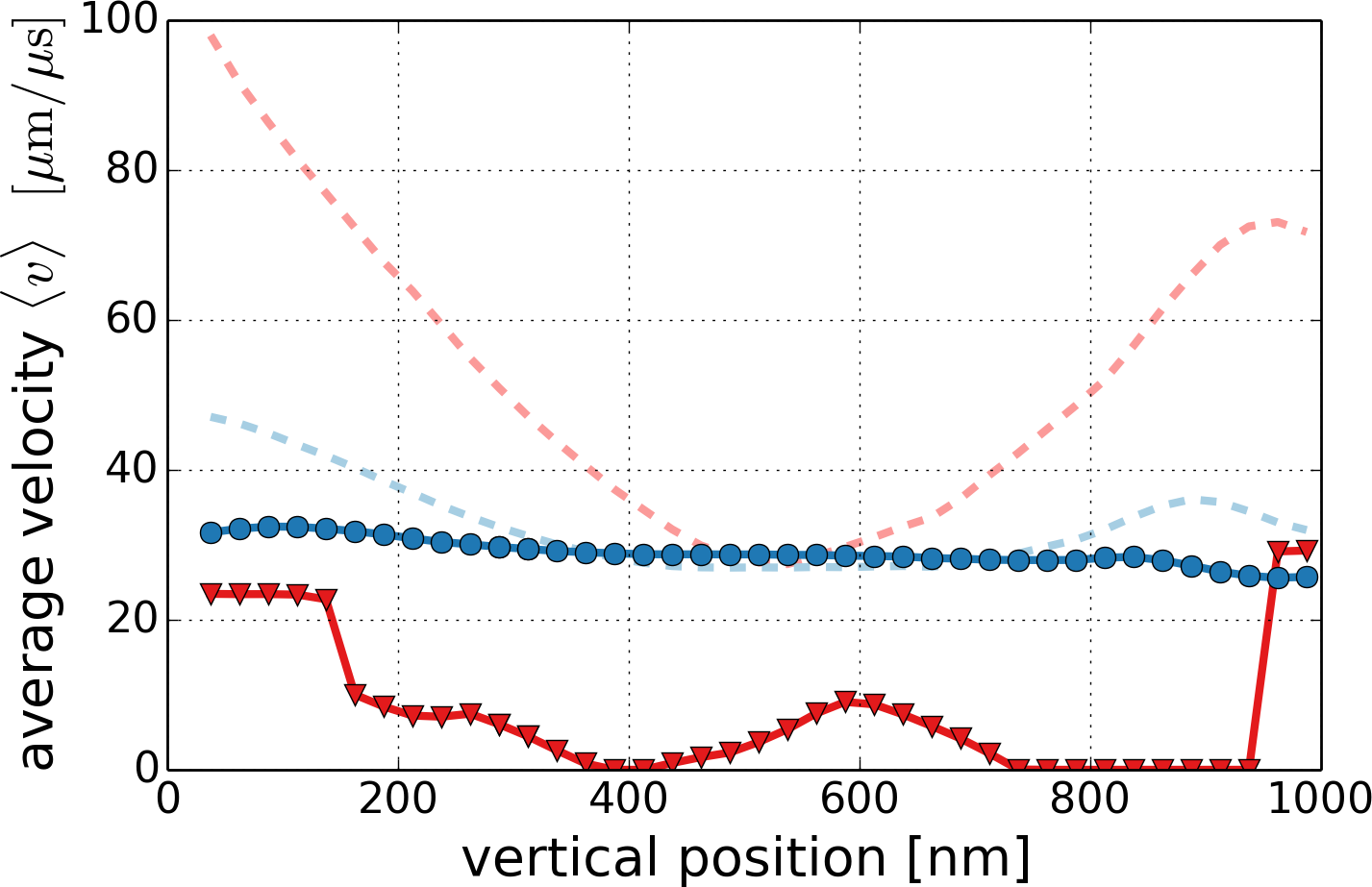}}
\hfill\hbox{}\\
\vspace{3mm}%
\textbf{\footnotesize profiles for the double slip configuration}\\\vspace{-6mm}
\hbox{}\hfill
\subfloat[total density]{\includegraphics[width=0.3\textwidth]{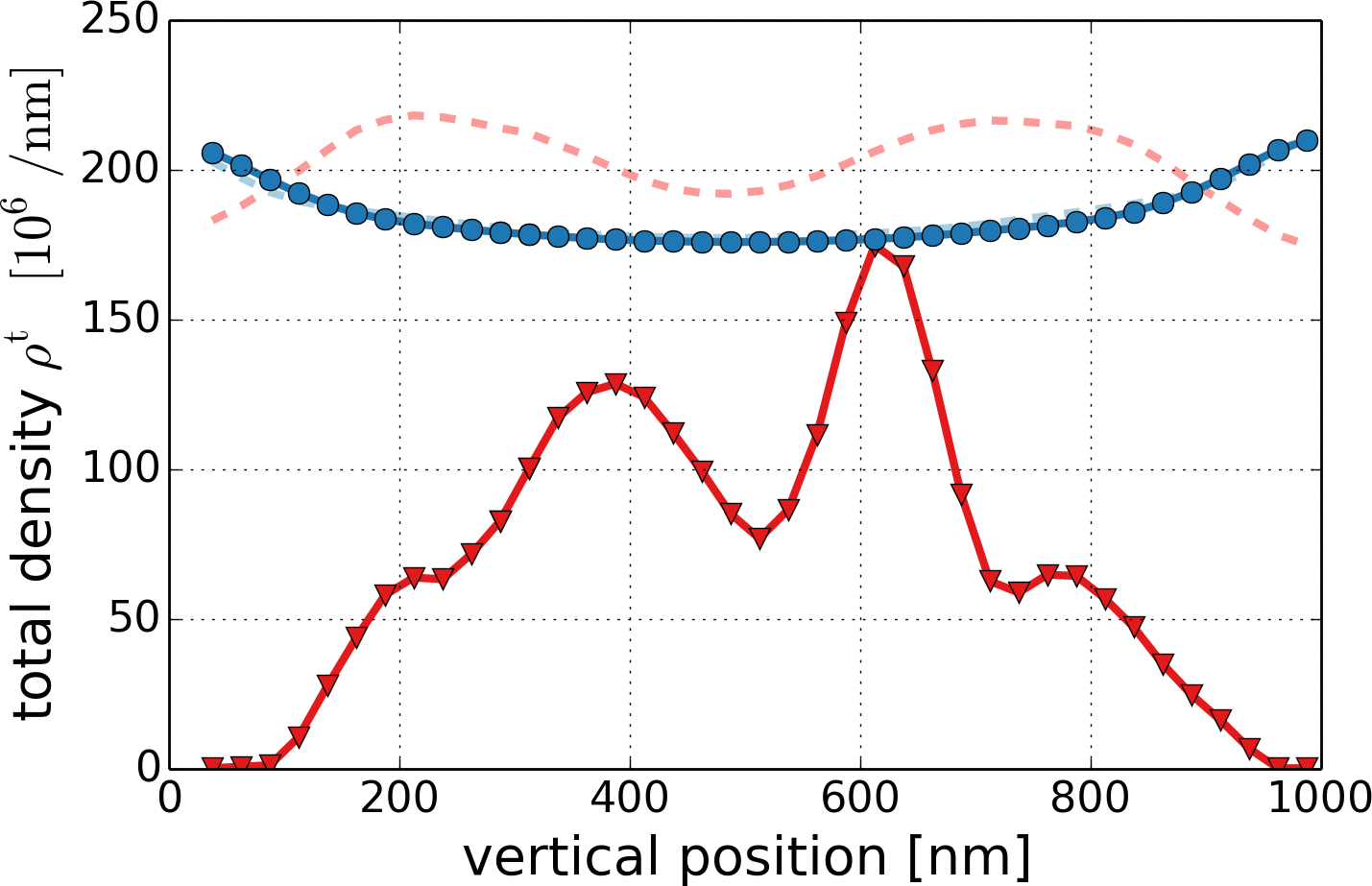}}\hfill
\subfloat[plastic slip]{\includegraphics[width=0.3\textwidth]{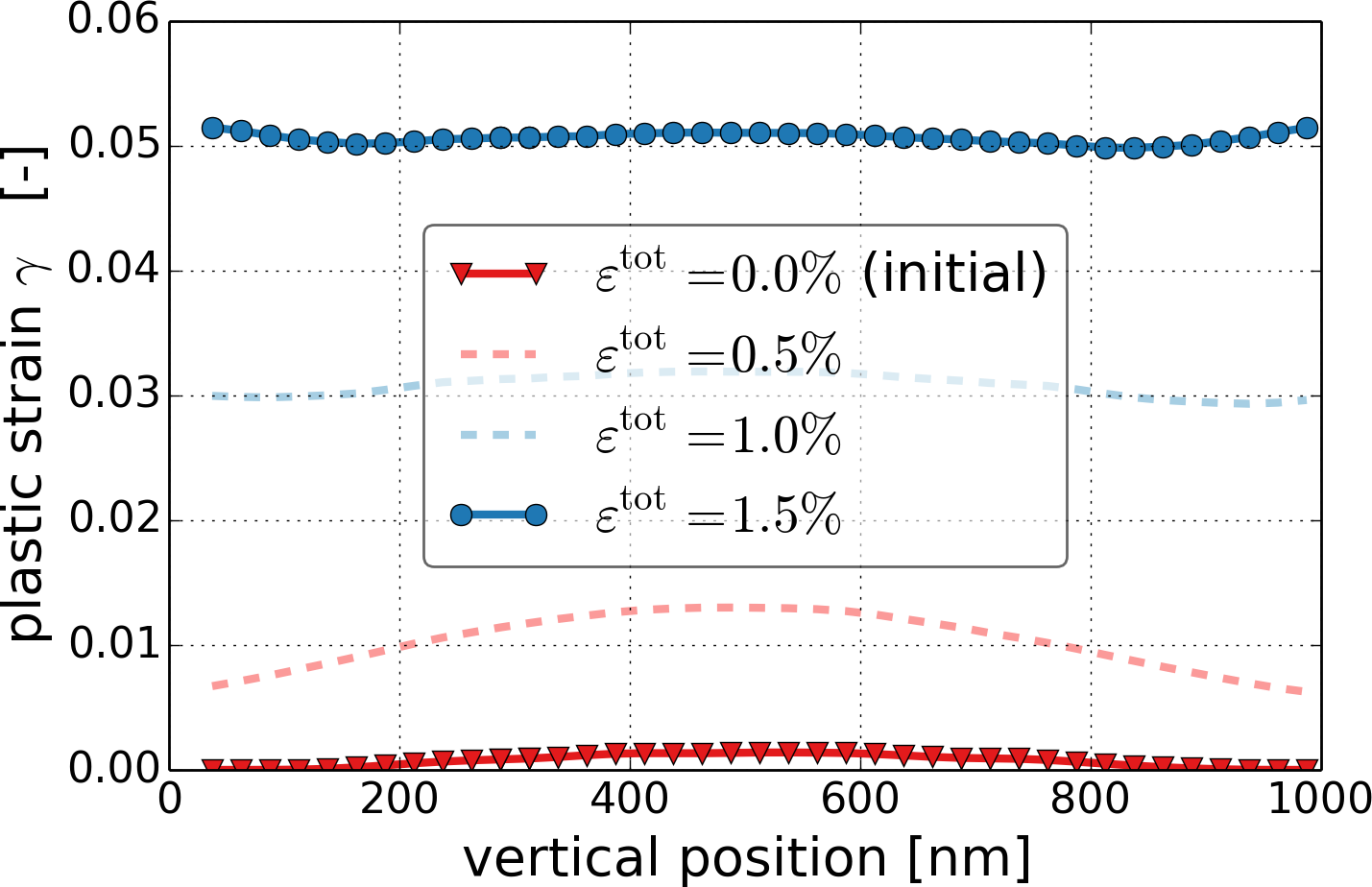}}\hfill
\subfloat[velocity] {\includegraphics[width=0.3\textwidth]{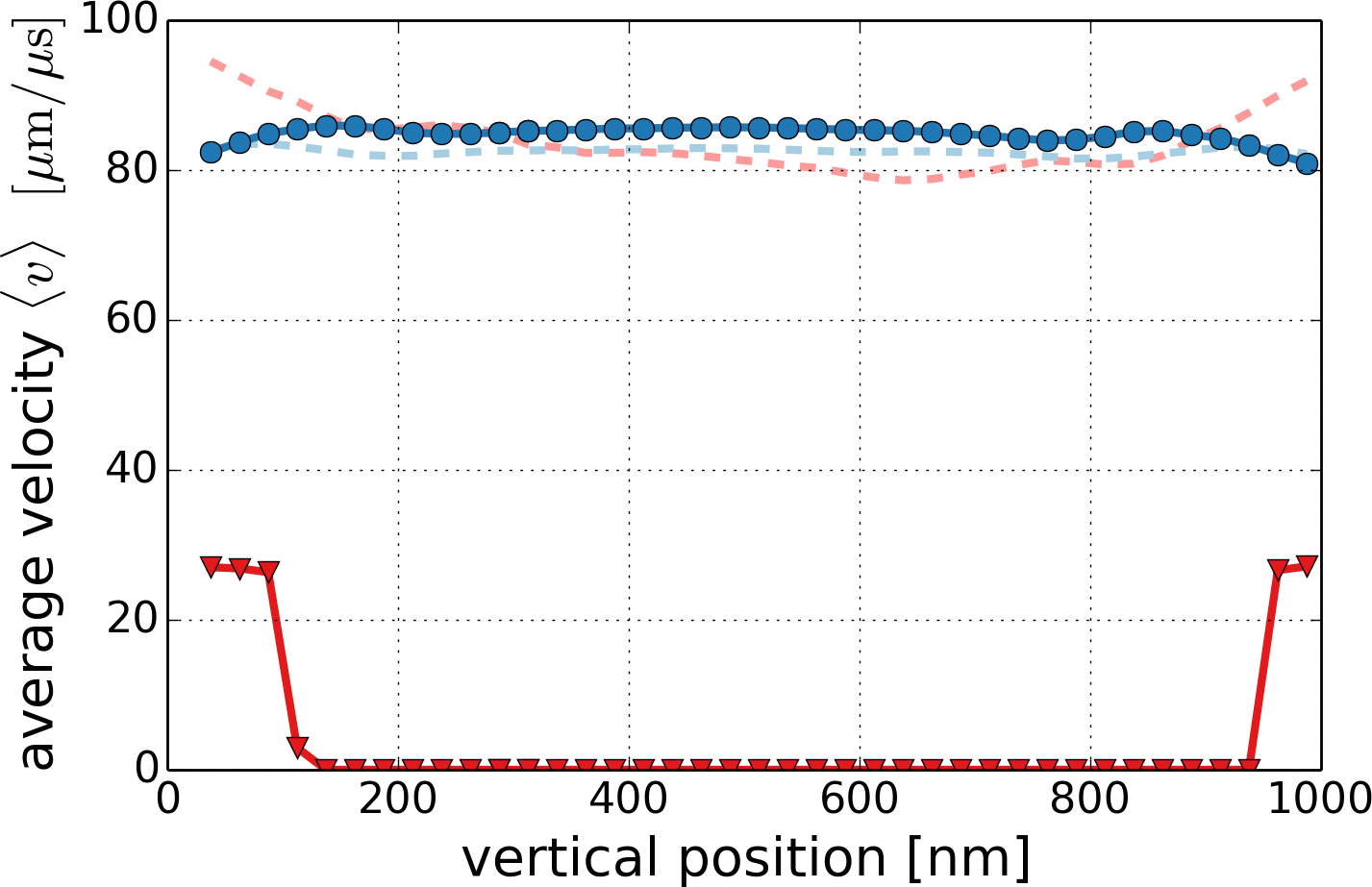}}
\hfill\hbox{}\\
\caption{Profiles of CDD field variables along a central slip plane for the single slip (a-c) and double slip configuration (d-f) for 4 different time steps. Double slip density values and plastic slip (d-f) were multiplied  by a factor of 2 to make them comparable with the single slip situation.}
\label{fig:profiles_ds}
\end{figure}

\begin{figure}[htb]
\centering
\subfloat[\label{fig:study3_sig_eps} average resolved shear stress vs. total strain]{\includegraphics[width=.49\textwidth]{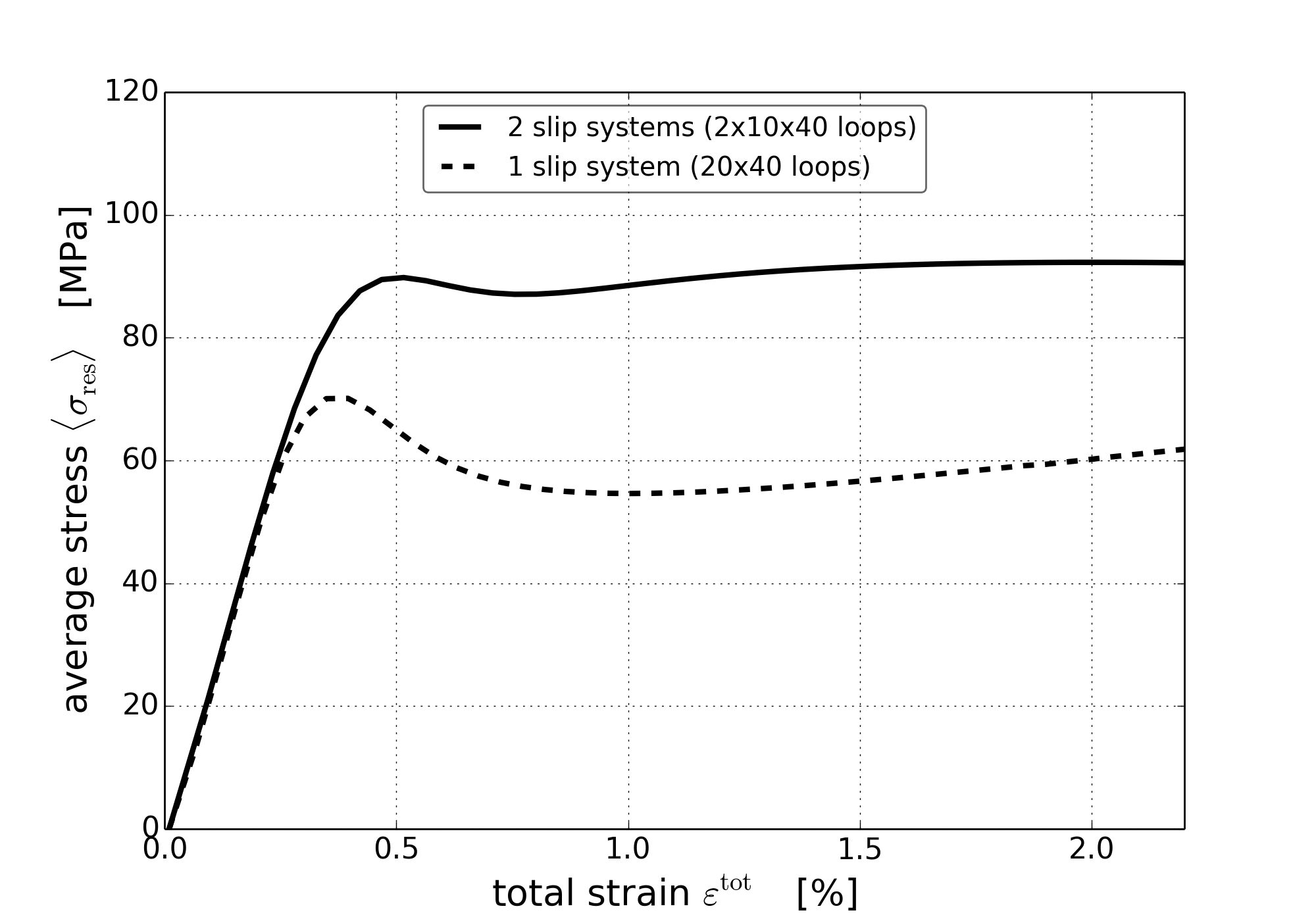}}\hfill
\subfloat[\label{fig:study3_rho_gamma} average total density vs. total strain]{\includegraphics[width=.49\textwidth]{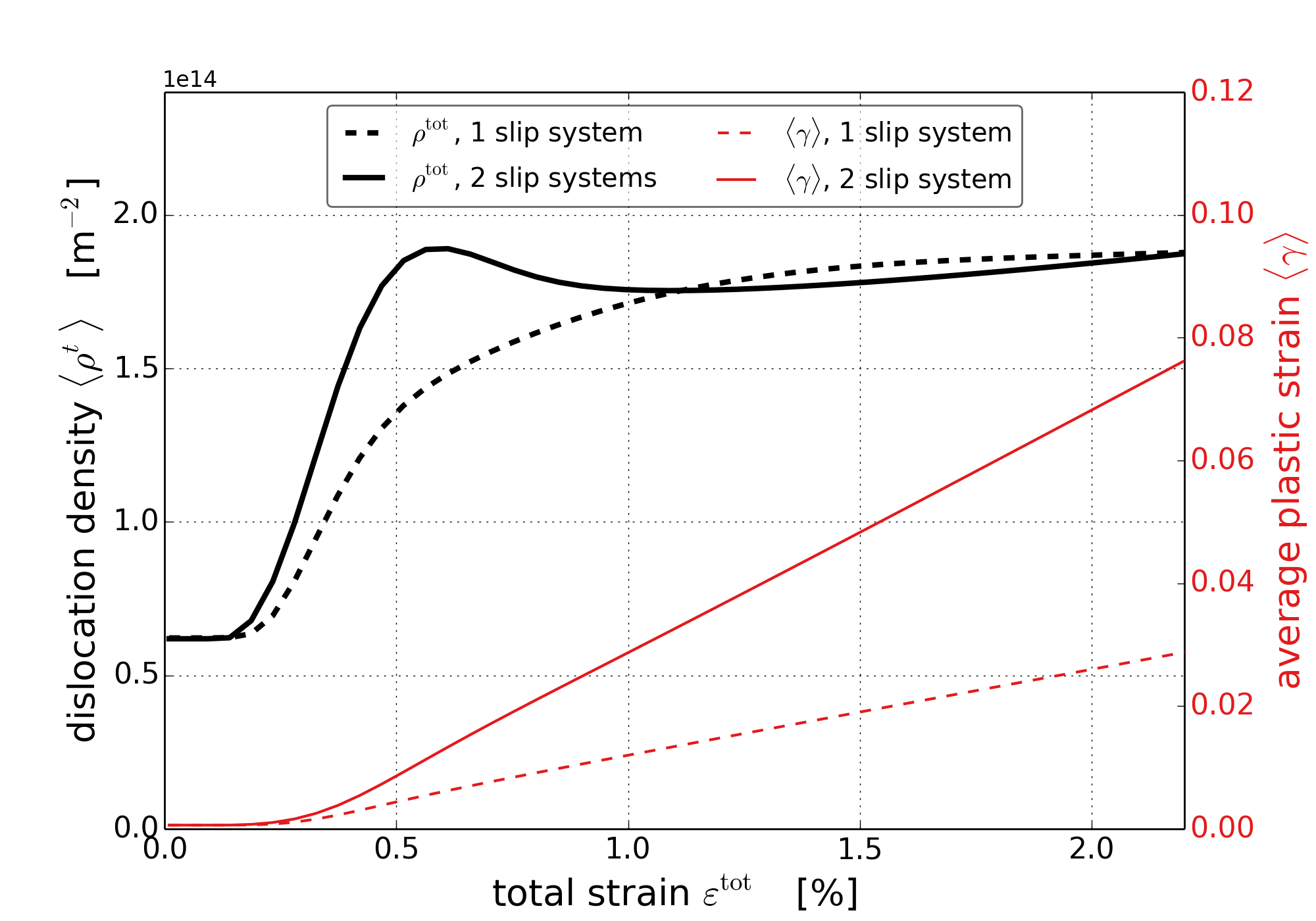}}
\footnotesize
\caption{\label{fig:study3_sig_rho}
Stress-strain plot and evolution of total density and plastic slip for the 
single slip and double slip configuration.}
\end{figure}

\begin{figure}[htb]
\centering\footnotesize
\hbox{}\hspace{3cm}\makebox(11,16)[tl]{\textbf{single slip, open boundaries}} \hfill %
\makebox(20,16)[tr]{\textbf{double slip, open boundaries}} \hspace{1.2cm}\hbox{}\vspace{-6mm}\\
\hbox{}\hspace{1.cm}{$\rho(\xi,\varphi)\;[1/{\rm nm}^2]$ \hspace{0.6cm}$q(\xi,\varphi)\;[1/{\rm nm}^3]$ \hspace{.7cm}$k(\xi,\varphi)\;[1/{\rm nm}]$}  \hfill %
{$\rho(\xi,\varphi)\;[1/{\rm nm}^2]$ \hspace{.6cm}$q(\xi,\varphi)\;[1/{\rm nm}^3]$ \hspace{.8cm}$k(\xi,\varphi)\;[1/{\rm nm}]$} \hspace{0.1cm}\hbox{}\\\vspace{-4mm}
\begin{sideways}\parbox{85mm}{%
${}\quad\; t=0.02\,$\textmu$s$ \hspace{1.cm} $t=0.01\,$\textmu$s$ \hspace{1.1cm} $t=0\,$\textmu$s$ \\%
${}\quad \varepsilon^{\rm tot}=1.0\% \hspace{1.2cm} \varepsilon^{\rm tot}=0.5\% \hspace{1.1cm} \varepsilon^{\rm tot}=0.0\%$}\end{sideways}$\;$%
\includegraphics[width=.46\textwidth]{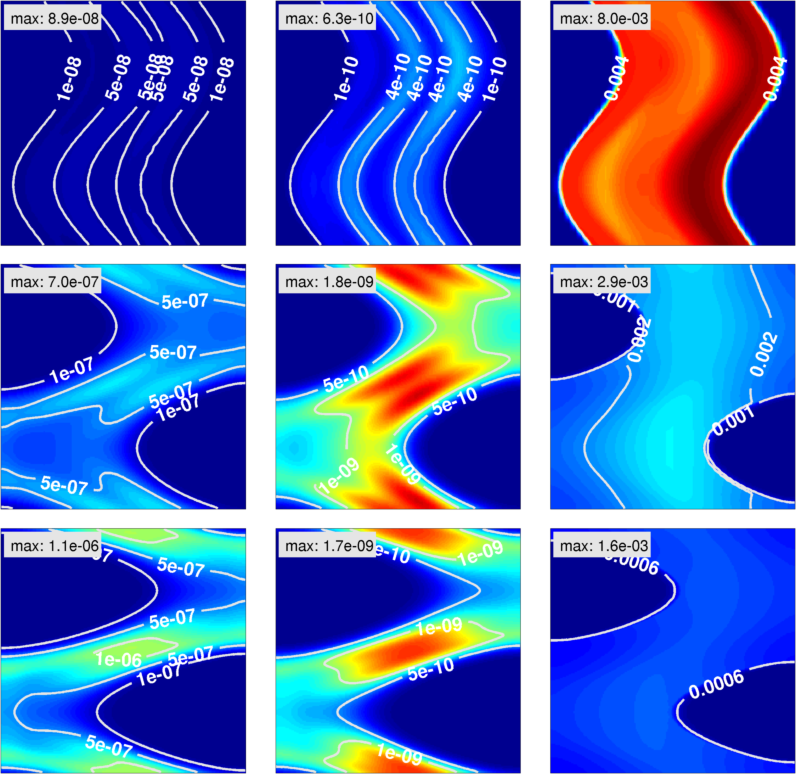}%
\hfill
\includegraphics[width=.46\textwidth]{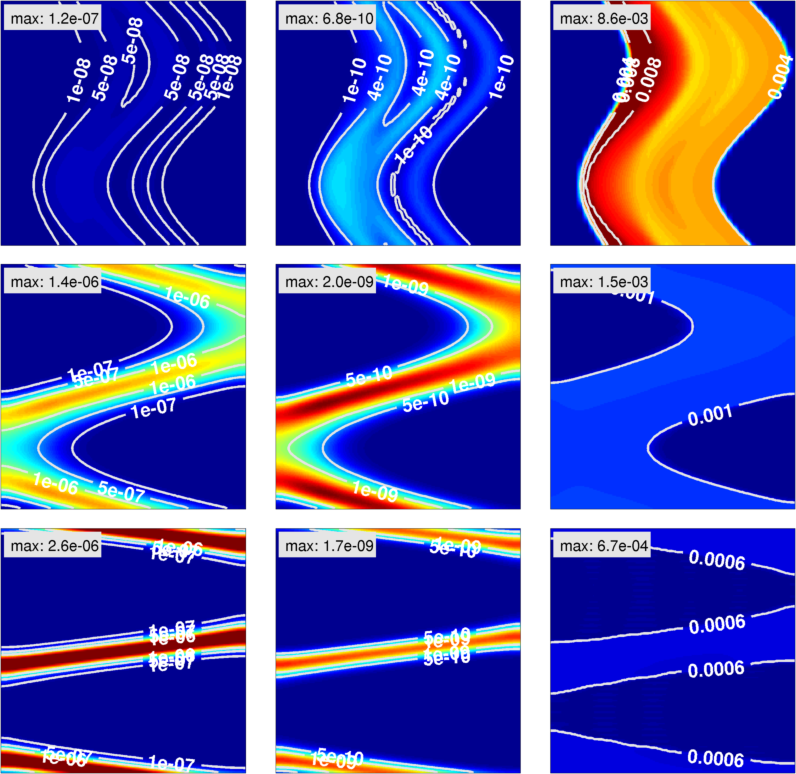}%
\hbox{}\\\vspace{-1mm}
\hbox{}\hspace{7.5mm}
\includegraphics[viewport=0 0 600 60, clip, width=0.45\textwidth]{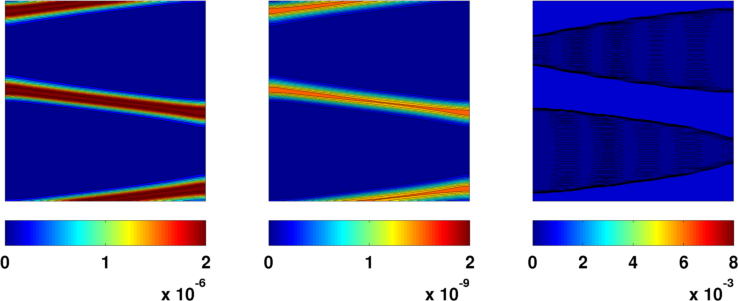}
\hfill%
\includegraphics[viewport=0 0 600 60, clip, width=0.45\textwidth]{figures/fig9/cbar.png}%
\hspace{0.5mm}\hbox{}\\\vspace{-3mm}
\caption{\label{fig:study32} Time evolution of hdCDD density
  $\rho(\xi,\varphi)$, curvature density $q(\xi,\varphi)$ and curvature
  $k(\xi,\varphi)=q(\xi,\varphi)/\rho(\xi,\varphi)$ for open boundaries, left:
  in single slip configuration (compare Study~1,
  \Figref{fig:hdrho_open_closed_case1}) and right: in double slip
  configuration with random initial values taken from the center region of
  \Figref{fig:study31}. On the vertical axis is the line orientation
  $\varphi\in[0,2\pi]$, on the horizontal axis is the local $\xi$ coordinate,
  the small text label indicate the maximum field value (comparison with
  \Figref{fig:hdrho_open_closed_case1} of Study~1).}
\end{figure}

We now investigate the macroscopic elasto-plastic response together with the
microstructural evolution in a configuration with the two slip systems
\eqref{eq:slips}. We use fully averaged dislocation distributions (Case~2 with
${\vartriangle} \overline s=200$\,nm), and the interaction of the dislocation
densities in the two systems is described by the yield stress
\begin{equation}\label{eq:Taylor}
  \tau^{\text{y}}_{s,g}=  a \mu b_s \sqrt{\rho^\text{tot}_1+\rho^\text{tot}_2}\,,
    \end{equation}
where $\rho^\text{tot}_s$ is reconstructed in $\mathcal B$ from
$\rho^\text{tot}_{s,g}$ by averaging and interpolation using the construction \eqref{eq:Case2}. \Insert{This particular form of the yield stress represents that dislocations on the one slip system act as forest for the family of dislocations on the other slip system and vice versa \citep[see e.g.][]{Kubin2008_ActaMater56}.}
The line tension is obtained in full analogy to Study 1 for
each slip system separately. Subsequently, we compare a single slip and double
slip scenario with the following initial distribution of the dislocation
density: in both cases we have altogether 800 dislocation loops in an
averaging volume for which we again assume an out-of-plane averaging length of
$L_z=1.15$\,\textmu m. The loops' radii are taken from a uniform random
distribution in the range of $[100,200]$\,nm. We either distribute them across
the two slip systems or only across a single slip system such that the average
dislocation density in the full body is always the same. In this study we only
consider open boundaries since we focus here on the investigation of the
hardening/softening effects introduced by the interaction of the slip systems
and the concomitant change in dislocation microstructure. The results are
illustrated in \Figref{fig:profiles_ds} --\ref{fig:study32}.

Initially ($\varepsilon^{\rm tot}\leq 0.2\%$) both systems respond nearly
perfectly elastic because the resolved shear stress is in almost all regions
smaller than the yield stress. This can be seen in the linear increase in
\Figref{fig:study3_sig_eps} which is a consequence of the (nearly) zero
velocity in \Figref{fig:profiles_ds} (c) and (f).

Eventually ($0.2\% < \varepsilon^{\rm tot} < 0.5\%$), starting from the outer
regions of the density distribution where $\rho^{\rm tot}$ is smaller, the
yield stress will be overcome. This results in a non-zero dislocation velocity
mainly in the surface-near regions (\Figref{fig:profiles_ds} (c) and (f) and
further plots shown in appendix B).  Already shortly after $\varepsilon^{\rm
  tot}=0.2\%$ it becomes visible that the single slip and double slip systems
behave very differently. The reason for this lies in the crystallography of
the model systems: the Burgers vectors of the symmetrically inclined slip
systems are such that under the prescribed shear deformation \Insert{the
  diagonal components in the plastic strain tensors will cancel out}. This
results in a higher resolved shear stress since the plastic softening
contribution is smaller. At the same time, however, the resulting velocity is
higher, giving rise to more dislocation activity which can be observed in the
plastic strain profile (compare \Figref{fig:profiles_ds} (b) and (e)). For the
single slip situation these relations are just the other way around: the
plastic slip reduces the resolved shear stress and thus the dislocation
velocity. The reduced plastic activity also shows in the evolution of the
dislocation density and plastic slip which happens at a lower rate than for
the double slip system, cf.~\Figref{fig:study3_rho_gamma}.

We will now take a closer look at details of the stress-strain curve
(\Figref{fig:study3_sig_eps}).  What causes the 'humps' and different maxima
for single/double slip following the elastic regime at $\varepsilon^{\rm
  tot}\approx 0.5\%$? \Figref{fig:study32} shows the higher-dimensional
density, curvature density and curvature fields. There it can be seen, that
for the single slip system the lower velocity broadens the density
distribution but does not allow for a more significant expansion of loops and
thus retards the density production. As a consequence of the reduced loop
expansion, the loops' curvature is also much higher for the single slip system,
giving rise to a more pronounced influence of the line tension (cf.\ plots in
Appendix~B); as a consequence of the retarded density production the yield
stress is effectively lower. The latter shows in \Figref{fig:study3_sig_eps}
in the lower maximum of the hump as compared to the double slip situation; the
former shows in the steeper inclination following the hump.  While loops
expand the influence of the line tension becomes smaller and tends to zero
with edge segments leaving the film and screw dislocations becoming straight
lines. The influence of the yield stress approaches a constant value once the
density from the nearly straight edge dislocations saturates. Therefore,
towards larger total strains ($>2\%$) the only difference between the single
and double slip system stems from the plastic strain, while microscopic,
short-range stresses had a stronger influence on early details of the 
stress-strain response.

\section{Summary and Conclusion}

Modeling and prediction of crystal-plasticity on the micro-meter scale
requires a faithful representation of the underlying physical mechanisms. We
introduced the higher-dimensional CDD theory as a mathematical description of
the kinematic behavior of statistically averaged ensembles of dislocations. A
special emphasis was put on a consistent geometric description of the CDD
field equations for the general case of arbitrarily oriented slip systems. In
particular the transfer of information from two-dimensional crystallographic
slip planes to the three-dimensional continuous body was concisely
formulated. Furthermore, a conservative discontinuous Galerkin scheme suitable
for the dislocation problem was derived.
These model formulations were then applied to simulate different plain-strain
slip geometries under tensile and shear loading conditions together with
different physical boundary conditions. We analyzed in detail how systems of
dislocation evolve while they interact with each other due to short-range and
long-range stress components. In doing so we could directly link the
dislocation microstructure -- represented by the higher-dimensional density
and curvature density -- to the macroscopic behavior (e.g.\ stress-strain
response or average density). By comparing to systems of straight edge
dislocations we observed that \Insert{there the line curvature has a strong influence on the initial hardening behavior which is then complemented by a sustained influence of mobile screw dislocations: the quantitative and qualitative change in the stress-strain curve could be well explained by hdCDD.} Furthermore, we showed how physical boundary conditions influence the
orientation distribution of dislocations and again, how this impacts the
system response. Finally, we studied a double slip configuration where we
could attribute the distinctly different hardening behavior for
single/double slip to microstructural aspects.

The hdCDD model contains microstructural information which otherwise is only
available in DDD simulations. These microstructural details have a strong
influence on the system response, which is why a comparison with other models
is difficult: no other standard continuum plasticity model is able to
represent e.g.\ the conversion of SSDs into GNDs and the dislocation line
length production accompanied by expansion of dislocation loops.  For this, direct
comparisons with DDD simulation will be extremely interesting and helpful to
further benchmark our model. The necessary extraction of information from DDD
simulations and conversion into continuous fields, however, is a non
trivial task which needs to be undertaken with care in future
work. 
\Insert{More realistic CDD systems will additionally require to incorporate
  further dislocation interactions and reactions as well as mechanisms for dislocation
  sources or annihilation which are not a priori included in the CDD theory
  (see e.g.\ \cite{Sandfeld2011_ICNAAM} and \cite{Zhu201419} for steps into
  this direction). DDD simulations will be very useful there as well}.  With
this we hope to go beyond what up to date is possible with DDD and apply CDD
to realistic three-dimensional systems with large numbers of interacting
dislocations and plastic strain.

\section*{Acknowledgment}

We gratefully acknowledge financial support from the Deutsche
Forschungsgemeinschaft (DFG) through Research Unit FOR1650 'Dislocation-based
Plasticity' (DFG grants Sa 2292/1-1 and Wi 1430/7-1).  The second author has
been partly supported by 'Strategic Scholarships for Frontier Research
Network' of Thailand's Commission on Higher Education (CHE-SFR), Royal Thai
Government.


\section*{Appendix A: Computation of consistent initial values}
\label{sec:appendix_IVs}

Special care has to be taken with creating initial values for CDD
simulations. 'Consistent initial values' are those which represent averaged
systems of dislocations such that the solenoidality of $\alphaII$ is not
violated, i.e.\ $\div\alphaII=0$. An additional constraint is the compatibility
between a resulting GND density and the plastic shear,
$\vec\kappa^\perp=-\frac{1}{b}\nabla\gamma_s$. One way of guaranteeing these
conditions is to create initial values in a 2-dimensional slip plane by
superposition of objects which a priori fulfill these conditions. E.g.\ one
may choose a closed dislocation loop of radius $R$ with center point $\mathbf
r_0$; a point $\mathbf r$ in $\Gamma_{s,g}$ then has the distance $d= |R(-\sin
\varphi, \cos \varphi) + \mathbf r_0 - \mathbf r|$ from the loops' line and
the density and the curvature can be obtained by 'smearing-out' the line such
that the total line length $2\pi R$ is preserved, i.e.,
\begin{eqnarray*}
  \rho(\mathbf r,\varphi) = 
  \begin{cases}
      \frac{R \exp(-1 /(1-(d/d_0)^2))}{2\pi\int_0^{d_0} t 
        \exp(-1/(1-(t/d_0)^2)) \, \mathrm dt } & 
      \quad{\rm if}\quad d < d_0\,,\\
      0 &  \quad\text{else,}
    \end{cases}
\end{eqnarray*}
where the parameter $d_0$ governs the width of the compact density
distribution around the line. Note, that this results in an area density
i.e.\ $\rho$ has the unit of line length per area. The curvature density can
be obtained as
%
$\displaystyle
    q(\mathbf r,\varphi) = \frac{\rho(\mathbf r, \varphi)}{R}
$.
%
The corresponding plastic slip generated by the expanded loop is given by
%
$\displaystyle
\gamma(\mathbf r) = b\int_{|\mathbf r - \mathbf r_0|}^{R + d_0} \int_0^{2\pi}
\rho(\eta,0,\varphi) \,{\rm d}\varphi\, {\rm d}\eta
$.
%
We superimpose $N_d$ such dislocation loops with randomly chosen centers of
the loops with a uniform distribution and such that the compact function for
the CDD fields fits completely into the slip plane. Additionally, the loops'
radii 
are chosen from a uniform random distribution. The 1D-slip planes
represent a homogeneous distribution into the second direction, so that we
integrate the CDD fields along the width for $\rho_{s,g}(\xi), q_{s,g}(\xi)$
and $\gamma_{s,g}(\xi)$, see \Figref{fig:IVs} for an visualization.


\section*{Appendix B: Evolution of CDD field quantities for Study 3}
\label{sec:appendix_profiles}

\begin{figure}[htb]
\centering
\footnotesize
\hbox{}\hspace{2cm}\makebox(15,16)[tl]{$\rho^{\text{tot}}$, single slip, $(s=1)$} 
\hspace{5cm}\includegraphics[width=0.20\textwidth]{figures/colorscale.png}\makebox(15,16)[tl]{$\times 10^{14}/{\rm m}^2$}\hfill%
\makebox(80,16)[tr]{\hspace{0.25cm}{$\rho^{\text{tot}}$, double slip $(s=1,2)$}} \hspace{0.5cm}\hbox{}\vspace{-15mm}\\
\begin{sideways}\parbox{35mm}{$\quad t=0\mu $s\\${}\;\varepsilon^{\rm tot}=0.0\%$}\end{sideways}$\;$%
\includegraphics[ viewport=0 0 700 155, clip, width=0.46\textwidth]{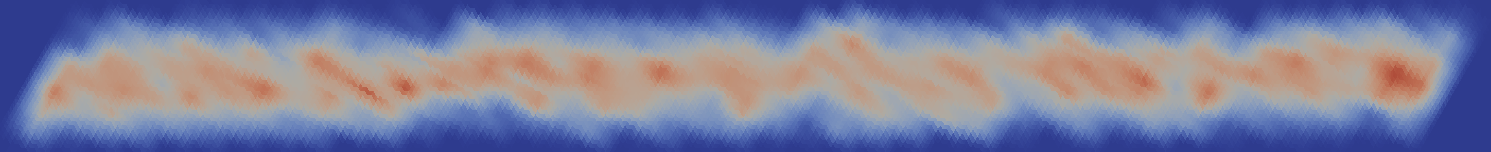}\quad
\includegraphics[viewport=0 0 700 155, clip, width=0.46\textwidth]{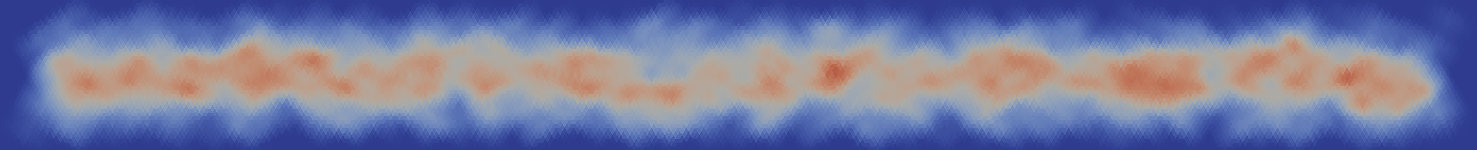}\vspace{-15mm}\\
\begin{sideways}\parbox{35mm}{$\;t=0.01\mu$ s\\${}\varepsilon^{\rm tot}=0.5\%$}\end{sideways}$\;$%
\includegraphics[viewport=0 0 700 155, clip, width=0.46\textwidth]{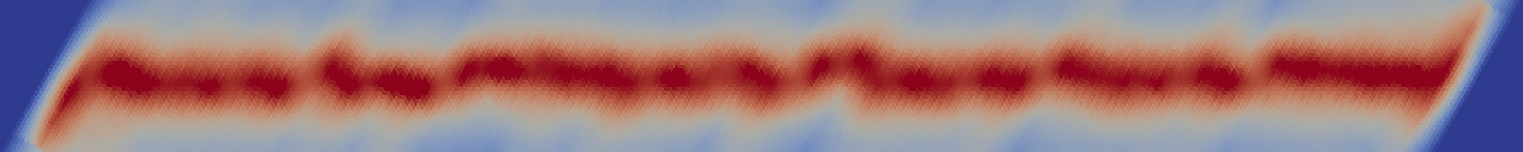}\quad
\includegraphics[viewport=0 0 700 155, clip, width=0.46\textwidth]{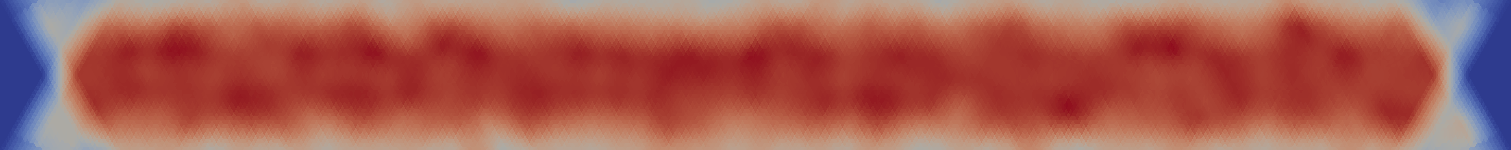}\vspace{-15mm}\\
\begin{sideways}\parbox{35mm}{$t=0.02\mu s$\\${}\varepsilon^{\rm tot}=1.0\%$}\end{sideways}$\;$%
\includegraphics[viewport=0 0 700 155, clip, width=0.46\textwidth]{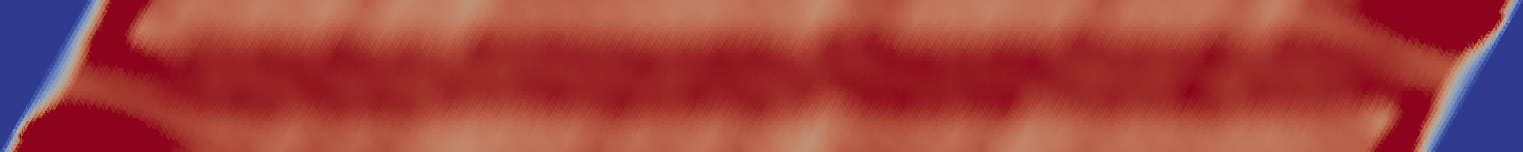}\quad
\includegraphics[viewport=0 0 700 155, clip,  width=0.46\textwidth]{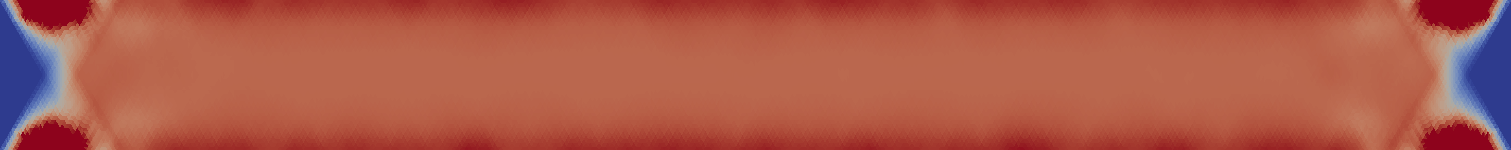}\\
\footnotesize
\caption{\label{fig:study31}
Initial distribution and time evolution of total density
$\rho_{s,g}^{\text{tot}}$ for Case 2, Study~3, for system with $s=1$ and $s=1,2$ at $t=0.01$~\textmu
s and $t = 0.02$~\textmu s with open boundary conditions for the dislocation
density.}
\end{figure}
\begin{figure}[htpb]
\centering
\begin{minipage}{0.95\textwidth}
\centering
\textbf{\footnotesize profiles for the single slip configuration}\\\vspace{-5mm}
\hbox{}\hfill
\subfloat[total density] {\includegraphics[width=0.31\textwidth]{./figures/study3_profiles/1ss/rho_tot.png}}\hfill
\subfloat[GND density (edges)] {\includegraphics[width=0.31\textwidth]{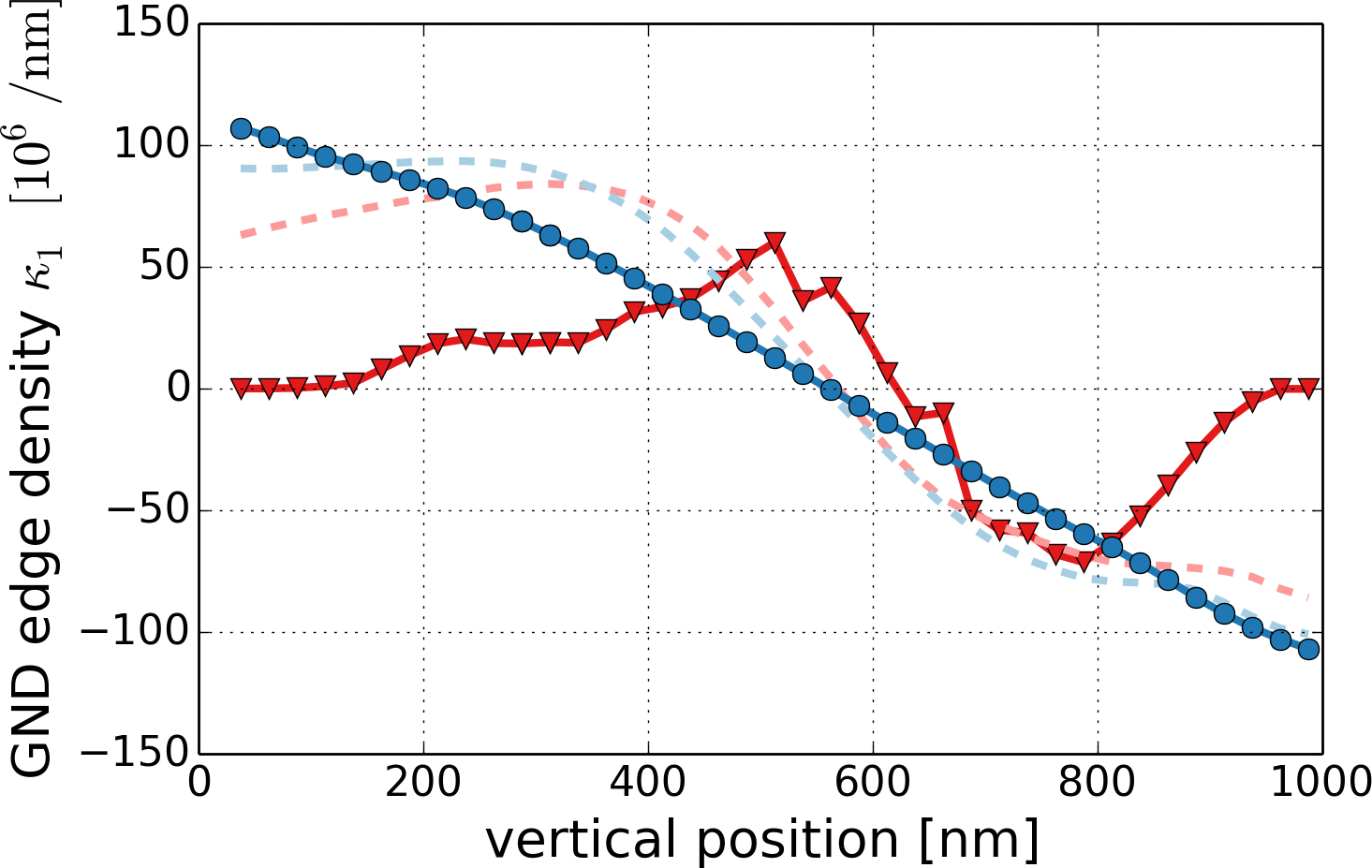}}\hfill
\subfloat[plastic slip] {\includegraphics[width=0.31\textwidth]{./figures/study3_profiles/1ss/gamma.png}}
\hfill\hbox{}\\
\hbox{}\hfill
\subfloat[line tension] {\includegraphics[width=0.31\textwidth]{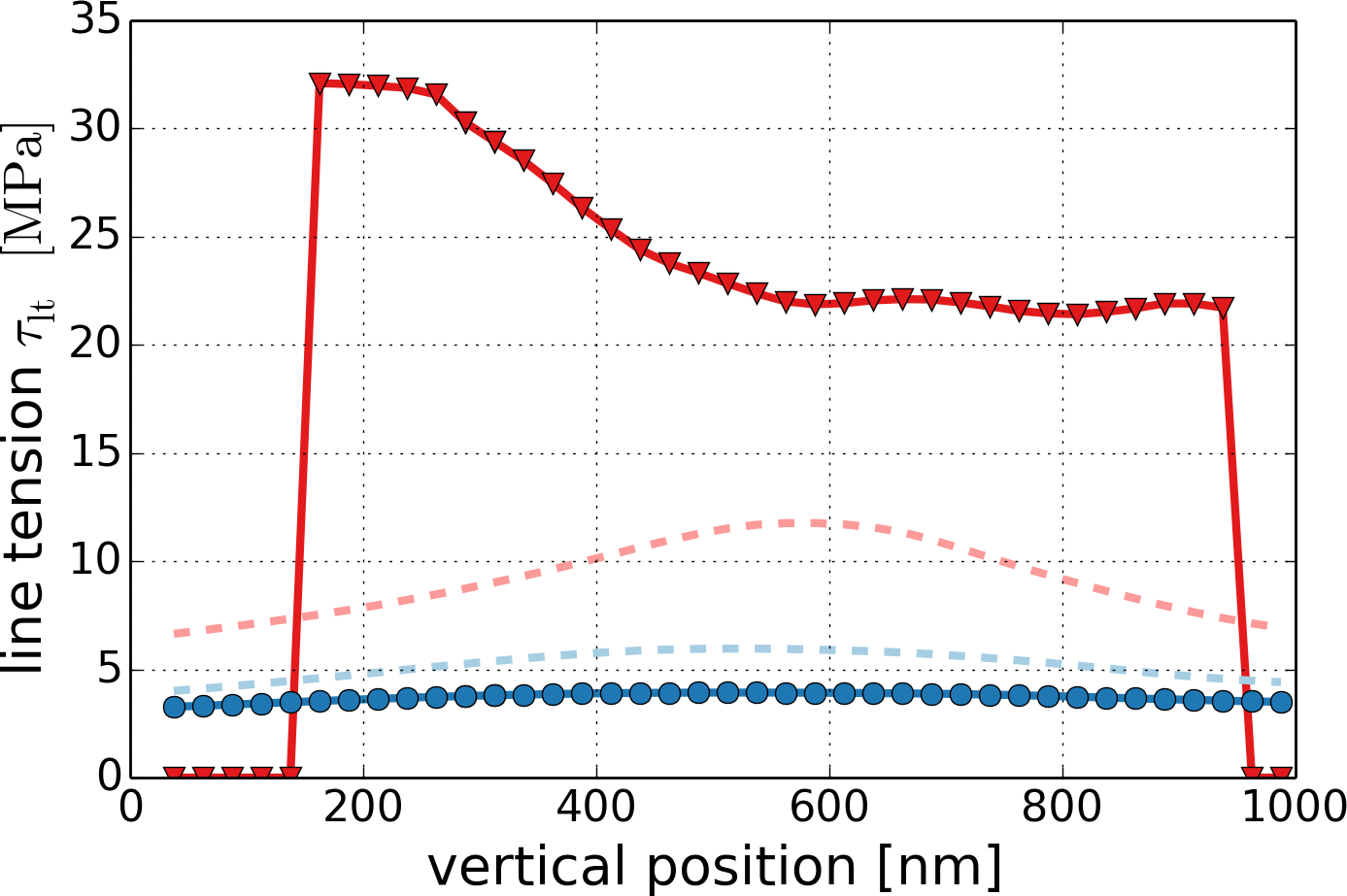}}\hfill
\subfloat[yield stress] {\includegraphics[width=0.31\textwidth]{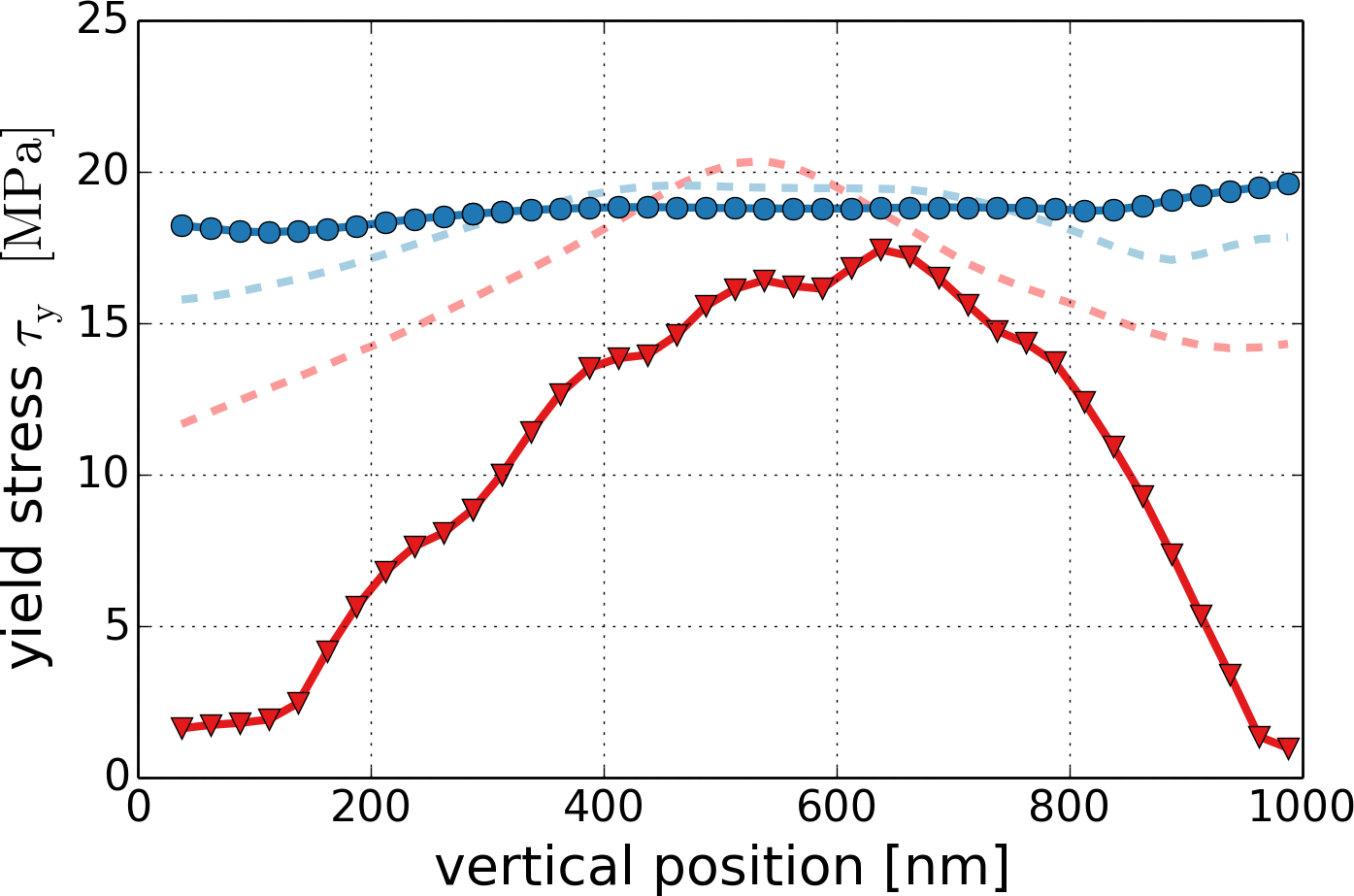}}\hfill
\subfloat[velocity]{\includegraphics[width=0.31\textwidth]{./figures/study3_profiles/1ss/velocity.png}}
\hfill\hbox{}\\
\vspace{4mm}
\textbf{\footnotesize profiles for the double slip configuration}\\\vspace{-5mm}
\hbox{}\hfill
\subfloat[total density]{\includegraphics[width=0.31\textwidth]{./figures/study3_profiles/2ss/rho_tot.png}}\hfill
\subfloat[GND density (edges)]{\includegraphics[width=0.31\textwidth]{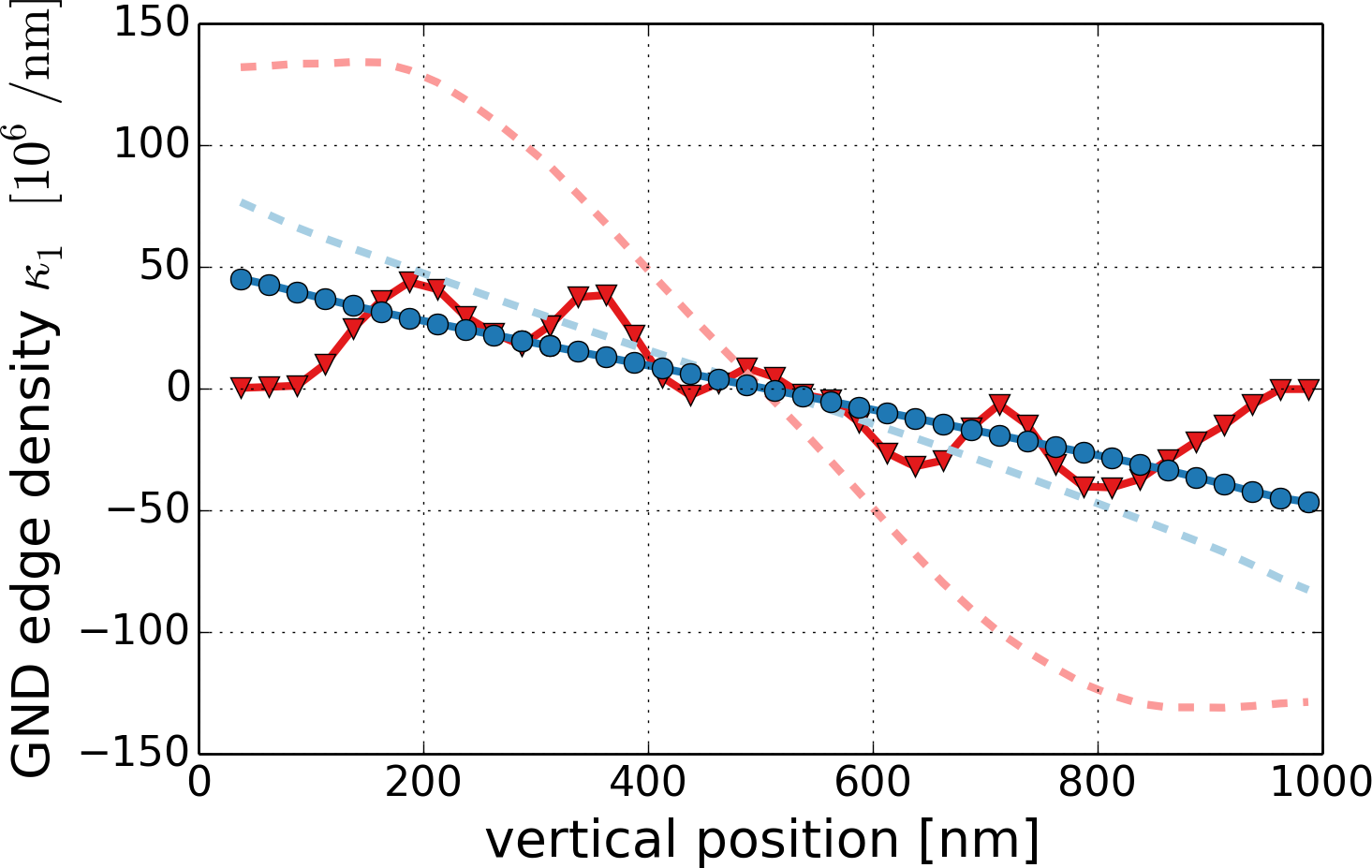}}\hfill
\subfloat[plastic slip]{\includegraphics[width=0.31\textwidth]{./figures/study3_profiles/2ss/gamma.png}}
\hfill\hbox{}\\
\hbox{}\hfill
\subfloat[line tension] {\includegraphics[width=0.31\textwidth]{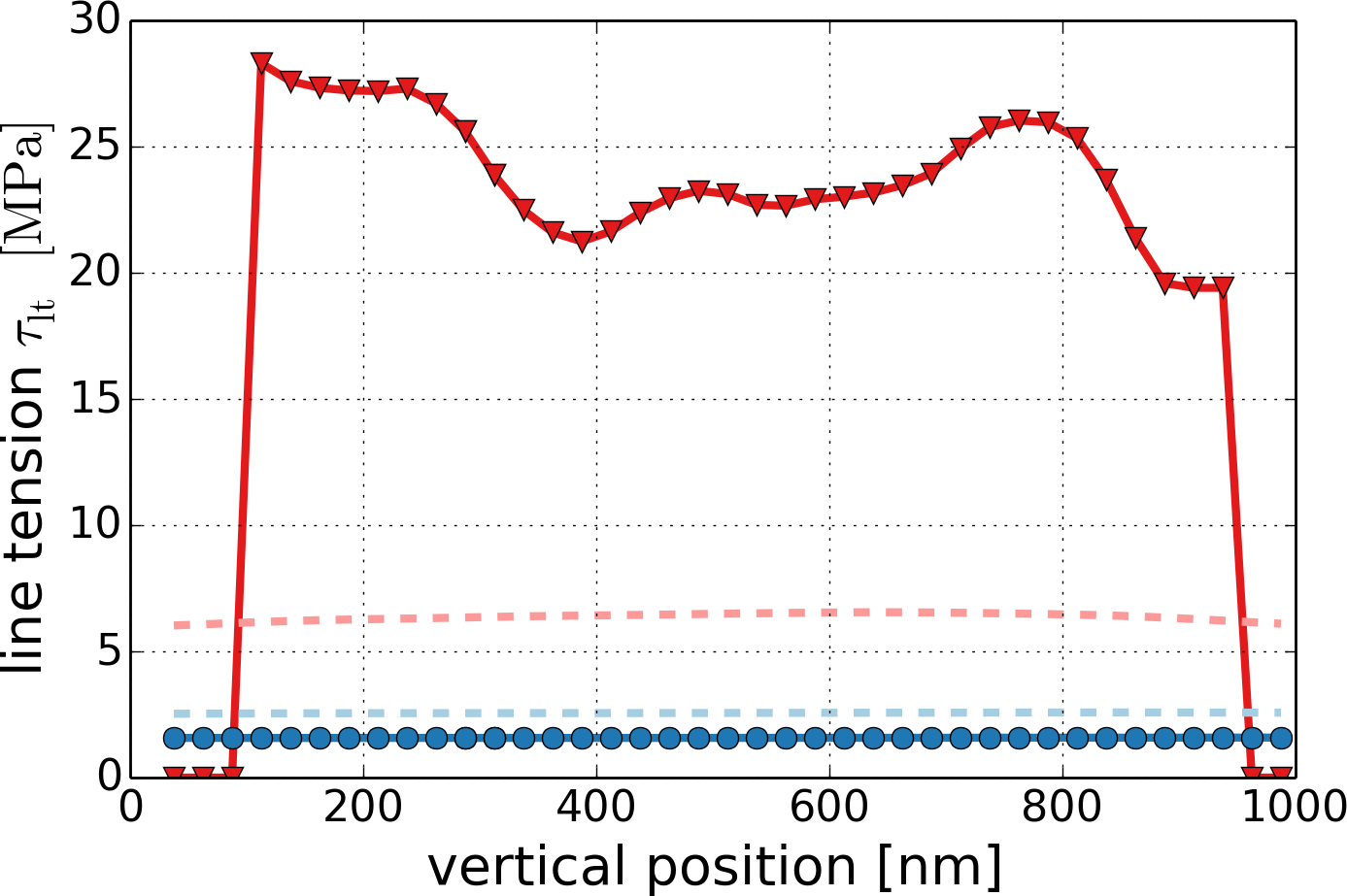}}\hfill
\subfloat[yield stress] {\includegraphics[width=0.31\textwidth]{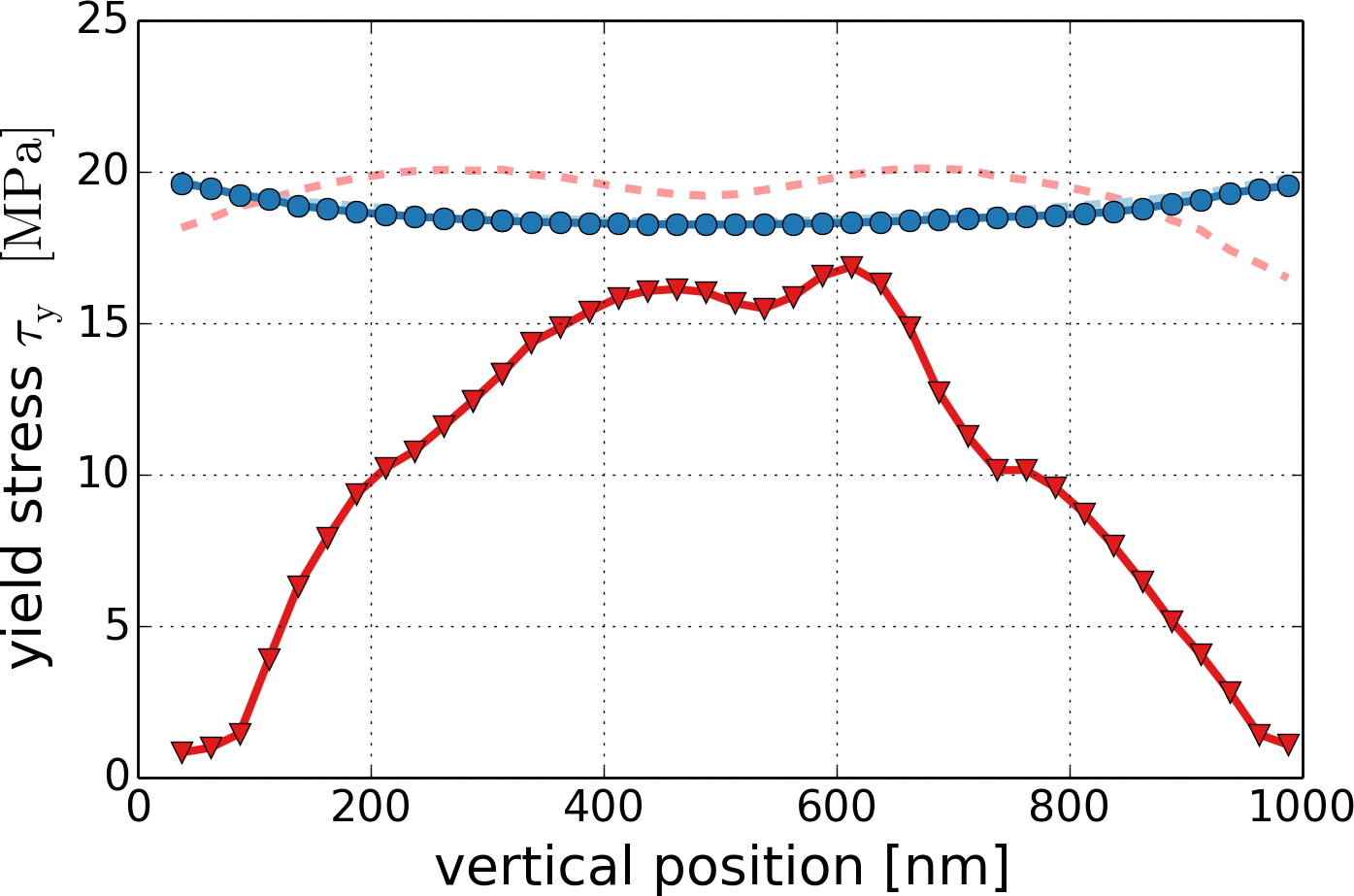}}\hfill
\subfloat[velocity] {\includegraphics[width=0.31\textwidth]{./figures/study3_profiles/2ss/velocity.png}}
\hfill\hbox{}\\

\caption{\footnotesize Profiles of CDD field variables along a center slip plane for single slip (a-f) and double slip (g-l) configuration. Density values and plastic slip for the double slip configuration (g)-(i) were multiplied  by a factor of 2 to make them comparable with the single slip configuration.}
\label{fig:app_profiles}
\end{minipage}
\end{figure}

The following figures show the evolution of total density for the single and
double slip system (\Figref{fig:study31}). The largest difference in these two
situations is that due to the different stress state in double slip the
dislocation activity is much more pronounced. In \Figref{fig:app_profiles}
also the profiles for all relevant CDD field quantities and stress components
are shown. One of the key feature of hdCDD is that the distinction into GNDs
and SSDs is obtained naturally from the higher-dimensional configuration
space. This can be also observed in the evolution of the integrated quantities
$\rho^{\rm tot}$ and $\kappa^{\rm edge}$ in \Figref{fig:app_profiles} (a)+(b)
and (g)+(h). Therein, the difference $\rho^{\rm tot} - |\kappa^{\rm edge}|$
would yield the SSD density of edges.

\small

\section*{Bibliography}
\bibliographystyle{elsarticle-harv}
\bibliography{literature_sandfeld}

\providecommand{\noopsort}[1]{}\providecommand{\singleletter}[1]{#1}%
\begin{thebibliography}{82}
\expandafter\ifx\csname natexlab\endcsname\relax\def\natexlab#1{#1}\fi
\expandafter\ifx\csname url\endcsname\relax
  \def\url#1{\texttt{#1}}\fi
\expandafter\ifx\csname urlprefix\endcsname\relax\def\urlprefix{URL }\fi

\bibitem[{Acharya and Fressengeas(2012)}]{Acharya2012}
Acharya, A., Fressengeas, C., 2012. Coupled phase transformations and
  plasticity as a field theory of deformation incompatibility. International
  Journal of Fracture 174, 87--94.

\bibitem[{Arsenlis et~al.(2007)Arsenlis, Cai, Tang, Rhee, Oppelstrup, Hommes,
  Pierce, and Bulatov}]{arsenlis2007}
Arsenlis, A., Cai, W., Tang, M., Rhee, M., Oppelstrup, T., Hommes, G., Pierce,
  T.~G., Bulatov, V.~V., 2007. Enabling strain hardening simulations with
  dislocation dynamics. Modelling and Simulation in Materials Science and
  Engineering 15, 553--595.

\bibitem[{Arsenlis et~al.(2004)Arsenlis, Parks, Becker, and
  Bulatov}]{Arsenlis2004_JMPS52}
Arsenlis, A., Parks, D.~M., Becker, R., Bulatov, V.~V., 2004. On the evolution
  of crystallographic dislocation density in non-homogeneously deforming
  crystals. Journal of the Mechanics and Physics of Solids 52~(6), 1213--1246.

\bibitem[{Arzt(1998)}]{Arzt1998_ActaMater_p5611}
Arzt, E., 1998. Size effects in materials due to microstructural and
  dimensionalconstraints: a comparative review. Acta Materialia 46~(16),
  5611--5626.

\bibitem[{Ashby(1970)}]{Ashby1970_PhilMag_p399}
Ashby, M., 1970. The deformation of plastically non-homogeneous materials.
  Philosophical Magazine 21, 399--424.

\bibitem[{Basinski and Basinski(1979)}]{Basinski79}
Basinski, S., Basinski, Z., 1979. Dislocations in Solids: Plastic Deformation
  and Work Hardening. Vol.~4. North-Holland, Amsterdam,.

\bibitem[{Bilby et~al.(1955)Bilby, Bullough, and
  Smith}]{Bilby1955_ProcRoySocLondonSerA_p263}
Bilby, B.~A., Bullough, R., Smith, E., 1955. Continuous distributions of
  dislocations: A new application of the methods of non-riemannian geometry.
  Proceedings of the Royal Society of London. Series A 231, 263--273.

\bibitem[{Bulatov and Cai(2002)}]{bulatov2002}
Bulatov, V.~V., Cai, W., 2002. Nodal effects in dislocation mobility. Physical
  Review Letters 89~(11), 115501 1--4.

\bibitem[{Deshpande et~al.({2005})Deshpande, Needleman, and Van~der
  Giessen}]{Deshpande2005}
Deshpande, V., Needleman, A., Van~der Giessen, E., {2005}. {Plasticity size
  effects in tension and compression of single crystals}. Journal of the
  Mechanics and Physics of Solids {53}~({12}), {2661--2691}.

\bibitem[{Devincre and Kubin(1997)}]{Devincre1997}
Devincre, B., Kubin, L.~P., 1997. Mesoscopic simulations of dislocations and
  plasticity. Materials Science and Engineering: A 8, 234--236.

\bibitem[{Ebrahimi et~al.(2014)Ebrahimi, Monavari, and
  Hochrainer}]{Ebrahimi2014}
Ebrahimi, A., Monavari, M., Hochrainer, T., 2014. Numerical implementation of
  continuum dislocation dynamics with the discontinuous {G}alerkin method.
  Materials Research Society Symposium Proceedings 1651.

\bibitem[{El-Azab(2000)}]{El-Azab2000_PhysRevB}
El-Azab, A., 2000. Statistical mechanics treatment of the evolution of
  dislocation distributions in single crystals. Physical Review B 61~(18),
  11956--11966.

\bibitem[{El-Azab et~al.(2007)El-Azab, Deng, and Tang}]{El-Azab_PhilMag}
El-Azab, A., Deng, J., Tang, M., 2007. Statistical characterization of
  dislocation ensembles. Philosophical Magazine 87, 1201--1223.

\bibitem[{Engels et~al.(2012)Engels, Ma, and Hartmaier}]{Engels2012159}
Engels, P., Ma, A., Hartmaier, A., 2012. Continuum simulation of the evolution
  of dislocation densities during nanoindentation. International Journal of
  Plasticity 38~(0), 159--169.

\bibitem[{Fertig and Baker(2009)}]{Fertig2009874}
Fertig, R.~S., Baker, S.~P., 2009. Simulation of dislocations and strength in
  thin films: A review. Progress in Materials Science 54~(6), 874--908.

\bibitem[{Fivel et~al.(1997)Fivel, Verdier, and Ganova}]{fivel1997}
Fivel, M., Verdier, M., Ganova, G., 1997. 3d simulation of a nanoindentation
  test at a mesoscopic scale. Materials Science and Engineering: A 234-236,
  923--926.

\bibitem[{Fleck et~al.(1994)Fleck, Muller, Ashby, and
  Hutchinson}]{Fleck1994_ActaMetallMater_p475}
Fleck, N., Muller, G., Ashby, M., Hutchinson, J., 1994. Strain gradient
  plasticity - theory and experiment. Acta Metallurgica et Materialia 42~(2),
  475--487.

\bibitem[{Foreman(1967)}]{Foreman1967_PhilMag15_p1011}
Foreman, A., 1967. Bowing of a dislocation segment. Philosophical Magazine
  15~({137}), 1011--1021.

\bibitem[{Fredriksson and Gudmundson(2005)}]{Fredriksson20051834}
Fredriksson, P., Gudmundson, P., 2005. Size-dependent yield strength of thin
  films. International Journal of Plasticity 21~(9), 1834--1854.

\bibitem[{Gao and Huang(2003)}]{Gao2003_ScrMater_p113}
Gao, H., Huang, Y., 2003. Geometrically necessary dislocation and
  size-dependent plasticity. Scripta Materialia 48~(2), 113--118.

\bibitem[{Ghoniem et~al.(2000)Ghoniem, Tong, and Sun}]{Ghoniem2000}
Ghoniem, N.~M., Tong, S., Sun, L., 2000. Parametric dislocation dynamics: A
  thermodynamics-based approach to investigations of mesoscopic plastic
  deformation. Physical Review B 61~(2), 913.

\bibitem[{Greer and De~Hosson(2011)}]{Greer:2011bb}
Greer, J.~R., De~Hosson, J. T.~M., 2011. {Plasticity in small-sized metallic
  systems: Intrinsic versus extrinsic size effect}. Progress in Materials
  Science 56~(6), 654--724.

\bibitem[{Groma(1997)}]{Groma1997_PhysRevB_p5807}
Groma, I., 1997. Link between the microscopic and mesoscopic length-scale
  description of the collective behavior of dislocations. Physical Review B
  56~(10), 5807--5813.

\bibitem[{Groma et~al.(2003)Groma, Csikor, and Zaiser}]{Groma2003_ActaMater}
Groma, I., Csikor, F., Zaiser, M., 2003. Spatial correlations and higher-order
  gradient terms in a continuum description of dislocation dynamics. Acta
  Materialia 51, 1271--1281.

\bibitem[{Gurtin(2002)}]{Gurtin2002_JMPS50_p5}
Gurtin, M.~E., 2002. A gradient theory of single-crystal viscoplasticity that
  accounts for geometrically necessary dislocations. Journal of the Mechanics
  and Physics of Solids 50~(1), 5--32.

\bibitem[{Hesthaven and Warburton(2008)}]{HesW08}
Hesthaven, J.~S., Warburton, T., 2008. Nodal discontinuous {G}alerkin methods.
  Vol.~54 of Texts in Applied Mathematics. Springer, New York.

\bibitem[{Hirschberger et~al.(2011)Hirschberger, Peerlings, and
  Brekelmans}]{Hirschberger2011_MSMSE19}
Hirschberger, C.~B., Peerlings, R. H.~J., Brekelmans, W A M~andGeers, M. G.~D.,
  2011. On the role of dislocation conservation in single-slip crystal
  plasticity. Modelling and Simullation in Materials Scicience and Engineering
  19.

\bibitem[{Hochbruck et~al.(2015)Hochbruck, Pa\v{z}ur, Schulz, Thawinan, and
  Wieners}]{Hochbruck13}
Hochbruck, M., Pa\v{z}ur, T., Schulz, A., Thawinan, E., Wieners, C., 2015.
  Efficient time integration for discontinuous {G}alerkin approximations of
  linear wave equations. Zeitschrift fur Angewandte Mathematik und Mechanik
  95~(3), 237--259.

\bibitem[{Hochrainer(2006)}]{Hochrainer2006_Dissertation}
Hochrainer, T., 2006. Evolving systems of curved dislocations: Mathematical
  foundations of a statistical theory. Ph.D. thesis, University of Karlsruhe,
  IZBS.

\bibitem[{Hochrainer et~al.(2014)Hochrainer, Sandfeld, Zaiser, and
  Gumbsch}]{Hochrainer2013_JMPS}
Hochrainer, T., Sandfeld, S., Zaiser, M., Gumbsch, P., 2014. Continuum
  dislocation dynamics: towards a physical theory of crystal plasticity.
  Journal of the Mechanics and Physics of Solids 63, 167--178.

\bibitem[{Hochrainer et~al.(2007)Hochrainer, Zaiser, and
  Gumbsch}]{Hochrainer2007_PhilMag}
Hochrainer, T., Zaiser, M., Gumbsch, P., 2007. A three-dimensional continuum
  theory of dislocation systems: kinematics and mean-field formulation.
  Philosophical Magazine 87~(8-9), 1261--1282.

\bibitem[{Hochrainer et~al.(2009)Hochrainer, Zaiser, and
  Gumbsch}]{Hochrainer2009_ICNAAM}
Hochrainer, T., Zaiser, M., Gumbsch, P., 2009. Dislocation transport and line
  length increase in averaged descriptions of dislocations. AIP Conference
  Proceedings 1168~(1), 1133--1136.

\bibitem[{Jerusalem et~al.(2012)Jerusalem, Fernandez, Kunz, and
  Greer}]{Jerusalem201293}
Jerusalem, A., Fernandez, A., Kunz, A., Greer, J.~R., 2012. Continuum modeling
  of dislocation starvation and subsequent nucleation in nano-pillar
  compressions. Scripta Materialia 66~(2), 93--96.

\bibitem[{Kondo(1952)}]{Kondo1952_Proc2JapanNatCongressofApplMech_p41}
Kondo, K., 1952. On the geometrical and physical foundations of the theory of
  yielding. Proceedings of the 2nd Japan National Congress for Applied
  Mechechanics, 41--47.

\bibitem[{Kosevich(1979)}]{Kosevich79}
Kosevich, A.~M., 1979. Crystal dislocations and the theory of elasticity. In:
  Nabarro, F.~R. (Ed.), Dislocations in Solids. Vol.~1. North Holland,
  Amsterdam, pp. 33--141.

\bibitem[{Kratochvil et~al.(2007)Kratochvil, Kruzik, and
  Sedlacek}]{Kratochvil2007_PhysRevB}
Kratochvil, J., Kruzik, M., Sedlacek, R., 2007. Statistically based continuum
  model of misoriented dislocation cell structure formation. Physical Reviews B
  75~(6), 064104--1 -- 064104--14.

\bibitem[{Kr\"oner(1955)}]{Kroener1955_ZPhys}
Kr\"oner, E., 1955. Der fundamentale zusammenhang zwischen versetzungsdichte
  und spannungsfunktion. Zeitschrift f\"ur Physik 142, 463--475.

\bibitem[{Kr{\"o}ner(1958)}]{Kroner1958}
Kr{\"o}ner, E., 1958. Kontinuumstheorie der Versetzungen und Eigenspannungen.
  Springer.

\bibitem[{Kubin and Canova(1992)}]{Kubin1992}
Kubin, L., Canova, G., 1992. The modeling of dislocation patterns. Scripta
  Metallurgica et Materialia 27~(8), 957--962.

\bibitem[{Kubin et~al.(2008)Kubin, Devincre, and Hoc}]{Kubin2008_ActaMater56}
Kubin, L., Devincre, B., Hoc, T., 2008. Modeling dislocation storage rates and
  mean free paths in face-centered cubic crystals. Acta Materialia 56,
  6040--6049.

\bibitem[{Le and G\"unther(2014)}]{Le2014164}
Le, K., G\"unther, C., 2014. Nonlinear continuum dislocation theory revisited.
  International Journal of Plasticity 53~(0), 164--178.

\bibitem[{Leung et~al.(2015)Leung, Leung, Cheng, and Ngan}]{Leung20151}
Leung, H., Leung, P., Cheng, B., Ngan, A., 2015. A new
  dislocation-density-function dynamics scheme for computational crystal
  plasticity by explicit consideration of dislocation elastic interactions.
  International Journal of Plasticity 67~(0), 1--25.

\bibitem[{Li et~al.(2014)Li, Zbib, Sun, and Khaleel}]{Li20143}
Li, D., Zbib, H., Sun, X., Khaleel, M., 2014. Predicting plastic flow and
  irradiation hardening of iron single crystal with mechanism-based continuum
  dislocation dynamics. International Journal of Plasticity 52~(0), 3--17, in
  Honor of Hussein Zbib.

\bibitem[{Liu et~al.(2011)Liu, Zhuang, Liu, Zhao, and Zhang}]{Liu2011201}
Liu, Z., Zhuang, Z., Liu, X., Zhao, X., Zhang, Z., 2011. A dislocation dynamics
  based higher-order crystal plasticity model and applications on confined
  thin-film plasticity. International Journal of Plasticity 27~(2), 201--216.

\bibitem[{Monavari et~al.(2014)Monavari, Zaiser, and Sandfeld}]{Monavari2014}
Monavari, M., Zaiser, M., Sandfeld, S., 2014. Comparison of closure
  approximations for continuous dislocation dynamics. Materials Research
  Society Symposium Proceedings 1651.

\bibitem[{Nix and Gao(1998)}]{Nix1998_JMechandPhysSolids_p411}
Nix, W.~D., Gao, H., 1998. Indentation size effects in crystalline materials: A
  law for strain gradient plasticity. Journal of the Mechanics and Physics of
  Solids 46~(3), 411--425.

\bibitem[{Nye(1953)}]{Nye1953_ActaMetall}
Nye, J., 1953. Some geometrical relations in dislocated crystals. Acta
  Metallurgica 1, 153--162.

\bibitem[{Orowan(1934)}]{Orowan1934_ZPhys_p605}
Orowan, E., 1934. {Zur Kristallplastizit\"at}. Zeitschrift f\"ur Physik 89,
  605--659.

\bibitem[{Orowan(1940)}]{Orowan1940_ProcPhysSoc52}
Orowan, E., 1940. Problems of plastic gliding. Proceedings of the Physical
  Society 52~(1), 8.

\bibitem[{Po et~al.(2014)Po, Mohamed, Crosby, Erel, El-Azab, and
  Ghoniem}]{Po:2014tc}
Po, G., Mohamed, M.~S., Crosby, T., Erel, C., El-Azab, A., Ghoniem, N.~M.,
  2014. Recent progress in discrete dislocation dynamics and its applications
  to micro plasticity. The Journal of The Minerals, Metals \& Materials
  Society, 1--13.

\bibitem[{Polanyi(1934)}]{Polanyi1934_ZPhys_p660}
Polanyi, M., 1934. {\"U}ber eine {A}rt {G}itterst{\"o}rung, die einen
  {K}ristall plastisch machen k{\"o}nnte. Zeitschrift f\"ur Physik 89,
  660--664.

\bibitem[{Reuber et~al.(2014)Reuber, Eisenlohr, Roters, and
  Raabe}]{Reuber2014333}
Reuber, C., Eisenlohr, P., Roters, F., Raabe, D., 2014. Dislocation density
  distribution around an indent in single-crystalline nickel: Comparing
  nonlocal crystal plasticity finite-element predictions with experiments. Acta
  Materialia 71~(0), 333--348.

\bibitem[{Sandfeld(2010)}]{Sandfeld_diss}
Sandfeld, S., 2010. The evolution of dislocation density in a higher-order
  continuum theory of dislocation plasticity. Shaker Verlag Aachen, Germany.

\bibitem[{Sandfeld and Hochrainer(2011)}]{Sandfeld2011_ICNAAM}
Sandfeld, S., Hochrainer, T., 2011. Towards {F}rank-{R}ead sources in the
  continuum dislocation dynamics theory. AIP Conference Proceedings 1389.

\bibitem[{Sandfeld et~al.(2010)Sandfeld, Hochrainer, Zaiser, and
  Gumbsch}]{sandfeld_etal10}
Sandfeld, S., Hochrainer, T., Zaiser, M., Gumbsch, P., 2010. Numerical
  implementation of a 3d continuum theory of dislocation dynamics and
  application to microbending. Philosophical Magazine 90~(27), 3697--3728.

\bibitem[{Sandfeld et~al.(2011)Sandfeld, Hochrainer, Zaiser, and
  Gumbsch}]{Sandfeld_JMR}
Sandfeld, S., Hochrainer, T., Zaiser, M., Gumbsch, P., 2011. Continuum modeling
  of dislocation plasticity: Theory, numerical implementation and validation by
  discrete dislocation simulations. Journal of Materials Research 26~(5),
  623--632.

\bibitem[{Sandfeld et~al.(2013)Sandfeld, Monavari, and Zaiser}]{Sandfeld2013}
Sandfeld, S., Monavari, M., Zaiser, M., 2013. From systems of discrete
  dislocations to a continuous field description: stresses and averaging
  aspects. Modelling and Simulation in Materials Science and Engineering
  21~(8), 085006.

\bibitem[{Sandfeld et~al.(2015)Sandfeld, Verbeke, and
  Devincre}]{Sandfeld2015_MRS}
Sandfeld, S., Verbeke, V., Devincre, B., 2015. Orientation-dependent pattern
  formation in a 1.5d continuum model of curved dislocations. Materials
  Research Society Symposium Proceedings 1755.

\bibitem[{Scardia et~al.(2014)Scardia, Peerlings, Peletier, and
  Geers}]{Scardia2014_JMPS70}
Scardia, L., Peerlings, R., Peletier, M., Geers, M., 2014. Mechanics of
  dislocation pile-ups: A unification of scaling regimes. Journal of the
  Mechanics and Physics of Solids 70~(1), 42--61.

\bibitem[{Schulz et~al.(2014)Schulz, Schmitt, Dickl, Sandfeld, Weygand, and
  Gumbsch}]{Schulz2013}
Schulz, K., Schmitt, S., Dickl, D., Sandfeld, S., Weygand, D., Gumbsch, P.,
  2014. Analysis of dislocation pile-ups using a dislocation-based continuum
  theory. Modelling and Simulation in Materials Science and Engineering 22~(2),
  025008.

\bibitem[{Schwarz et~al.({2005})Schwarz, Sedlacek, and
  Werner}]{Schwarz2005_MSEA}
Schwarz, C., Sedlacek, R., Werner, E., {2005}. {Application of a continuum
  dislocation-based model to a tensile test on a thin film}. Materials Science
  and Engineering: A {400}, {443--447}, {Intenational Conference on
  Fundamentals of Plastic Deformation, La Colle sur Loup, France, Sep. 13-17,
  2004}.

\bibitem[{Sedl{\'a}{\v{c}}ek et~al.(2003)Sedl{\'a}{\v{c}}ek, Kratochv\'{\i}l,
  and Werner}]{sedlacek_kw03}
Sedl{\'a}{\v{c}}ek, R., Kratochv\'{\i}l, J., Werner, E., 2003. The importance
  of being curved. Philosophical Magazine 83~(31-34), 3735--3752.

\bibitem[{Sedl\'a\v{c}ek et~al.(2003)Sedl\'a\v{c}ek, Kratochv\'il, and
  Werner}]{Sedlacek2003_PhilMag}
Sedl\'a\v{c}ek, R., Kratochv\'il, J., Werner, E., 2003. The importance of being
  curved: bowing dislocations in a continuum description. Philosophical
  Magazine 83~(31-34), 3735--3752.

\bibitem[{Shan et~al.(2008)Shan, Mishra, Syed~Asif, Warren, and
  Minor}]{Shan2008}
Shan, Z.~W., Mishra, R.~K., Syed~Asif, S.~A., Warren, O.~L., Minor, A.~M.,
  2008. Mechanical annealing and source-limited deformation in
  submicrometre-diameter ni crystals. Nature Materials 7~(2), 115--119.

\bibitem[{Stolken and Evans(1998)}]{Stolken1998_ActaMater_p5109}
Stolken, J., Evans, A., 1998. A microbend test method for measuring the
  plasticity length scale. Acta Materialia 46, 5109--5115.

\bibitem[{Taupin et~al.(2013)Taupin, Capolungo, Fressengeas, Das, and
  Upadhyay}]{Taupin2013370}
Taupin, V., Capolungo, L., Fressengeas, C., Das, A., Upadhyay, M., 2013. Grain
  boundary modeling using an elasto-plastic theory of dislocation and
  disclination fields. Journal of the Mechanics and Physics of Solids 61~(2),
  370--384.

\bibitem[{Taylor(1934{\natexlab{a}})}]{Taylor1934_ProcRoySocA_p362}
Taylor, G., 1934{\natexlab{a}}. The mechanism of plastic deformation of
  crystals. Proceedings of the Royal Society of London. Series A, Mathematical
  and Physical Sciences 145, 362--415.

\bibitem[{Taylor(1934{\natexlab{b}})}]{Taylor1934_ProcRoySocA145}
Taylor, G.~I., 1934{\natexlab{b}}. The mechanism of plastic deformation of
  crystals. part {I}. theoretical. Proceedings of the Royal Society of London
  A: Mathematical, Physical and Engineering Sciences 145~(855), 362--387.

\bibitem[{Wallin et~al.(2008)Wallin, Curtin, Ristinmaa, and
  Needleman}]{Wallin20083167}
Wallin, M., Curtin, W., Ristinmaa, M., Needleman, A., 2008. Multi-scale
  plasticity modeling: Coupled discrete dislocation and continuum crystal
  plasticity. Journal of the Mechanics and Physics of Solids 56~(11),
  3167--3180.

\bibitem[{Weygand et~al.(2002)Weygand, Friedman, van~der Giessen, and
  Needleman}]{weygand2002}
Weygand, D., Friedman, L.~H., van~der Giessen, E., Needleman, A., 2002. Aspects
  of boundary-value problem solutions with three-dimensionaldislocation
  dynamics. Modelling and Simulation in Materials Science and Engineering 10,
  437--468.

\bibitem[{Weygand and Gumbsch(2005)}]{weygand2005}
Weygand, D., Gumbsch, P., 2005. Study of dislocation reactions and
  rearrangements under different loading conditions. Modelling and Simulation
  in Engineering 400-401, 158--161.

\bibitem[{Wulfinghoff and B\"ohlke(2015)}]{Wulfinghoff2015}
Wulfinghoff, S., B\"ohlke, T., 2015. Gradient crystal plasticity including
  dislocation-based work-hardening and dislocation transport. International
  Journal of Plasticity, http://dx.doi.org/10.1016/j.ijplas.2014.12.003.

\bibitem[{Xiang(2009)}]{Xiang2009728}
Xiang, Y., 2009. Continuum approximation of the {P}each-{K}oehler force on
  dislocations in a slip plane. Journal of the Mechanics and Physics of Solids
  57~(4), 728--743.

\bibitem[{Xiong et~al.(2012)Xiong, Deng, Tucker, McDowell, and
  Chen}]{Xiong2012899}
Xiong, L., Deng, Q., Tucker, G., McDowell, D.~L., Chen, Y., 2012. A concurrent
  scheme for passing dislocations from atomistic to continuum domains. Acta
  Materialia 60~(3), 899--913.

\bibitem[{Yefimov et~al.(2004)Yefimov, Groma, and van~der
  Giessen}]{Yefimov2004_JMechandPhysSolids52}
Yefimov, S., Groma, I., van~der Giessen, E., 2004. A comparison of a
  statistical-mechanics based plasticity model with discrete dislocation
  plasticity calculations. J. Mech. and Phys. Solids 52~(2), 279 -- 300.

\bibitem[{Zaiser and Hochrainer(2006)}]{Zaiser2006_ScriptaMater_p717}
Zaiser, M., Hochrainer, T., 2006. Some steps towards a continuum representation
  of 3d dislocation systems. Scripta Materialia 54~(5), 717--721.

\bibitem[{Zaiser et~al.(2001)Zaiser, Miguel, and Groma}]{Zaiser2001_PhysRevB}
Zaiser, M., Miguel, M.-C., Groma, I., 2001. Statistical dynamics of dislocation
  systems: The influence of dislocation-dislocationcorrelations. Physical
  Review B 64, 224102.

\bibitem[{Zaiser et~al.(2007)Zaiser, Nikitas, Hochrainer, and
  Aifantis}]{Zaiser2007_PhilMag_p1283}
Zaiser, M., Nikitas, N., Hochrainer, T., Aifantis, E., 2007. Modelling size
  effects using 3d density-based dislocation dynamics. Philosophical Magazine
  87, 1283--1306.

\bibitem[{Zaiser and Sandfeld(2014)}]{Zaiser2014_ModSimMat}
Zaiser, M., Sandfeld, S., 2014. Scaling properties of dislocation simulations
  in the similitude regime. Modelling and Simulation in Materials Science and
  Engineering 22~(6).
\newline\urlprefix\url{http://iopscience.iop.org/0965-0393/22/6/065012/}

\bibitem[{Zhang et~al.(2014)Zhang, Aifantis, and Ngan}]{Zhang201438}
Zhang, X., Aifantis, K.~E., Ngan, A.~H., 2014. Interpreting the stress-strain
  response of al micropillars through gradient plasticity. Materials Science
  and Engineering: A 591~(0), 38--45.

\bibitem[{Zhou et~al.(2010)Zhou, Biner, and LeSar}]{Zhou2010}
Zhou, C.~Z., Biner, S.~B., LeSar, R., 2010. Discrete dislocation dynamics
  simulations of plasticity at small scales. Acta Materialia 58, 1565--1577.

\bibitem[{Zhu et~al.(2014)Zhu, Wang, Zhu, and Xiang}]{Zhu201419}
Zhu, Y., Wang, H., Zhu, X., Xiang, Y., 2014. A continuum model for dislocation
  dynamics incorporating {F}rank-{R}ead sources and {H}all-{P}etch relation in
  two dimensions. International Journal of Plasticity 60~(0), 19--39.

\end{thebibliography}

\end{document}